\documentclass[fleqn,usenatbib]{mnras}

\usepackage{newtxtext,newtxmath}

\usepackage[T1]{fontenc}

\DeclareRobustCommand{\VAN}[3]{#2}
\let\VANthebibliography\thebibliography
\def\thebibliography{\DeclareRobustCommand{\VAN}[3]{##3}\VANthebibliography}

\usepackage{graphicx}	
\usepackage{amsmath}	
\usepackage{multirow}

\title[Braginskii Viscosity in Cosmological Simulations]{Braginskii Viscosity in Cosmological Simulations of Galaxy Clusters: Implementation, Validation, and First Application}

\author[T. Marin-Gilabert et al.]{
Tirso Marin-Gilabert,$^{1,2}$\thanks{E-mail: tmarin@usm.lmu.de}
Ulrich P. Steinwandel,$^{3}$
Milena Valentini,$^{4,5,6,7}$
John A. ZuHone,$^{1}$
and Klaus Dolag$^{2,3}$
\\
$^{1}$Center for Astrophysics | Harvard \& Smithsonian, 60 Garden St. Cambridge, MA 02138, USA\\
$^{2}$Universit{\"a}ts-Sternwarte, Fakult{\"a}t f{\"u}r Physik,  Ludwig-Maximilians-Universit{\"a}t  M{\"u}nchen, Scheinerstr. 1, 81679 M{\"u}nchen, Germany\\
$^{3}$Max-Planck-Institut f{\"u}r Astrophysik, Karl-Schwarzschild-Str. 1, D-85741 Garching, Germany\\
$^{4}$Astronomy Unit, Department of Physics, University of Trieste, via Tiepolo 11, I-34131 Trieste, Italy\\
$^{5}$INAF - Osservatorio Astronomico di Trieste, via Tiepolo 11, I-34131 Trieste, Italy\\
$^{6}$INFN, Instituto Nazionale di Fisica Nucleare, Via Valerio 2, I-34127, Trieste, Italy\\
$^{7}$ICSC - Italian Research Center on High Performance Computing, Big Data and Quantum Computing, via Magnanelli 2, 40033, Casalecchio di Reno, Italy\\
}

\date{Accepted XXX. Received YYY; in original form ZZZ}

\pubyear{\the\year{}}

\begin{document}
\label{firstpage}
\pagerange{\pageref{firstpage}--\pageref{lastpage}}
\maketitle

\begin{abstract}
We present the implementation of an anisotropic viscosity solver within the magnetohydrodynamics (MHD) framework of the TreeSPH code \textsc{OpenGadget3}. The solver models anisotropic viscous transport along magnetic field lines following the Braginskii formulation and includes physically motivated limiters based on the mirror and firehose instability thresholds, which constrain the viscous stress in weakly collisional plasmas. To validate the implementation, we performed a suite of standard test problems—including two variants of the sound wave test, circularly and linearly polarized Alfvén waves, fast magnetosonic wave, and the Kelvin–Helmholtz instability—both with and without the plasma-instability limiters. The results show excellent agreement with the \textsc{AREPO} implementation of a similar anisotropic viscosity model, confirming the accuracy and robustness of our method. Our formulation integrates seamlessly within the individual adaptive timestepping framework of \textsc{OpenGadget3}, avoiding the need for subcycling. This provides efficient and stable time integration while maintaining physical consistency. Finally, we applied the new solver to a cosmological zoom-in simulation of a galaxy cluster as a proof-of-concept application, demonstrating its capability to model anisotropic transport and plasma microphysics in realistic large-scale environments. Our implementation offers a versatile and computationally efficient tool for studying anisotropic viscosity in magnetized astrophysical systems.
\end{abstract}




\section{Introduction}

Galaxy clusters are the largest gravitationally bound objects in the Universe and constitute crucial laboratories for understanding both astrophysical processes and cosmological evolution \citep{Allen_2011, Kravtsov_2012}. Their baryonic content is dominated by the intracluster medium (ICM), a hot, diffuse plasma with temperatures reaching $T \sim 10^7 - 10^8$~K and densities of $n \sim 10^{-2}- 10^{-3}$ cm$^{-3}$ \citep[see review by][]{Carilli_2002}, where thermal particle velocities are high and Coulomb collisions are infrequent. In massive clusters, this corresponds to ion mean free paths of order $\lambda_i \sim 1$–$10$~kpc and high plasma beta, $\beta \equiv 8\pi P_{\rm th}/B^2 \gg 1$, consistent with X-ray–inferred thermodynamics and $\mu$G magnetic fields \citep{Sarazin_1986, Schekochihin_2006}. In such a weakly collisional environment, the particle mean free path becomes comparable to—or even exceeds—typical macroscopic length scales such as pressure or temperature gradients \citep{Sarazin_1986}. Under these conditions, individual particles can carry momentum over large distances, making collisional transport processes, like viscosity, effective even in a plasma that is not strongly collisional in the classical sense. Viscosity arises from the momentum exchange between particles moving along different velocity gradients \citep{Schekochihin_2005, Kunz_2012}. In the ICM, where the ion mean free path can be on the order of several kiloparsecs, ions from one fluid element can traverse significant distances before scattering. This means that even weak velocity gradients can lead to substantial momentum flux, resulting in non-negligible viscous stresses on the fluid. 

In the unmagnetized, fully collisional limit, viscosity is isotropic and follows the Spitzer form \citep{Spitzer_1962, Braginskii_1965}. However, in the weakly collisional, magnetized ICM, the ion gyro-radius is much smaller than the mean free path ($r_i \ll \lambda_i$), so momentum transport becomes anisotropic with respect to the magnetic field \citep[e.g.,][]{Braginskii_1965, Schekochihin_2006}. Magnetic fields thus impose a directional bias that shapes macroscopic dynamics and stability \citep{Schekochihin_2005, Squire_2016}, affecting both thermal conduction and viscosity \citep[e.g., ][]{Quataert_2008, Squire_2023}. The resulting pressure anisotropy, i.e. the difference between the pressure field components parallel and perpendicular to the magnetic field direction, can, in high-$\beta$ plasmas, drive kinetic microinstabilities—most notably the firehose (for excess parallel pressure) and mirror (for excess perpendicular pressure)—which rapidly generate pitch-angle scattering and regulate the anisotropy toward marginal stability \citep[e.g.,][]{Schekochihin_2008, Bale_2009, Rincon_2014}. This anisotropic (Braginskii) viscosity profoundly alters plasma dynamics—shaping magnetic field geometry or the development of turbulence—with consequences that cannot be captured by isotropic models alone \citep[e.g.,][]{Kunz_2011, Squire_2016}.

Recent high-resolution X-ray spectroscopy has provided new constraints on turbulence in galaxy clusters. Hitomi observations of the Perseus galaxy cluster revealed relatively low velocity dispersions of $\sim160\ \mathrm{km\,s^{-1}}$ \citep{Hitomi_2016, Hitomi_2018}. XRISM finds low dispersion levels in systems such as Ophiuchus, Centaurus, Coma, and Abell 2029, with velocity dispersions in the range $\sim100$–$200\ \mathrm{km\,s^{-1}}$ and non-thermal pressure fractions of only a few percent \citep{XRISM_2025_centaurus, XRISM_2025_abell, XRISM_2025_Coma, XRISM_2025_Ophiuchus}. In the particular case of Abell 2029 \citep{XRISM_2025_Abell2029_profile}, XRISM finds a low and radially decreasing non-thermal pressure fraction out to $\sim R_{\rm 2500}$\footnote{$R_{\mathrm{2500}}$ is defined as the radius enclosing the region of the cluster with a mean density 2500 times larger than the critical density of the universe.}.

Cosmological and idealized simulations predict the existence of higher velocity dispersions and turbulent pressure fractions in galaxy clusters than this, across a range of setups and driving mechanisms \citep{Schmidt_2014, Miniati_2015, Schmidt_2016, Vazza_2018b}. At larger radii, the non-thermal pressure fraction in simulated clusters tends to increase \citep[e.g.][]{Nelson_2014}, contrary to the trend in Abell 2029. Part of this tension plausibly arises from projection effects. Emissivity weighting and multiphase structure along the line of sight biases measurements of line widths, which results in an underestimation of the velocity dispersion in 3D turbulence models \citep{Vazza_2025, XRISM_2025_LoS}. However, as shown by \citet{XRISM_2025_sims}, in the centers of cool core clusters, the discrepancy between observations and three recent simulations persists even after accounting for projection effects. One possibility is that the models for AGN feedback in these simulations are too ejective. However, viscosity offers a microphysical route to the same outcome. Viscous stresses damp small-scale eddies and shear, lowering non-thermal support and matching simulations with observations \citep{Kunz_2011}. In the case of Abell 2029 \citep{XRISM_2025_Abell2029_profile}, a more viscous ICM—possibly alongside AGN driving—can keep flows closer to laminar, smooth mixing layers, and dissipate residual shear before it cascades \citep{Marin-Gilabert_2025}, reducing observed line widths.

The effects of viscosity on the dynamics of the ICM have been investigated in a number of idealized studies. Braginskii viscosity has been shown to suppress the development of Kelvin–Helmholtz instabilities at cold fronts \citep{Suzuki_2013, Zuhone_2015}, to modify the nonlinear evolution of buoyancy-driven instabilities such as the magnetothermal instability (MTI) and the heat-flux-driven buoyancy instability (HBI) \citep{Quataert_2008, McCourt_2012, Parrish_2012}, and to alter the morphology and the growth of AGN-driven bubbles \citep{Dong_2009, Kingsland_2019}. Furthermore, it changes the propagation and damping of magnetohydrodynamics (MHD) waves, leading to the decay of fast magnetosonic modes and the destabilization of linearly polarized Alfvén waves \citep{Squire_2017, Berlok_2019}. These studies provide valuable benchmarks for testing numerical implementations of anisotropic viscosity.

Cosmological simulations offer the opportunity to assess viscous effects on cluster scales. Several works have focused on the generation of turbulence in the course of structure formation \citep{Dolag_2005, Iapichino_2008, Lau_2009, Vazza_2012, Iapichino_2017, Groth_2025}, on the role of AGN feedback in regulating the cooling and heating balance in cluster cores \citep{Sijacki_2007, Battaglia_2010, Gaspari_2011}, and on the amplification and distribution of magnetic fields during cluster assembly \citep{Dolag_2009, Vazza_2018, Steinwandel_2022, Steinwandel_2024, Tevlin_2025}. However, none of them included viscosity in their formulations. \citet{Marin-Gilabert_2024} studied the effects of isotropic viscosity in cosmological simulations, finding that isotropic viscosity leads to the suppression of small-scale turbulence and modifies the thermal structure of the ICM. To date, no cosmological-scale simulation has yet included anisotropic (Braginskii) viscosity, leaving open the question of how anisotropic momentum transport affects the evolution of galaxy clusters.

In this paper, we present the first implementation of Braginskii viscosity within the SPH framework of \textsc{OpenGadget3}. We validate our method using the benchmark suite established by \citet{Berlok_2019}, and we subsequently apply it to cosmological simulations of galaxy cluster formation. This approach allows us to demonstrate, for the first time, that Braginskii viscosity can be evolved self-consistently in cosmological SPH simulations of galaxy-cluster formation.

The goal of this paper is not only to describe the Braginskii viscosity implementation and the robustness of the scheme, but also to show the capability of the code to model galaxy clusters in a cosmological context. Future work will undertake a detailed physical analysis of the effects of Braginskii viscosity in the ICM from cosmological simulations.

The paper is structured as follows. In \S \ref{sec:methods}, we introduce the theory of weakly collisional plasmas and the implementation in \textsc{OpenGadget3}. \S \ref{sec:tests_results} shows the different tests performed and the results obtained. In \S \ref{sec:cosmo_sims} we show the results of a cosmological simulation including Braginskii viscosity. Finally, we discuss the results obtained and the conclusions in \S \ref{sec:conclusions}.

\section{Methods} \label{sec:methods}

\subsection{Theoretical Considerations}

The large difference in magnitude between the gyro-radius and the viscous scale allows treating the ICM as a magnetized plasma, where the dynamic interactions significantly impact the evolution of magnetic fields. The magnetic field applies forces to the plasma, while the flow advects and distorts the field. 
These complex interactions are captured by the equations of MHD, which describe the continuum evolution of a conducting fluid coupled to magnetic fields. This framework is particularly well-suited for describing the behavior of the ICM, where magnetic fields, though dynamically subdominant to thermal pressure in high-$\beta$ regions, still regulate plasma dynamics through both large-scale forces and the mediation of microscale instabilities.

In the Lagrangian form, the ideal MHD equations (conservation of mass, momentum, energy, and magnetic field) can be written as:
\begin{equation}
    \frac{\mathrm{d} \rho}{\mathrm{d} t} + \rho \nabla \cdot \mathbf{v} = 0 ,
    \label{eqn:mass_eq}
\end{equation}
\begin{equation}
    \rho \frac{\mathrm{d} \mathbf{v}}{\mathrm{d} t} + \nabla P = -\rho \nabla \Phi - \nabla \cdot \boldsymbol{\Pi} + \frac{(\nabla \times \mathbf{B}) \times \mathbf{B}}{4\pi} \, ,
    \label{eqn:momentum_eq}
\end{equation}
\begin{equation}
    \frac{\mathrm{d} E}{\mathrm{d} t} + \mathbf{v} \cdot \nabla P + (E + P)\nabla \cdot \mathbf{v} - \nabla \cdot \frac{\mathbf{B} (\mathbf{v} \cdot \mathbf{B})}{4 \pi} = -\rho \mathbf{v} \cdot \nabla \Phi - \nabla \cdot \mathbf{Q} - \nabla \cdot \left( \boldsymbol{\Pi} \cdot \mathbf{v} \right) \, ,
    \label{eqn:energy_eq}
\end{equation}
\begin{equation}
    \frac{{\rm d} \mathbf{B}}{{\rm d} t} = \left( {\bf B} \cdot \nabla \right) {\bf v} - {\bf B} \nabla \cdot {\bf v} \, ,
\end{equation}
where
\begin{equation}
    \frac{\mathrm{d}}{\mathrm{d}t} = \frac{\partial}{\partial t} + (\mathbf{v} \cdot \nabla)
\end{equation}
is the Lagrangian derivative. $\rho$ is the gas density, $\mathbf{v}$ the velocity of the fluid, $\mathbf{Q}$ is the heat flux, $\boldsymbol{\Pi}$ is the viscous stress tensor and $\Phi$ is the gravitational potential. $E$ is the energy per unit volume (kinetic + thermal + magnetic)
\begin{equation}
    E = \frac{\rho v^2}{2} + \rho u + \frac{B^2}{8 \pi} \, .
\end{equation}
The total pressure $P$ is equal to the thermal pressure $P_{\rm th} = (\gamma - 1) \rho u$ (ion+electron pressure), with $u$ being the specific internal energy, plus the magnetic pressure $P_{\rm mag} = B^2/8 \pi$
\begin{equation}
    P = P_{\rm th} + P_{\rm mag} = (\gamma - 1) \rho u + \frac{B^2}{8 \pi}\, ,
    \label{eqn:thermal_pressure}
\end{equation}
with an adiabatic index of $\gamma = 5/3$ for a monoatomic gas.

The plasma beta is given by the ratio of thermal to magnetic pressure:
\begin{equation}
    \beta = \frac{P_{\rm th}}{P_{\rm mag}} = \frac{8\pi P_{\rm th}}{B^2} = \frac{8\pi (\gamma - 1) \rho u}{B^2}  \, ,
\end{equation}
and gives an estimate of the magnetic field strength in a magnetized fluid. 

When considering a weakly collisional plasma where the gyro-radius is smaller than the mean free path ($\lambda_i \gg r_i$), the plasma can develop different pressures parallel ($p_{\parallel}$) and perpendicular ($p_{\perp}$) to the magnetic field, resulting in a pressure anisotropy. The anisotropy arises naturally due to the conservation of the first and second adiabatic invariants\footnote{The first adiabatic invariant $\mu$ corresponds to the magnetic moment of a gyrating particle, while the second invariant $\mathcal{J}$ is associated with the magnetic field-aligned particle action.} \citep{Squire_2017, Kunz_2022}. In the weakly collisional magnetized regime considered here, their conservation is only weakly broken by collisions. As a result, changes in magnetic field strength and density lead to differences between $p_{\perp}$ and $p_{\parallel}$ \citep{Schekochihin_2005}.

Fluid motions drive the pressure anisotropy through changes in magnetic field strength and compression, which satisfy:
\begin{equation}
    \frac{1}{B} \frac{{\rm d}{B}}{{\rm d}t} = \hat{b}\hat{b} : \nabla \mathbf{v} - \nabla \cdot \mathbf{v} \, ,
\end{equation}
where $\hat{b} = {\bf B}/B$ indicates the magnetic field direction. The notation ``$:$'' denotes the Frobenius inner product and is defined as the sum of the products of their components: $\hat{b}\hat{b} : \nabla \mathbf{v} = \Sigma_i \Sigma_j b_i b_j \partial_i v_j$.

The resulting pressure anisotropy can affect the stability of the plasma, leading to various instabilities such as the firehose and mirror instabilities \citep[e.g.,][]{Kunz_2014, Rincon_2014}, which can further influence the dynamics and evolution of the plasma. The total thermal pressure can be split into its parallel and perpendicular components:
\begin{equation}
    P_{\rm th} = \frac{2}{3}p_{\perp} + \frac{1}{3}p_{\parallel} \, .
\end{equation}

The Braginskii viscosity accounts for these anisotropies by introducing an anisotropic viscous stress, whose dynamical effect is strongest when the velocity gradient has a component parallel to the magnetic field lines. In the Braginskii regime, where the ion mean free path is smaller than the macroscopic dynamical scale and collisions dominate over the rate of strain ($|\nabla \mathbf{v}| \ll \nu_{ii}$) \citep{Braginskii_1965, Squire_2017}, the pressure anisotropy can be written as:
\begin{equation}
   \Delta p = p_{\perp} - p_{\parallel} = \eta\left( 3\hat{b}\hat{b} : \nabla \mathbf{v} - \nabla \cdot \mathbf{v} \right) = 0.960 \frac{p_i}{\nu_{ii}} \frac{{\rm d}}{{\rm d}t} \ln \left( \frac{B^3}{\rho^2} \right) \, , 
    \label{eqn:pressure_aniso}
\end{equation}
where $p_i$ is the ion thermal pressure\footnote{Ions dominate the momentum transfer (i.e., viscosity) due to their higher mass compared to electrons \citep{Sarazin_1986}.} \citep{ZuHone_2016}, and $\nu_{ii}$ is the ion collision frequency
\begin{equation}
    \nu_{ii} = \frac{4 \sqrt{\pi} \, n_i (Z\,e)^4 \ln \Lambda}{3m_i^{1/2} (k_{\rm B} T_i)^{3/2}} \, .
\end{equation}
$n_i$ is the number density, $m_i$ is the ion mass, $T_i$ is the temperature of the plasma, and $\ln{\Lambda} = 37.8$ is the Coulomb logarithm.
In this regime, the pressure anisotropy remains small compared to the ion thermal pressure, i.e. $|\Delta p| \ll p_i$.
Eq. \eqref{eqn:pressure_aniso} describes the creation of pressure anisotropy due to plasma motions: through the parallel rate of strain ($\hat{b}\hat{b} : \nabla \mathbf{v}$) and through the compression of the fluid ($\nabla \cdot \mathbf{v}$). In subsonic, weakly compressible flows, positive $\Delta p$ is created in regions of increasing field strength, while negative $\Delta p$ is created in regions of decreasing field strength \citep{Schekochihin_2005, Squire_2023}. The shear viscosity (Spitzer) coefficient ($\eta$) is the same as in the isotropic case:
\begin{equation}
    \eta = 0.960 \frac{n_i k_{\rm B} T_i}{\nu_{ii}} = 0.406 \frac{m_i^{1/2}(k_B T_i)^{5/2}}{(Z\,e)^4 \ln{\Lambda}} \, .
    \label{eqn:viscosity}
\end{equation} 
The numerical prefactor and scaling arise from solving the Boltzmann equation with a Fokker–Planck collision operator under the assumption of a small deviation from a Maxwellian distribution \citep{Spitzer_1962, Braginskii_1965}.

Parallel to magnetic fields, the plasma can move freely along the magnetic field lines, and viscosity behaves similarly to the isotropic Spitzer viscosity. However, perpendicular to the magnetic field lines, the movement is restricted and viscosity becomes significantly suppressed due to the smaller gyro-radius compared to $\lambda_i$. When the magnetic field is aligned with the velocity gradient, the parallel rate of strain ($\hat{b}\hat{b} : \nabla \mathbf{v}$) becomes maximum, while if they are perpendicular, the parallel rate of strain becomes zero.

The anisotropic viscous stress tensor accounts for the pressure anisotropy and can be written as:
\begin{equation}
   \boldsymbol{\Pi}_{\rm Aniso} = -\Delta p \left( \hat{b}\hat{b} - \frac{1}{3} \mathbf{I} \right) \, .
    \label{eqn:aniso_stress_tensor}
\end{equation}

\subsection{Plasma Microinstabilities} \label{sec:microins}

When $\beta \gg 1$, weakly collisional plasmas might become unstable against plasma microinstabilities, such as firehose and mirror instabilities. They arise from pressure anisotropies and are crucial for understanding plasma dynamics in a magnetized medium. These instabilities manifest when the differences between each pressure direction reach thresholds that destabilize the otherwise stable Alfvén and entropy modes. 
This keeps the pressure anisotropy at marginally stable levels, significantly affecting the plasma's macroscopic behavior and transport properties \citep{Schekochihin_2006, Kunz_2014}. The plasma microinstabilities grow, and when they reach a critical value, the resulting microscale fluctuations increase particle scattering and thereby regulate the pressure anisotropy toward marginal stability. \citep{Rappaz_2024}. For the large-scale plasma dynamics considered here, this regulation is effectively instantaneous \citep[e.g.,][]{Squire_2023}. 

The firehose instability occurs when the parallel pressure component significantly exceeds the perpendicular component ($p_{\parallel} > p_{\perp}$, $\Delta p < 0$). In bent magnetic field lines, the excess parallel pressure acts as an effective centrifugal force that counteracts the restoring magnetic tension, leading to the growth of Alfvénic, oblique perturbations\footnote{There are two types of firehose instability: parallel and oblique. In this work, we only consider the parallel one \citep{Hellinger_2000, Bott_2021}, while a treatment of the oblique firehose instability is beyond the scope of this paper and will be explored in future work.}. These perturbations enhance ion scattering, increasing the effective collision frequency ($\nu_{\rm eff} = \nu_{ii} + \nu_{\rm scatt}$), reducing the Coulomb mean free path, the system's effective viscosity, and maintaining the pressure anisotropy at marginally stable levels (eq. \eqref{eqn:pressure_aniso}) \citep{Kunz_2014, Arzamasskiy_2023}:
\begin{equation}
    \Delta p = p_{\perp} - p_{\parallel} < -\frac{B^2}{4\pi} \, .
    \label{eqn:firehose_inst}
\end{equation}

In contrast, the mirror instability arises from perpendicular pressure being larger than the parallel component ($p_{\perp} > p_{\parallel}$, $\Delta p > 0$). This generates compressive magnetic fluctuations that form magnetic mirrors, producing regions of enhanced and reduced magnetic field strength. In the nonlinear regime, these structures primarily trap particles and thereby regulate the pressure anisotropy toward marginal stability, effectively limiting the anisotropic viscous transport \citep{Southwood_1993, Kunz_2014, Rincon_2014}:
\begin{equation}
    \Delta p = p_{\perp} - p_{\parallel} > \frac{B^2}{8\pi} \, .
    \label{eqn:mirror_inst}
\end{equation}
The pressure anisotropy is stable within the firehose and mirror instability limits; therefore, we can write the stability criterion as \citep{Kunz_2012, Kunz_2014}:
\begin{equation}
    -\frac{B^2}{4\pi} < \Delta p < \frac{B^2}{8\pi} \, .
    \label{eqn:micro_limiters}
\end{equation}
This means that the stability of $\Delta p$ depends on the magnetic field strength. For weaker magnetic fields, the field is more susceptible to distortion, thus triggering the instabilities more easily. The threshold for triggering plasma microinstabilities is reduced, limiting the pressure anisotropy and suppressing the effect of viscosity. On the other hand, for stronger magnetic fields, magnetic tension stabilizes the field lines against perturbations, and the plasma microinstabilities are not triggered so easily. Therefore, the plasma can sustain larger pressure anisotropies before microinstability regulation sets in.

Both instabilities play a key role in the energy distribution and stability of high-$\beta$ plasmas, influencing large-scale plasma behavior. They act as self-regulating mechanisms to maintain the anisotropy within marginally stable limits, thereby fundamentally affecting the evolution and transport properties of magnetized astrophysical plasmas. We note that other microinstabilities, including heat-flux–driven instabilities such as the gyrothermal instability, may also operate in weakly collisional plasmas and affect the effective transport properties, but they are not included in the present work \citep{Schekochihin_2010, Rosin_2011}.

\subsection{Numerical Methods}

We have implemented Braginskii viscosity in the smoothed particle magnetohydrodynamics (SPMHD) code \textsc{OpenGadget3} \citep{Springel_2005, Groth_2023}. SPH works by interpolating physical quantities among the closer neighbor particles ($N_{\rm ngb}$) using a Gaussian-like smoothing kernel with compact support. Throughout this paper, we used a Wendland $C^6$ kernel \citep{Wendland_1995, Dehnen_2012} and 295 neighbors, including artificial viscosity \citep{Balsara_1995, Cullen_2010} and artificial conductivity \citep{Price_2008}. A detailed description of the numerical methods employed in \textsc{OpenGadget3} can be found in \citet{Springel_2005} and \citet{Beck_2015}.

To be able to solve the MHD equations numerically, they need to be discretized. In the case of \textsc{OpenGadget3}, we discretised the pressure anisotropy (eq. \eqref{eqn:pressure_aniso}) of a particle $i$ as:
\begin{equation}
    \Delta p\Big{|}_i = \eta \left( 3\hat{b}_{\alpha}\hat{b}_{\beta} \frac{\partial v_{\alpha}}{\partial x_{\beta}}\Big{|}_i - \left. \delta_{\alpha\beta} \frac{\partial v_{\gamma}}{\partial x_{\gamma}} \right|_i \right) \, ,
    \label{eqn:pressure_aniso_gadget}
\end{equation}
where $\hat{b} = \mathbf{B / |\mathbf{B}|}$ is the magnetic unit vector. The velocity gradient tensor entering this expression is the SPH estimate of the local bulk-flow gradient at the position of particle $i$, computed from the neighboring particles using the higher-order gradient estimator implemented in \textsc{OpenGadget3} and described in \citet{Beck_2015}. The anisotropic viscous stress tensor (eq. \eqref{eqn:aniso_stress_tensor}) is given by:
\begin{equation}
    \boldsymbol{\Pi}_{\rm Aniso, \, \alpha \beta} \Big{|}_i = - \Delta p\Big{|}_i \left( \hat{b}_{\alpha}\hat{b}_{\beta}\Big{|}_i - \frac{1}{3} \delta_{\alpha\beta} \right) \, .
\end{equation}
Similar to the isotropic case described in \citet{Marin-Gilabert_2022}, the change in velocity and entropy due to viscosity is described as:
\begin{multline}
    \displaystyle\frac{\mathrm{d}v_{\alpha}}{\mathrm{d}t}\bigg{|}_{{i, \mathrm{\textrm{ shear}}}} = \displaystyle\sum\limits_{j=1}^{N_{\rm ngb}} m_j \, \left[ \frac{\boldsymbol{\Pi}_{\rm Aniso, \, \alpha \beta}\big{|}_i}{\rho_i^2} \left( \nabla_i W_{ij}(r, h_i) \right) \Big{|}_{\beta} \right. \\ 
    \left. + \frac{\boldsymbol{\Pi}_{\rm Aniso, \, \alpha \beta}\big{|}_j}{\rho_j^2} \left( \nabla_i W_{ij}(r, h_j) \right) \Big{|}_{\beta} \right] \, ,
\end{multline}
and:
\begin{equation}
    \displaystyle\frac{\mathrm{d} A_i}{\mathrm{d}t}\bigg{|}_{{\mathrm{\textrm{shear}}}} = \frac{\gamma - 1}{\rho_i^{\gamma - 1}} \frac{\Delta p_{i}^2}{3 \rho_i \eta_i} \, .
    \label{eqn:entropy_aniso}
\end{equation}
$W_{ij}$ denotes the Wendland $C^6$ smoothing kernel evaluated for the particle pair $i,j$.

When the plasma microinstability limiters are enabled, the pressure anisotropy is limited by imposing the mirror and firehose thresholds of eq.~\eqref{eqn:micro_limiters} as instantaneous hard walls. We do not evolve an explicit finite scattering rate.

\section{Tests and Results} \label{sec:tests_results}

In this section, we present the tests performed of the anisotropic (Braginskii) viscosity implementation in \textsc{OpenGadget3}, following the setups described by \cite{Berlok_2019}. These tests serve to validate the numerical approach and explore the fundamental differences arising from anisotropic transport relative to the isotropic approximation.

\subsection{Sound Wave I} \label{sec:soundwave_I}

The first test is the viscous damping of a simple sound wave perturbation in a 3D setup of sizes $L = L_{x,y,z}$ with an initial velocity of
\begin{equation}
    v(x, 0) = A \sin(\mathbf{k} \cdot x) \hat{x} \, ,
    \label{eqn:soundwave}
\end{equation}
where $A$ is the initial amplitude of the perturbation and $\mathbf{k} = 2\pi/L \, \hat{x}$ is the wavenumber. We start with a constant density $\rho$ and a resolution of $128^3$, given by the number of particles per unit length placed in a regular grid. We include a constant static magnetic field $\mathbf{B}$, first in the $\hat{x}$-direction (parallel to the velocity gradient) and then in the $\hat{y}$ direction (perpendicular to the velocity gradient).

In the simple case of a plane wave initialized along the $\bf k$ direction, the velocity time evolution is given by \citep{Berlok_2019}:
\begin{equation}
    v_x(x, t) = A \sin(\mathbf{k} \cdot x) \, {\rm e}^{-\gamma t} \, ,
    \label{eqn:soundwave_time}
\end{equation}
where $\gamma$ is the damping rate due to viscosity. The viscous heating after a time $t$ is given by
\begin{equation}
    u(x, t) = u_0 + \frac{\rho A^2}{2} \cos^2(\mathbf{k} \cdot x) \left( 1 - e^{-2\gamma t} \right) \, .
    \label{eqn:soundwave_energy_time}
\end{equation}

To study the diffusion effects of viscosity only, we initially switch off the hydro solver, following \citet{Berlok_2019}. This allows us to isolate the effect of viscosity without the hydro effects due to, for instance, compression or expansion of the fluid. Under these conditions, we can directly use the solution introduced in \citet{Berlok_2019}, where the damping rate is given by 
\begin{equation}
    \gamma_{\parallel} = \frac{4}{3}\nu k_{\parallel}^2 \, .
    \label{eqn:gamma_hydro_off}
\end{equation}
The kinematic viscosity $\nu$ is the ratio between the shear viscosity coefficient ($\eta$) and the density
\begin{equation}
    \nu = \frac{\eta}{\rho} \, ,
\end{equation}
although in these test cases we set the constant density field to unity, so $\nu = \eta$.

\begin{figure}
    \centering
	\includegraphics[width=\columnwidth]{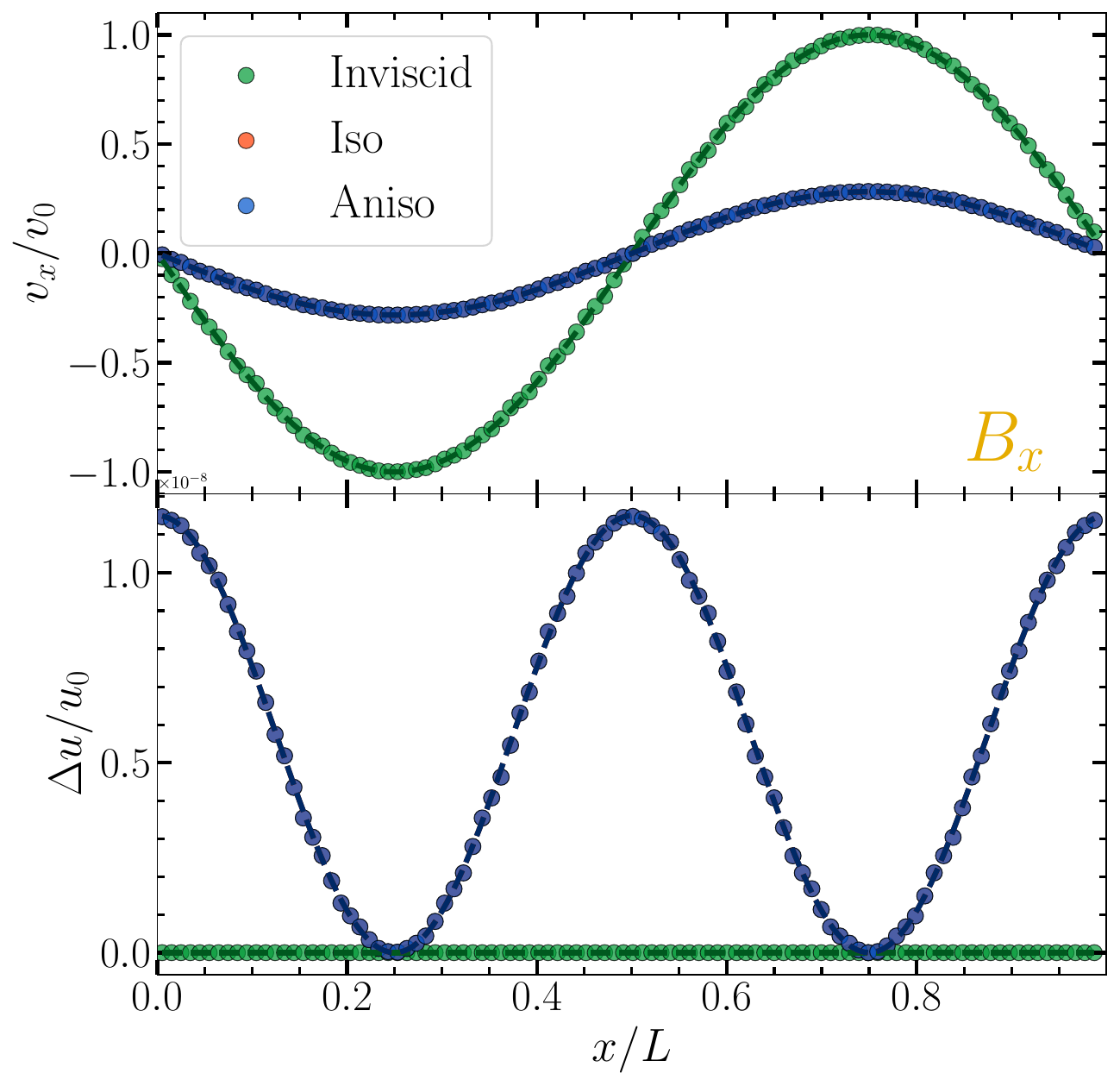}
    \caption{Velocity and viscous heating profiles of the sound wave described by eq. \eqref{eqn:soundwave} after $ct/L=1$ with the hydro solver off. The data points show the velocity profile for the different numerical simulations (green for the inviscid case, red for the isotropic viscosity case, and blue for the anisotropic case), while the dashed lines indicate the analytical solutions following the same color code as the data points. In this case, the magnetic field has only $x$-component, i.e., parallel to the velocity gradient. \textit{Top panel}: $v_x$ profile. \textit{Bottom panel}: Cumulative viscous heating. The isotropic and anisotropic viscous solutions overlap.}
    \label{fig:SoundwaveI_no_hydro}
\end{figure}

Fig.~\ref{fig:SoundwaveI_no_hydro} shows the velocity (top panel) and cumulative viscous heating (bottom panel) for the inviscid (green), isotropic (red), and anisotropic (blue) runs, compared with the analytical solution (dashed lines) after $ct/L=1$. In this case, the magnetic field is initialized only in the $\hat{x}$ direction, i.e., parallel to the velocity gradient; therefore, the damping rate of both the isotropic and anisotropic cases is given by \eqref{eqn:gamma_hydro_off}. All the results (dots) follow the expected analytical solution given by eq. \eqref{eqn:soundwave_time} (color-dashed lines), with the inviscid case keeping the initial velocity amplitude after $ct/L=1$, and the viscous runs damping the sound wave equally (top panel). Viscosity converts kinetic to internal energy (heating) following eq. \eqref{eqn:soundwave_energy_time}, being the viscous heating larger at the nodes of the velocity profile, where the velocity gradient is maximal (bottom panel). 

We also perform the same sound wave test, but in this case with the hydro solver on. Thus, in contrast to \citet{Berlok_2019}, we need to account for adiabatic compression for the derivation of the damping rate (see appendix \ref{app:gamma_soundwaveI}). The isotropic and anisotropic parallel damping rates are equal, while the perpendicular damping rate is given purely by compression
\begin{equation}
    \gamma_{\rm Iso} = \frac{2}{3}\nu k^2 \, , \hspace{0.7cm} \gamma_{\parallel} = \frac{2}{3}\nu k_{\parallel}^2 \, , \hspace{0.7cm} \gamma_{\perp} = \frac{1}{6}\nu k_{\perp}^2 \, .
    \label{eqn:damping_rates}
\end{equation}
In this case, we also ran a simulation with a magnetic field in the $\hat{y}$ direction, to test the anisotropic viscosity when $B \perp \nabla v$, and compare it with the analytical solution considering the compression and expansion of the fluid.

\begin{figure}
    \centering
	\includegraphics[width=\columnwidth]{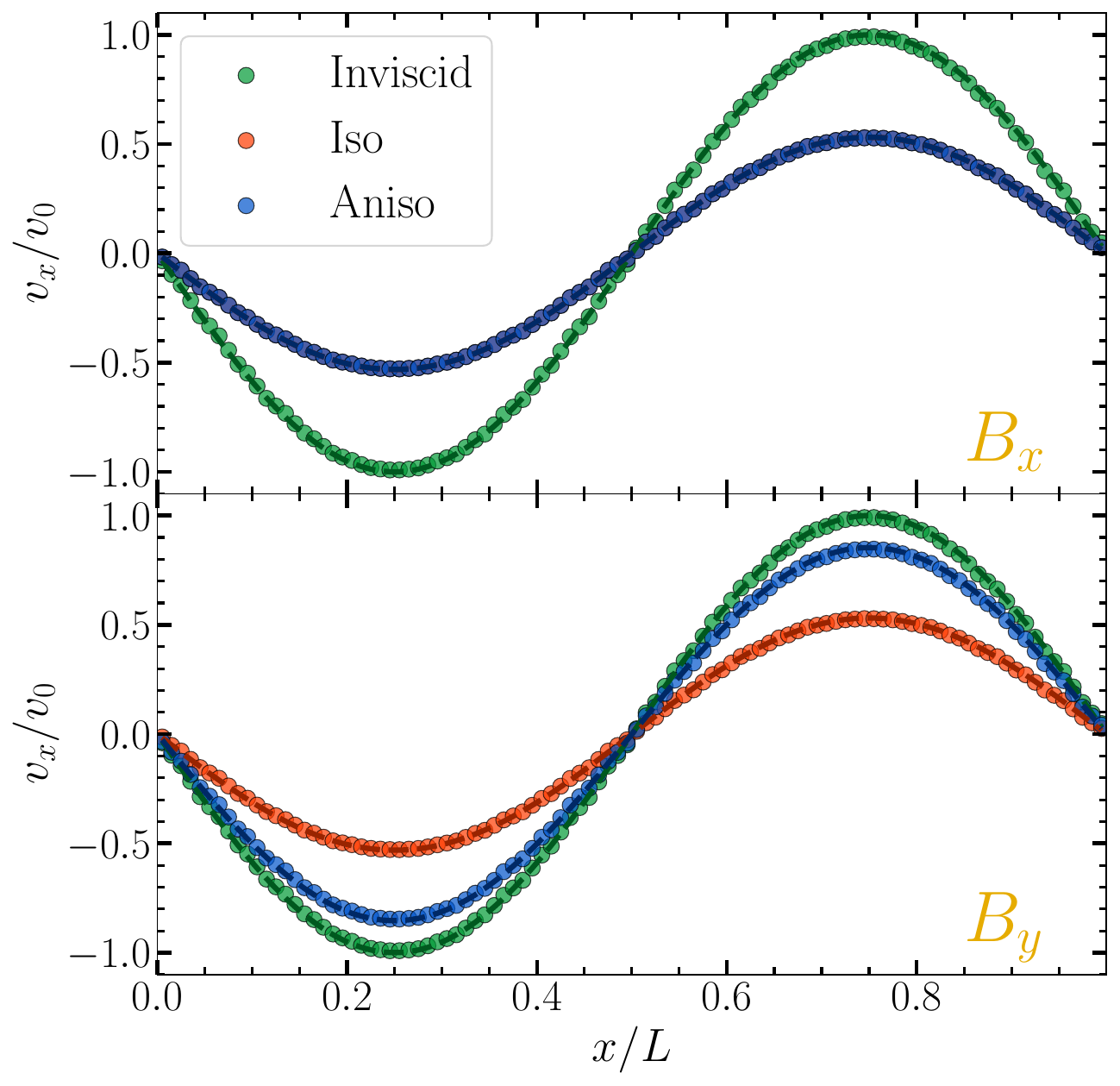}
    \caption{Velocity profile of the sound wave described by eq. \eqref{eqn:soundwave} after $ct/L=1$ with the hydro solver on. The data points show the velocity profile for the different numerical simulations (green for the inviscid case, red for the isotropic viscosity case, and blue for the anisotropic case), while the dashed lines indicate the analytical solutions following the same color code as the data points. \textit{Top panel}: Magnetic field in the $\hat{x}$ direction, parallel to the velocity gradient. In this case, the isotropic and anisotropic viscous solutions overlap. \textit{Bottom panel}: Magnetic field in the $\hat{y}$ direction, perpendicular to the velocity gradient.}
    \label{fig:Soundwave_vx}
\end{figure}

Fig.~\ref{fig:Soundwave_vx} shows the velocity profile at $ct/L=1$ for $B_x$ on the top panel, and $B_y$ on the bottom panel with the hydro solver on. In all cases, the numerical results match exactly the analytical solution, with the damping rates given by \eqref{eqn:damping_rates}. Regardless of the magnetic field direction, the inviscid run keeps the initial amplitude, showing that it has not been damped due to numerical viscosity. In the sound wave with a parallel magnetic field ($B_x \parallel \Delta v$, top panel), the isotropic and the anisotropic cases have the same amplitude after $ct/L=1$, as expected by the damping rates \eqref{eqn:damping_rates}. This shows that, in the direction of the magnetic field lines, the overall effect of the anisotropic viscosity is equal to the isotropic one. In contrast, when the magnetic field is perpendicular to the velocity gradient ($B_y \perp \Delta v$, bottom panel), the anisotropic case does not mimic the isotropic case. However, there is still viscosity due to compression. In an incompressible fluid, the overall effect of anisotropic viscosity should be zero, although this is not the case in a compressible fluid.

\begin{figure}
    \centering
	\includegraphics[width=\columnwidth]{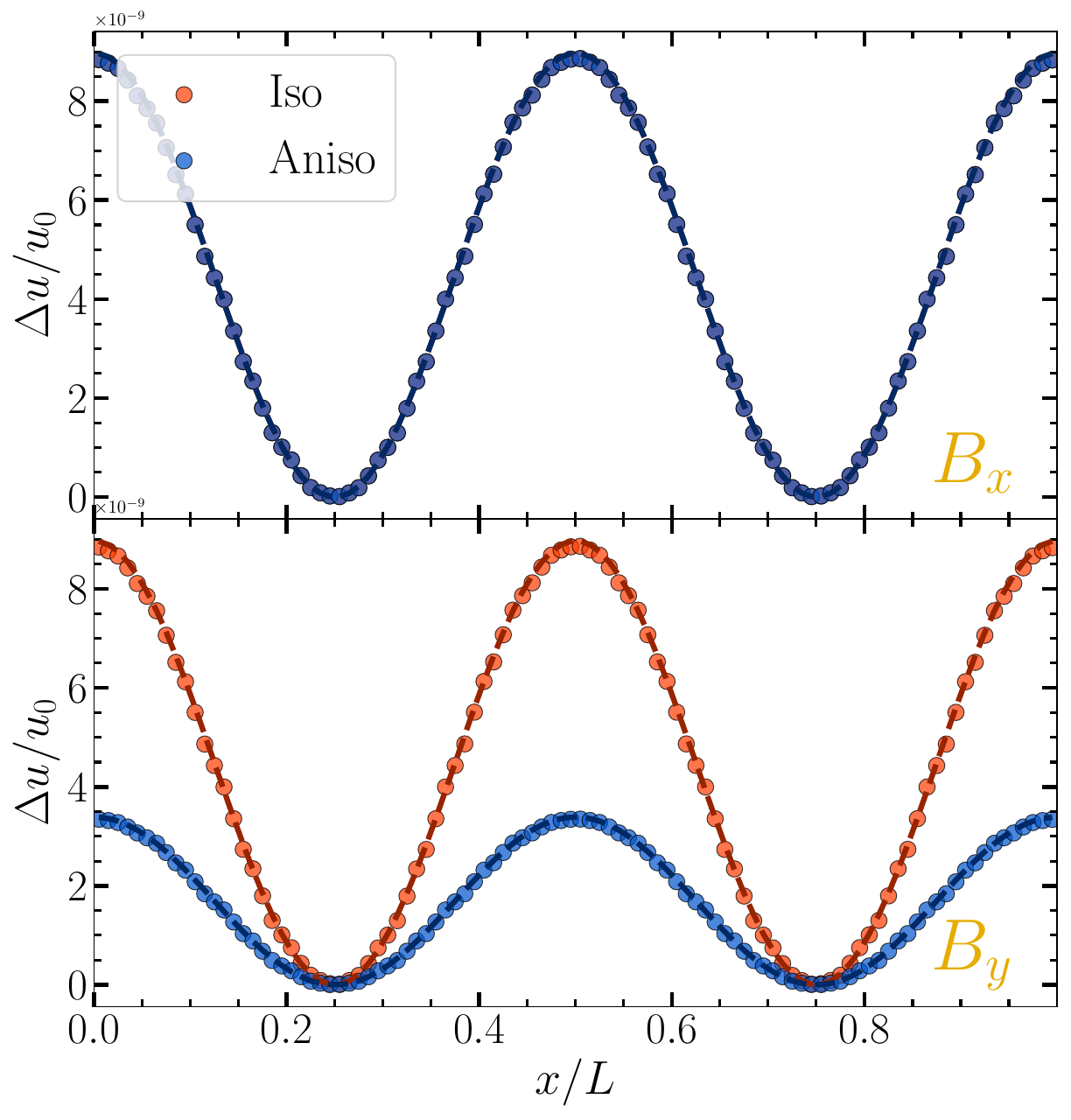}
    \caption{Cumulative viscous heating profile of the sound wave described by eq. \eqref{eqn:soundwave} at $ct/L=1$. The data points are the results from the simulations, and the dashed lines indicate the analytical solution (red for the isotropic viscosity case, and blue for the anisotropic case). \textit{Top panel}: Magnetic field parallel to the velocity gradient. The isotropic and anisotropic viscous solutions overlap. \textit{Bottom panel}: Magnetic field perpendicular to the velocity gradient.}
    \label{fig:Soundwave_vx_energy}
\end{figure}

The cumulative viscous heating at $ct/L=1$ is shown in Fig.~\ref{fig:Soundwave_vx_energy}. When the magnetic field is parallel to the velocity gradient (top panel), the viscous heating is equal in the isotropic and anisotropic cases, whereas in the perpendicular setup (bottom panel), the anisotropic viscous stress is smaller, which leads to a lower viscous heating.

\subsection{Sound Wave II} \label{sec:soundwave_II}

Following \citet{Hopkins_2017} and \citet{Berlok_2019}, for the second test we simulate another soundwave, where the constant static magnetic field has two components: ${\bf B} = B_0 \hat{b}$, with $\hat{b} = (\hat{x}+\hat{y})/\sqrt{2}$. The initial velocity is given by 
\begin{equation}
    v(x, 0) = A \, q(x) \hat{y} \,
    \label{eqn:ICs_soundwaveII}
\end{equation}
with 
\begin{equation}
    q(x) = \frac{3}{2} - \frac{1}{2} \left( {\rm erf}\left(\frac{x-x_0}{a}\right) - {\rm erf}\left(\frac{x+x_0}{a}\right) \right) \, ,
    \label{eqn:ICs_soundwaveII_q}
\end{equation}
where $x_0 = 1/4$ and $a = 0.05L$ in our setup. Due to the effect of anisotropic viscosity, the initial velocity profile is damped, following
\begin{equation}
    v_y(x, t) = A \sum_{n=0}^{\infty} \frac{a_n}{10} \cos{(k_n x)} \left( 1 + 9 {\rm e}^{-\gamma_n t} \right) \, ,
    \label{eqn:vy_soundwaveII}
\end{equation}
where $k_n = 2 \pi n / L$, and 
\begin{equation}
    a_n = 
    \begin{cases}
        \hfill 2 \hfill & {\textrm{for }} n=0 \, ,  \\
        -2 \frac{\sin(3n\pi/2)}{n\pi} \, {\rm e}^{-n^2\pi^2/400} & {\textrm{for }} n>0 \, ,
    \end{cases}
\end{equation}
see \citet{Berlok_2019} for derivation. 

Although initially the sound wave is only propagating in the $\hat{y}$ direction, after $ct/L=1$ there is also a flow in the $\hat{x}$ direction. The anisotropic viscosity couples the $y$-component to the $x$-component, due to the misalignment of the wave propagation and the magnetic field. This produces a transfer of kinetic energy from $v_y$ to $v_x$, while dissipating a fraction into heat. Since the velocity only depends on $x$, only $\partial_x$ is non-zero. Thus, $\hat{b}\hat{b}:\nabla v = b_x b_j \partial_x v_j$. With a magnetic field in the $\hat{x}$ and $\hat{y}$ directions, we can express $\hat{b}\hat{b}:\nabla v$ as
\begin{multline}
    \hat{b}\hat{b}:\nabla v = b_x \left( b_x \partial_x v_x + b_y \partial_x v_y \right) = \\= \frac{1}{\sqrt{2}} \left( \frac{1}{\sqrt{2}} \partial_x v_x + \frac{1}{\sqrt{2}} \partial_x v_y \right) = \frac{1}{2} \left( \partial_x v_x  + \partial_x v_y \right)\, .
\end{multline}
The divergence is given by $\nabla \cdot {\bf v} = \partial_x v_x$, therefore the pressure anisotropy \eqref{eqn:pressure_aniso} becomes
\begin{equation}
    \Delta p = \eta \left(\frac{3}{2} \left( \partial_x v_x + \partial_x v_y \right) - \partial_x v_x \right) = \frac{\eta}{2} \left( \partial_x v_x + 3 \partial_x v_y \right) \, .
\end{equation}
This leads to a force in the $\hat{x}$ direction, which is given by the viscous stress tensor \eqref{eqn:aniso_stress_tensor}:
\begin{equation}
    \Pi_{xx} = \Delta p \left( b_x^2 - \frac{1}{3} \right) = \frac{1}{6} \Delta p \, ,
\end{equation}
\begin{equation}
    \Pi_{xy} = \Delta p \, b_x b_y = \frac{1}{2} \Delta p \, .
\end{equation}
Since only $\partial_x \neq 0$, the force in the $\hat{x}$ direction is given by
\begin{multline}
    \left( \nabla \cdot {\bf \Pi} \right)_x = \partial_x \Pi_{xx} = \frac{1}{6} \partial_x \Delta p = \frac{\eta}{12} \partial_x \left( \partial_x v_x + 3 \partial_x v_y \right) = \\ = \frac{\eta}{12} \left( \partial_x^2 v_x + 3 \partial_x^2 v_y \right) \, ,
\end{multline}
showing that the misalignment of the magnetic field leads to an acceleration in the $\hat{x}$ direction. In the isotropic case, $\left( \nabla \cdot {\bf \Pi} \right)_x = 0$, since there is no initial velocity in $x$ and there is no anisotropic source, thus $v_x$ does not change (see Fig.~\ref{fig:soundwave_II_hydro}, in appendix \ref{app:soundwaveII}). In the anisotropic case, the $v_x$ evolves as
\begin{equation}
    v_x(x,t) = -A \sum_{n=0}^{\infty} \frac{3a_n}{10} \cos{(k_n x)} \left( 1 - {\rm e}^{-\gamma_n t} \right) \, ,
    \label{eqn:vx_soundwaveII}
\end{equation}
where $\gamma_n$ is the damping rate (see appendix \ref{app:damping_ratesII}).

In this test, the hydro forces lead to the development of acoustic waves, deviating the result from the analytical solution (see appendix \ref{app:soundwaveII}). Therefore, to be able to isolate the effects of the Braginskii viscosity and compare them with the analytical solution, we switch off the hydro solver as we did for the sound wave I (\S \ref{sec:soundwave_I}), following \citet{Berlok_2019} (in appendix \ref{app:soundwaveII} we discuss the results with the hydro solver on, and the creation of acoustic waves, which was already found in \citealp{Hopkins_2017}). In our setup, the box has dimensions of $L_x = 10L_y = 10L_z$, where the resolution is $N_x=128$ in $\hat{x}$, and $N_y = N_z =12$ in $\hat{y}$ and $\hat{z}$ respectively. 

\begin{figure}
    \centering
	\includegraphics[width=\columnwidth]{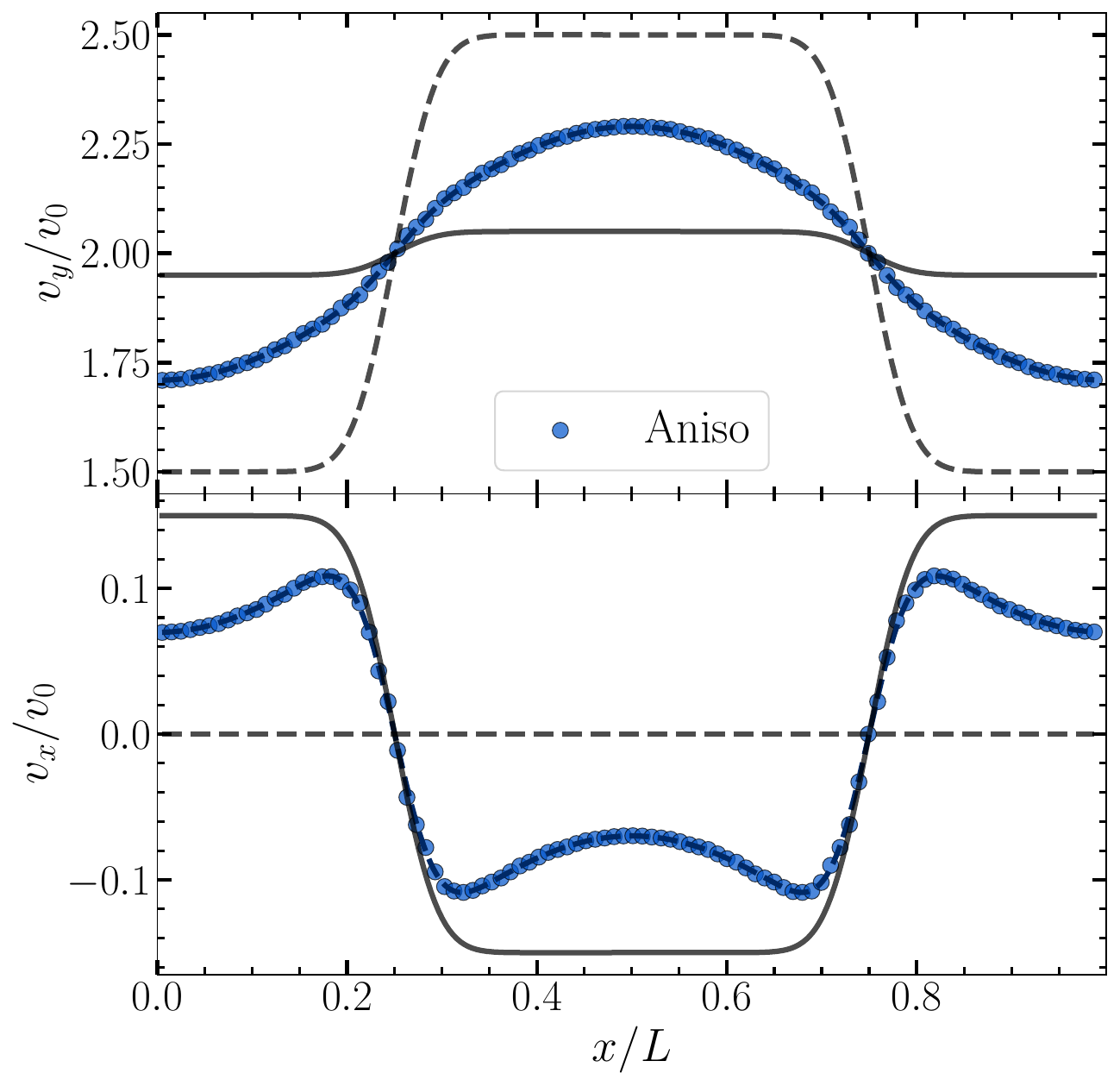}
    \caption{Velocity profiles of the sound wave described in eq. \eqref{eqn:ICs_soundwaveII} and \eqref{eqn:ICs_soundwaveII_q} after $ct/L=1$ for the anisotropic case. The black-dashed lines show the initial conditions and the solid lines the evolution after $t\gg1$. We compare our results (blue dots) with the exact analytical solution (blue-dashed lines). {\it Top panel}: $v_y$ profile. {\it Bottom panel}: $v_x$ profile triggered by $\Delta p$ due to the misalignment between the magnetic field and the wave propagation.}
    \label{fig:Soundwave_II_vel}
\end{figure}

The energy transfer from $v_y$ to $v_x$ due to the pressure anisotropy can be seen in Fig.~\ref{fig:Soundwave_II_vel} after $ct/L = 1$ (upper panel for $v_y$ and lower panel for $v_x$). The initial conditions are indicated by the black-dashed line, and the black-solid lines indicate the solution when $t \gg 1$. The results match exactly the analytical solution, showing how the pressure anisotropy generates a force in $\hat{x}$ that progressively leads to a $v_x$ flow.

\begin{figure}
    \centering
	\includegraphics[width=\columnwidth]{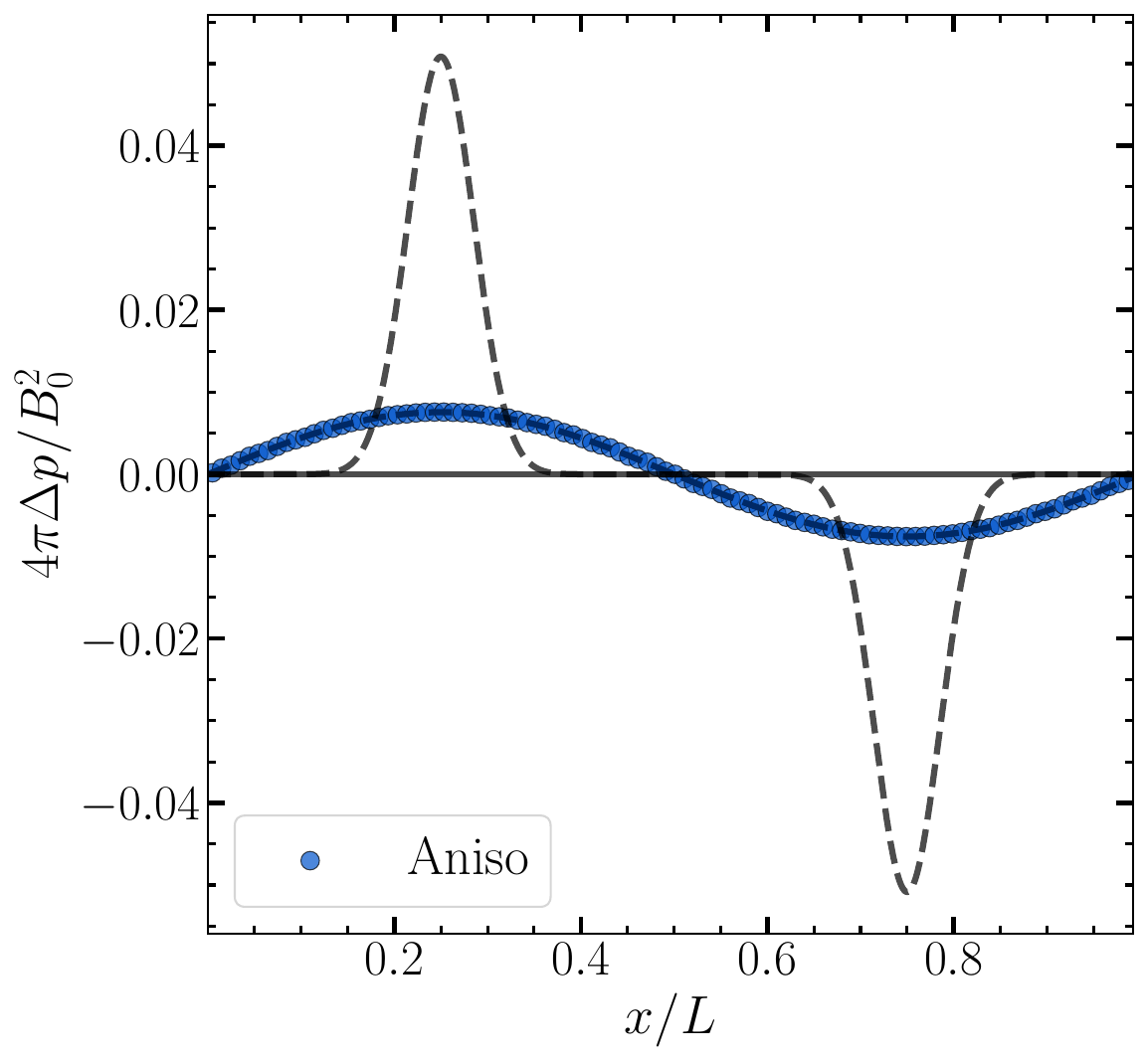}
    \caption{$\Delta p$ profile of the sound wave described by eq. \eqref{eqn:ICs_soundwaveII} and \eqref{eqn:ICs_soundwaveII_q} after $ct/L=1$ for the anisotropic case (blue dots). Black-dashed line shows the initial conditions, and the solid line shows the evolution after $t\gg1$. The exact analytical solution is given by the blue-dashed line.}
    \label{fig:Soundwave_II_PANI}
\end{figure}

The evolution of the pressure anisotropy triggered due to the misalignment of the magnetic field and the wavevectors is given by 
\begin{equation}
    \Delta p (x,t) = -\frac{3 \rho c \nu}{2} \sum_{n=1}^{\infty} k_n a_n \sin (k_n x) \, {\rm e}^{-\gamma_n t} \, .
\end{equation}
Fig.~\ref{fig:Soundwave_II_PANI} shows that the profile of $\Delta p$ in our simulations after $ct/L = 1$ matches exactly the profile predicted theoretically.

\begin{figure}
    \centering
	\includegraphics[width=\columnwidth]{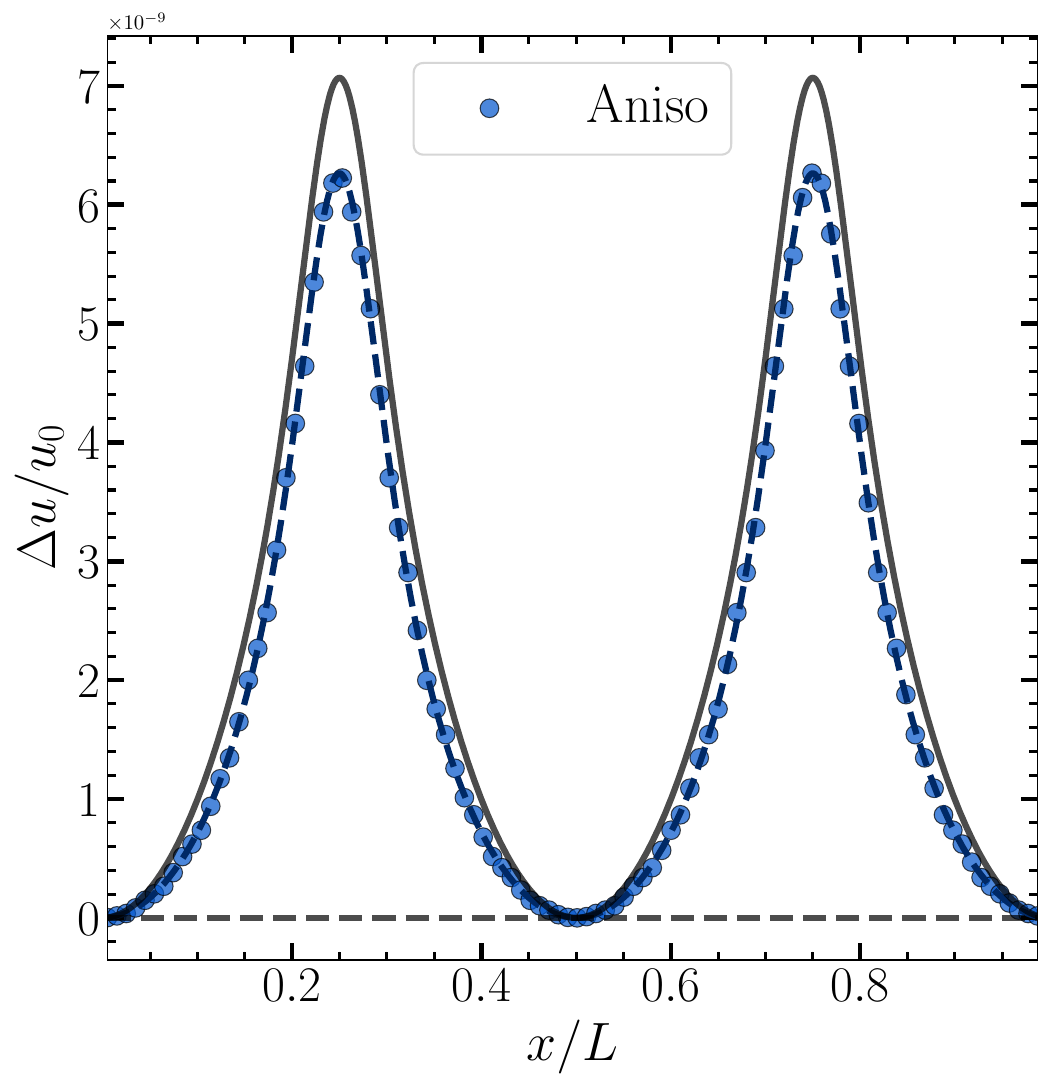}
    \caption{Cumulative viscous heating profile due to Braginskii viscosity of the sound wave described by eq. \eqref{eqn:ICs_soundwaveII} and \eqref{eqn:ICs_soundwaveII_q} after $ct/L=1$. The blue dots show the results for the anisotropic case, and the blue-dashed line shows the analytical solution. The black-dashed line shows the initial conditions and the solid line the evolution after $t\gg1$.}
    \label{fig:Soundwave_II_E}
\end{figure}

In the process of converting kinetic energy from $v_y$ to $v_x$, a fraction is dissipated into heat following the expression derived by \citet{Berlok_2019}:
\begin{multline}
    \Delta u(t) = u_0 + \frac{9 \rho c^2}{10} \sum_{n=1}^{\infty} \sum_{m=1}^{\infty} a_n a_m \frac{\sqrt{\gamma_n \gamma_m}}{\gamma_n + \gamma_m} \sin (k_n x) \sin(k_mx) \\ \times \left( 1 - {\rm e}^{-(\gamma_n + \gamma_m)t} \right) \, ,
\end{multline}
where $u_0$ is the initial internal energy. The evolution of the cumulative viscous heating over time can be seen in Fig.~\ref{fig:Soundwave_II_E}. Similarly to the test \ref{sec:soundwave_I}, viscosity heats the plasma mainly in the nodes of the wave ($x/L=0.25, 0.75$), where the velocity gradient is maximum.

\subsection{Circularly Polarized Alfvén Wave} \label{sec:circ_alfven}

\begin{figure}
    \centering
	\includegraphics[width=\columnwidth]{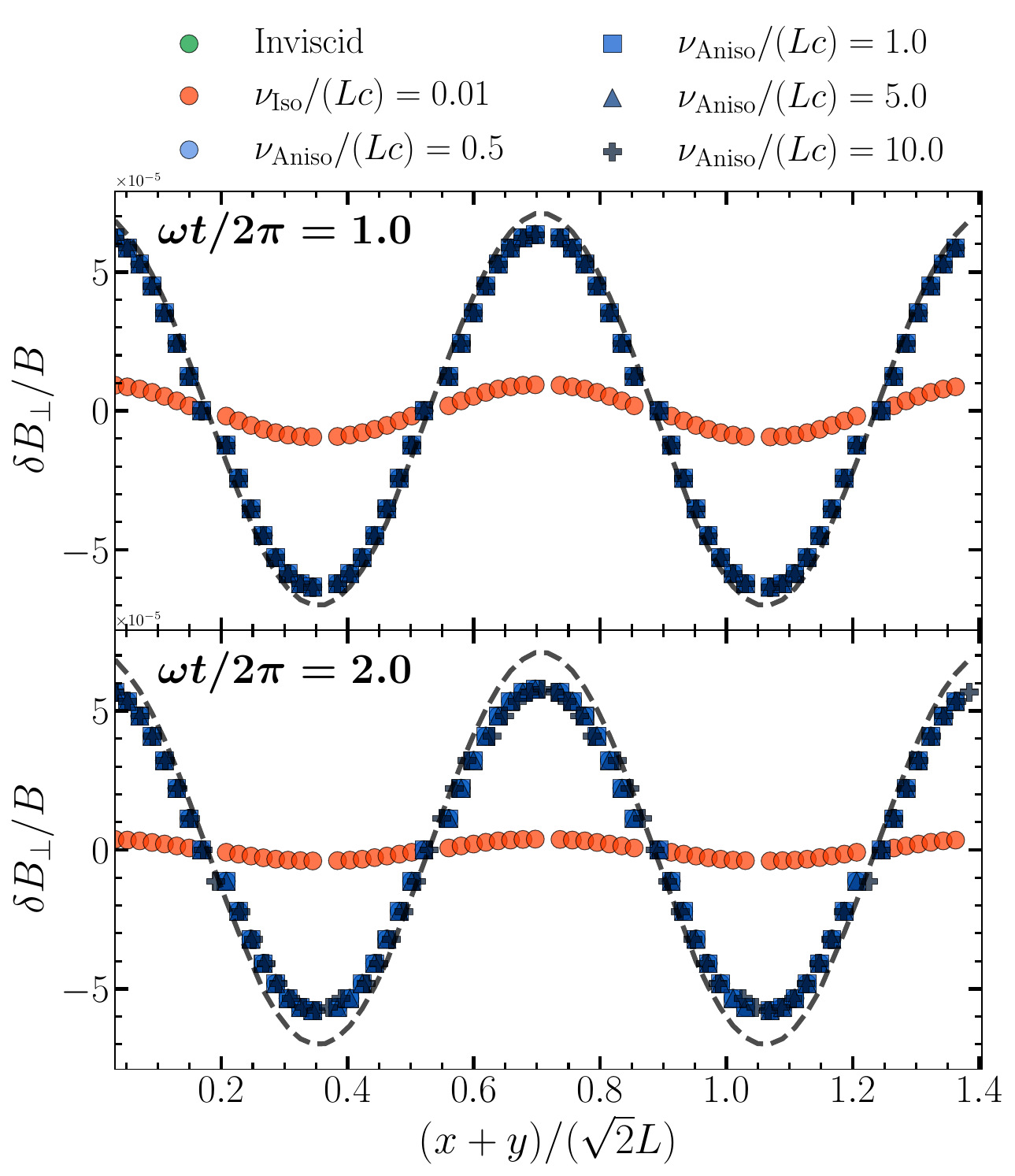}
    \caption{$B_{\perp}$ profile of the circularly polarized Alfvén wave described by eq. \eqref{eqn:ICs_circ_alfven_B} and \eqref{eqn:ICs_circ_alfven_v} for the inviscid case (green dots), isotropic case (red dots), and different amounts of anisotropic viscosity (blue markers). The different markers indicate different levels of anisotropic viscosity. The black-dashed lines show the initial amplitude. {\it Top panel}: Result after one period ($\omega t/2\pi = 1.0$). {\it Bottom panel}: Result after two periods ($\omega t/2\pi = 2.0$).}
    \label{fig:Circ_wave}
\end{figure}

A particularly useful test is the circularly polarized Alfvén wave, in which the density and the magnetic field strength stay constant, leading to a zero pressure anisotropy (see eq. \eqref{eqn:pressure_aniso}). This means that, in the presence of anisotropic viscosity, the circularly polarized Alfvén wave does not decay, in contrast to the isotropic case. To test this, we set a magnetic field ${\bf B} = B_0 \hat{b}$, with $\hat{b} = (\hat{x}+\hat{y})/\sqrt{2}$, with an initial perturbation
\begin{equation}
    \frac{\delta {\bf B}}{B_0} = A \left( \cos({\bf k \cdot r})  \frac{\hat{y} - \hat{x}}{\sqrt{2}} - \sin({\bf k \cdot r}) \hat{z}\right) \, ,
    \label{eqn:ICs_circ_alfven_B}
\end{equation}
and
\begin{equation}
    \delta v = - \frac{\omega_0}{k} \frac{\delta {\bf B}}{B_0} \, .
    \label{eqn:ICs_circ_alfven_v}
\end{equation}
The wavevector is set parallel to the magnetic field ${\bf k} = k_{\parallel} \, \hat{b}$, with $k_{\parallel} = 2 \sqrt{2} \pi / L$, and the frequency of the Alfvén wave $\omega_0 = k_{\parallel} v_A$, with $v_A$ the Alfvén velocity 
\begin{equation}
    v_A = \frac{B}{\sqrt{4 \pi \rho}} \, .
\end{equation}
We use a 3D setup, where the resolution of the box is $N = 32^3$, with a length $L$ on each side. Fig.~\ref{fig:Circ_wave} shows the results for the amplitude of the wave after one period ($\omega_0 t / 2 \pi = 1$, top panel) and after two periods ($\omega_0 t / 2 \pi = 2$, bottom panel) for the non-viscous case, isotropic viscosity with $\nu_{\rm Iso} / (Lc) = 0.01$, and four different amounts of anisotropic viscosity: $\nu_{\rm Aniso} / (Lc) = 0.5, 1, 5, 10$. The physical motivation for the different levels of viscosity employed is discussed in appendix \ref{app:ratio_visc_icm}. After one period, the inviscid and anisotropic cases retain exactly the initial amplitude (black dashed line), whereas the isotropic viscosity strongly damps the wave. After two periods, the inviscid and anisotropic runs still follow the initial amplitude, although numerical viscosity starts to damp the wave in all cases. This residual damping is numerical and depends on the resolution and the details of the MHD discretization; it is not caused by the Braginskii viscous term, as shown by the agreement between the inviscid and anisotropic runs even for very large values of $\nu_{\rm Aniso}$. Compared with the results shown in \citet{Berlok_2019}, the case with $\nu_{\rm Aniso} / (Lc) = 10$ does not show numerical noise and follows the inviscid case solution even after $\omega_0 t / 2\pi = 2$. However, our simulations use a higher resolution and a smaller perturbation amplitude.
It is important to note that the isotropic case strongly damps the wave, although the Spitzer viscosity value (eq. \eqref{eqn:viscosity}) is $10^3$ smaller than the highest value of anisotropic viscosity. This shows the robustness and the accuracy of our Braginskii implementation. If $\Delta p$ were not strictly zero in the anisotropic cases, the wave would be strongly damped immediately due to the extremely high Spitzer viscosity value. 

\subsection{Linearly Polarized Alfvén Wave} \label{sec:lin_alfven}

\begin{figure}
    \centering
	\includegraphics[width=\columnwidth]{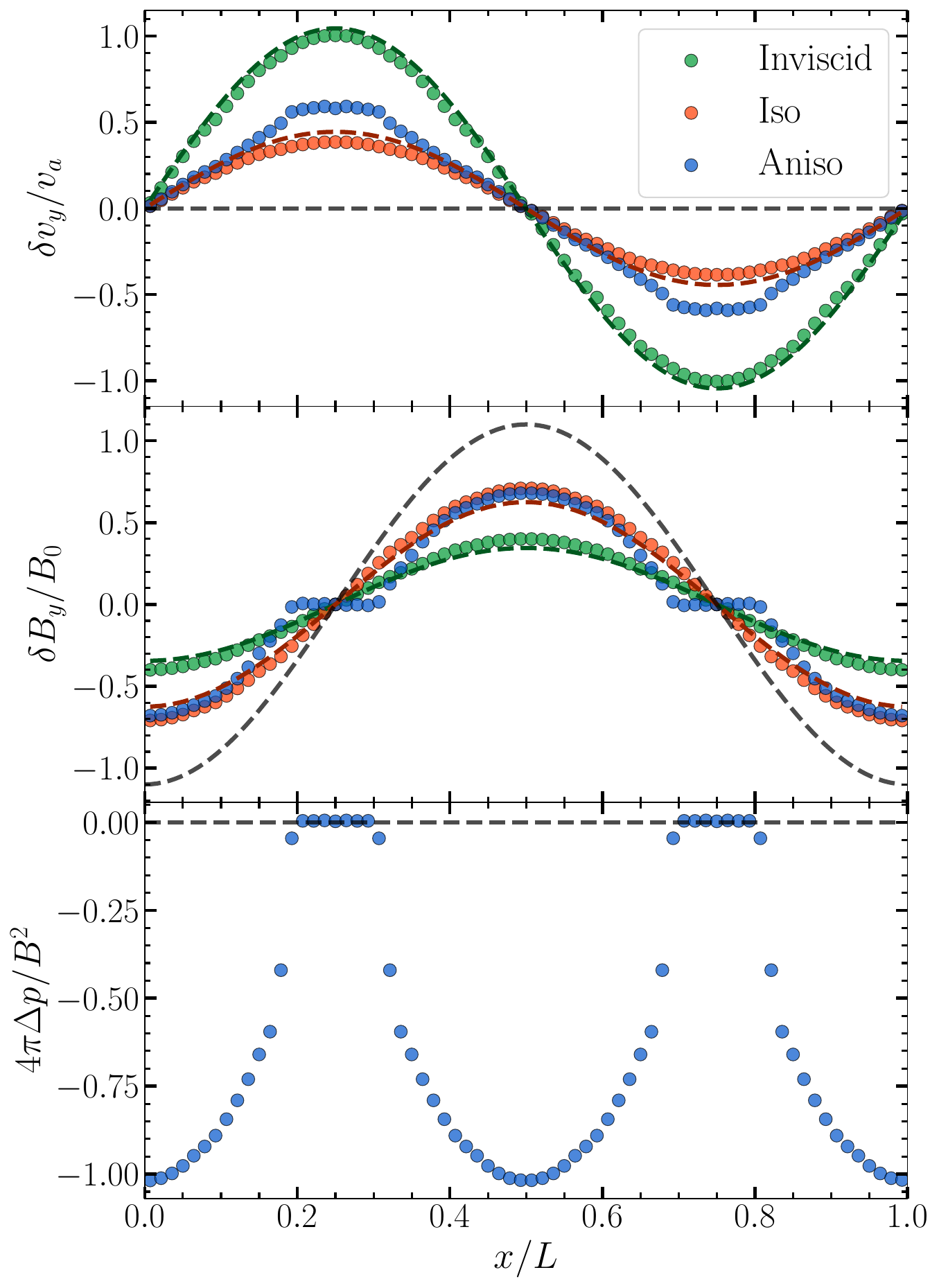}
    \caption{Linearly polarized Alfvén wave (eq. \eqref{eqn:linear_alfven_ICs}) after $\omega t = 0.2$ for the inviscid (green), isotropic (red), and anisotropic (blue) cases. The color-dash lines show the inviscid and isotropic analytical solutions, while the black-dashed lines show the initial conditions. {\it Top panel}: $\delta v_y$ profile. {\it Middle panel}: $\delta B_y$ profile. {\it Bottom panel}: $\Delta p$ profile normalized to $B^2/4\pi$ to highlight the firehose instability limit at $4\pi \Delta p / B^2 = -1$.}
    \label{fig:Lin_wave}
\end{figure}

While the Braginskii viscosity does not affect a circularly polarized Alfvén wave, it interrupts a linearly polarized standing Alfvén wave \citep{Squire_2016, Squire_2017, Squire_2017b}. In a linearly polarized Alfvén wave, $|\bf B|$ oscillates as the wave evolves, driving $\Delta p$ \eqref{eqn:pressure_aniso}. The wave is initialized at maximum magnetic perturbation and zero velocity, so $|{\bf B}|$ initially decreases. This produces negative pressure anisotropy, $\Delta p<0$, which reduces the magnetic tension. When $\Delta p = -B^2/4\pi$, the firehose instability is triggered \eqref{eqn:firehose_inst} and the effective magnetic tension vanishes. This mechanism interrupts the linearly polarized Alfvén wave if the initial amplitude of the wave is \citep{Squire_2017, Berlok_2019}
\begin{equation}
    A \gtrsim \sqrt{\frac{2v_A^2}{3 \nu \, \omega_0}} \, .
\end{equation}
To recreate this test, we set up a thin box of size $L_x = L_y = 20L_z$, with $N_x =N_y = 128$ and $N_z = 6$. The initial magnetic field is given by ${\bf B} = B_0 \hat{x}$, with a perturbation
\begin{equation}
    \frac{\delta {\bf B}}{B_0} = - A \cos (k x) \hat{y} \, ,
    \label{eqn:linear_alfven_ICs}
\end{equation}
with $k = 2\pi/L$. $B_0$ is chosen so $\beta = 10^3$, producing a minimum amplitude to interrupt the wave of $A_{\rm min} = 0.6$. To be able to trigger the firehose instability in our setup, we set an initial $A = 1.1$, with $\nu / (cL) = 0.01$, and zero initial velocity.
The evolution of the wave with isotropic viscosity is given by
\begin{equation}
    \frac{\delta {\bf B}}{B_0} = - A \mathrm{e}^{-\gamma t} \left[\cos(\Omega t)+\frac{\gamma}{\Omega}\sin(\Omega t)\right] \cos (k x) \, \hat{y} \, ,
    \label{eqn:evo_lin_alfven_B}
\end{equation}
\begin{equation}
    \frac{\delta {\bf v}}{v_A} = \frac{\omega_0}{\Omega} v_A A \,\mathrm{e}^{-\gamma t}\sin(\Omega t) \sin (kx) \, \hat{y} \, ,
    \label{eqn:evo_lin_alfven_v}
\end{equation}
where $\gamma$ is the damping rate, $\Omega = \sqrt{\omega_0^2 - \gamma^2}$, and $\omega_0 = kv_A$ (see appendix \ref{app:alfven_iso_visc} for the derivation). In the inviscid and isotropic cases, one recovers the analytical solutions given by eqs. \eqref{eqn:evo_lin_alfven_B} and \eqref{eqn:evo_lin_alfven_v}, without interruption of the wave (see Fig.~\ref{fig:Lin_wave}). However, in the presence of anisotropic viscosity, the wave evolution is interrupted due to the generation of pressure anisotropy. The velocity and magnetic field profiles (top and middle panel, respectively) are modified with respect to the analytical solutions \eqref{eqn:evo_lin_alfven_B} and \eqref{eqn:evo_lin_alfven_v}. The generation of negative pressure anisotropy triggers the firehose instability, which sets a lower limit (bottom panel), while in regions where the magnetic perturbation is zero, the pressure anisotropy is kept to zero.

This test highlights that, in contrast to isotropic viscosity, in high-$\beta$ weakly collisional plasmas, anisotropic viscosity imposes dynamical constraints that fundamentally alter Alfvénic dynamics. The wave self-organizes into tensionless and tension-balanced segments rather than decaying as a simple, phase-preserving mode.

\subsection{Fast Magnetosonic Wave}

\begin{figure}
    \centering
	\includegraphics[width=\columnwidth]{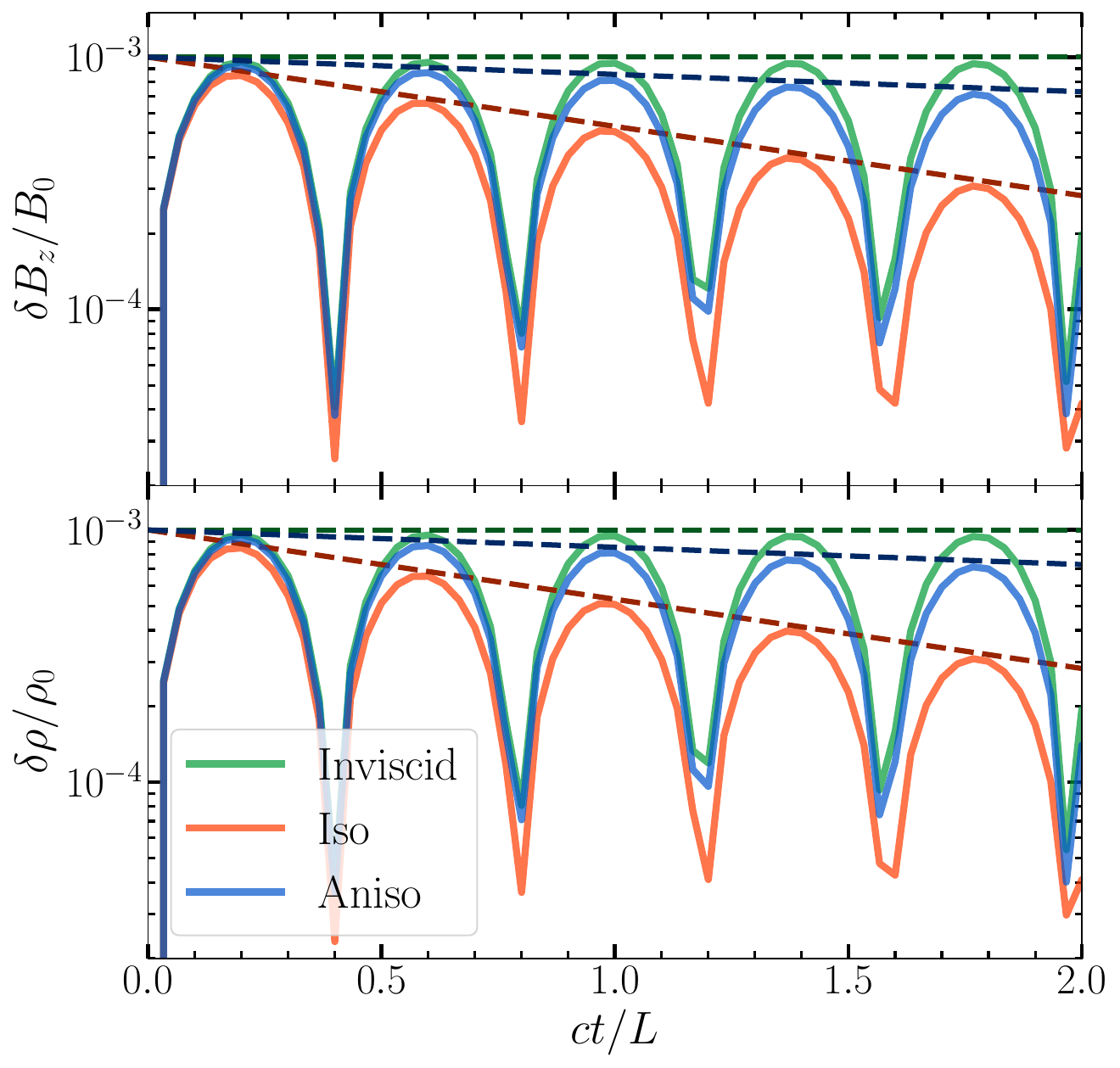}
    \caption{Decay of the fast magnetosonic wave initialized by eq.~\eqref{eqn:ICs_fast_wave} for the inviscid (green), isotropic (red), and anisotropic (blue) cases. The dashed lines show the theoretical decay of the amplitude for each case. {\it Top panel}: Evolution of $\delta B_z$. {\it Bottom panel}: Evolution of $\delta \rho$.}
    \label{fig:Fast_wave_evo}
\end{figure}

\begin{figure}
    \centering
	\includegraphics[width=\columnwidth]{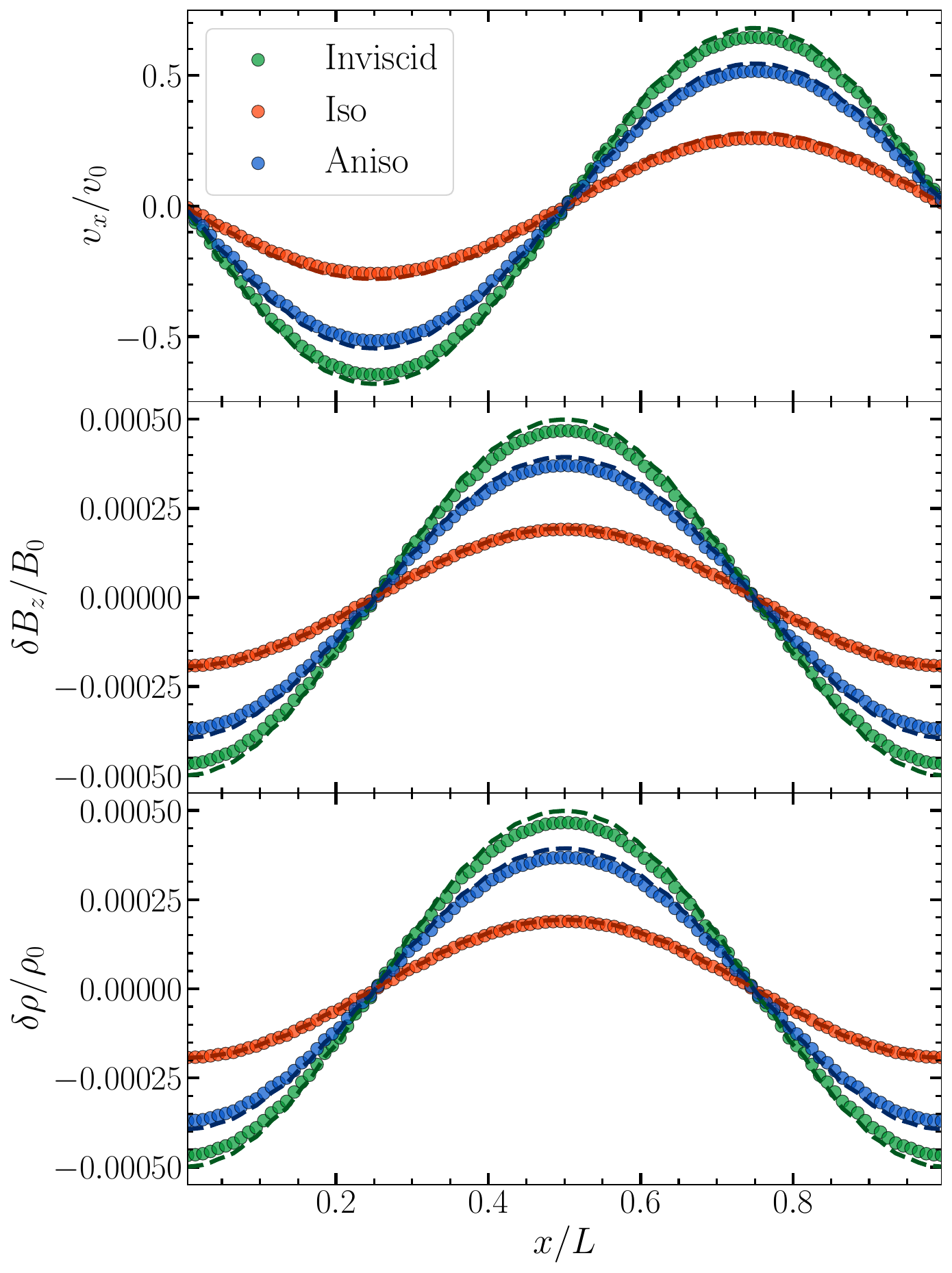}
    \caption{Profiles of the fast magnetosonic wave after $ct/L=1$. The data points show the results of our simulations, and the dashed lines show the analytical solutions: inviscid case in green, isotropic in red, and anisotropic in blue. {\it Top panel}: Profile of the $v_x$. {\it Middle panel}: Profile of the $\delta B_z$. {\it Bottom panel}: Profile of the $\delta \rho$.}
    \label{fig:Fast_wave}
\end{figure}

The fast magnetosonic wave is a compressive, propagating fluctuation. In the perpendicular setup considered here, its phase speed is given by
\begin{equation}
    v = \sqrt{c_s^2 + v_A^2} \, ,
\end{equation}
where the sound speed of the gas ($c_s$) and the Alfvén velocity ($v_A$) are of the same order of magnitude. 
We follow the initial conditions (ICs) for the fast magnetosonic wave test of \citet{Berlok_2019} to study how inviscid dynamics, isotropic viscosity, and Braginskii viscosity affect the compressible mode that couples density, velocity, and magnetic field. We use a 3D periodic domain of sides $L$ with a uniform magnetic field ${\bf B} = B_0 \hat{z}$, $\beta = 25$, and a resolution of $N = 128^3$. The wavevector is set perpendicular to ${\bf B}$ to excite a fast magnetosonic wave: ${\bf k} = k_x\hat{x} + k_y\hat{y}$, with $k_x = k_y = k_{\perp} / \sqrt{2} = 2 \pi / L$. An initial perturbation in velocity is used to trigger the wave:
\begin{equation}
    v({\bf r}, 0) = -A \sin({\bf k \cdot r}) \, \omega_0 \frac{\bf k}{k^2} \, ,
    \label{eqn:ICs_fast_wave}
\end{equation}
where $A = 10^{-3}$ is the initial amplitude, and $\omega_0$ is the real part of the dispersion relation: $\omega_0 = k_{\perp} \sqrt{v_A^2 + c_s^2}$ (see \citealp{Berlok_2019} for details). The evolution of the wave is described by
\begin{equation}
    v({\bf r}, t) = -A \sin({\bf k \cdot r}) \, \left[ \omega_0 \cos (\omega_0 t) - \gamma \sin(\omega_0 t) \right] {\rm e}^{-\gamma t} \frac{\bf k}{k^2} \, ,
    \label{eqn:fast_vel_analytic}
\end{equation}
where the damping rates are given by \eqref{eqn:damping_rates}. Since the fast magnetosonic wave is compressive, it triggers density and magnetic fluctuations, which evolve following
\begin{equation}
    \frac{\delta \rho}{\rho_0} = \frac{\delta B_z}{B_0} = A \cos ({\bf k \cdot r}) \sin (\omega_0 t) \, {\rm e}^{-\gamma t} \, .
    \label{eqn:fast_rho_B_evo}
\end{equation}
Fig.~\ref{fig:Fast_wave_evo} shows the amplitude evolution measured from the density and magnetic field fluctuation. The inviscid run keeps its initial amplitude over the entire simulation, while the isotropic and anisotropic runs follow an exponential decay with the damping rates predicted analytically. 
The velocity, magnetic field, and density profiles at $ct / L = 1$ are shown in Fig.~\ref{fig:Fast_wave}. In the inviscid case, the velocity profile (top panel) remains sinusoidal with the initial amplitude, following the analytical solution \eqref{eqn:fast_vel_analytic}. With both isotropic and anisotropic viscosity, the profile amplitude is reduced in agreement with the theoretical damping rates, where the damping of the anisotropic case is weaker compared to the isotropic case. These differences in velocity also lead to differences in the magnetic (middle panel) and density fluctuations (bottom panel), all three cases following the expected theoretical behavior. 

\subsection{Kelvin-Helmholtz Instability}

To perform a more complex test, we use the KHI setup described in \citet{Marin-Gilabert_2022}, although in this case we add a magnetic field $\mathbf{B}$ and increase the shear velocity. We initialize a periodic 3D box of size $\Delta x=\Delta y=256$ and $\Delta z=8$ kpc with two fluids in pressure equilibrium, following \citet{Marin-Gilabert_2022}. The initial density, temperature, and shear velocity are
\begin{equation}
    \rho, T, v_x =
    \begin{cases}
        \rho_{\rm c}, T_{\rm c}, v_{\rm c}, & |y| < 64, \\
        \rho_{\rm h}, T_{\rm h}, v_{\rm h}, & |y| > 64,
    \end{cases}
\end{equation}
with $\rho_{\rm c}=6.26\times10^{-8}$, $\rho_{\rm h}=3.13\times10^{-8}$, $T_{\rm c}=2.5\times10^6$, $T_{\rm h}=5\times10^6$, $v_{\rm c}=-80$, and $v_{\rm h}=80$ in code units. Thus, $\rho_{\rm c}/\rho_{\rm h}=T_{\rm h}/T_{\rm c}=2$ and $\Delta v_{\rm shear}=160$. The instability is seeded by adding a small perturbation to the $y$-velocity at the two interfaces, $y_{\rm Int}=\pm64$,
\begin{multline}
    v_y = -\delta v_y \left[
    \sin \left( \frac{2\pi (x+\lambda/2)}{\lambda}\right)
    \exp \left(-\left(\frac{y-y_{\rm Int}}{\sigma}\right)^2\right)
    \right. \\
    \left.
    + \sin \left( \frac{2\pi x}{\lambda}\right)
    \exp \left(-\left(\frac{y+y_{\rm Int}}{\sigma}\right)^2\right)
    \right] \, ,
    \label{eqn:khi_seed}
\end{multline}
with $\lambda=128$, $\sigma=0.2\lambda$, and $\delta v_y=|v_{\rm c}|/10=8$ in code units.
With this setup, the velocity gradient has only a $y$-component; therefore, we should see the maximum viscous effects when $\mathbf{B} = B_0\,\hat{y}$. However, in the case of a magnetic field in the $\hat{y}$ direction, the fluids' motion leads to a shear amplification due to the compression of the magnetic field lines, thus suppressing the instability \citep[][]{Das_2023}. For this reason, we can only include a magnetic field in the $\hat{z}$ direction ($B_z \perp \nabla v$) and in the $\hat{x}$ direction ($B_x \perp \nabla v$). In the presence of magnetic fields in the shear direction ($\hat{x}$), the KHI is fully suppressed due to magnetic tension when 
\begin{equation}
    B_{\rm h}^2 + B_{\rm c}^2 > 4\pi \frac{\rho_{\rm h} \rho_{\rm c}}{\rho_{\rm h} + \rho_{\rm c}} \Delta v_{\rm shear}^2 \, ,
    \label{eqn:mag_suppression}
\end{equation}
where $B_{\rm h}$ and $B_{\rm c}$ are the magnetic fields of the hot and cold medium, respectively \citep{Vikhlinin_2001}. In our setup, where $B_{\rm h} = B_{\rm c}$, $\rho_{\rm c} = 2\rho_{\rm h}$ and $\Delta v_{\rm shear} = 80$, the KHI is suppressed when $\beta \lesssim 16$. Therefore, we use a $\beta = 10^3$ to be able to study the growth of the instability depending on viscosity without suppression due to magnetic tension. We ran the KHI simulation for four different cases: inviscid, isotropic (Spitzer) viscosity, anisotropic (Braginskii) viscosity, and anisotropic (Braginskii) viscosity + plasma microinstability limiters. In all the viscous cases, the same constant value of the Spitzer coefficient is used: $\eta = 25\eta_{\rm Crit}$, where $\eta_{\rm Crit}$ is the critical viscosity needed to suppress the instability in this setup \citep{Marin-Gilabert_2022, Marin-Gilabert_2025}. We choose a high value of the Spitzer coefficient to highlight the differences between the isotropic and the anisotropic case.
\begin{figure*}
    \centering
	\includegraphics[width=\textwidth]{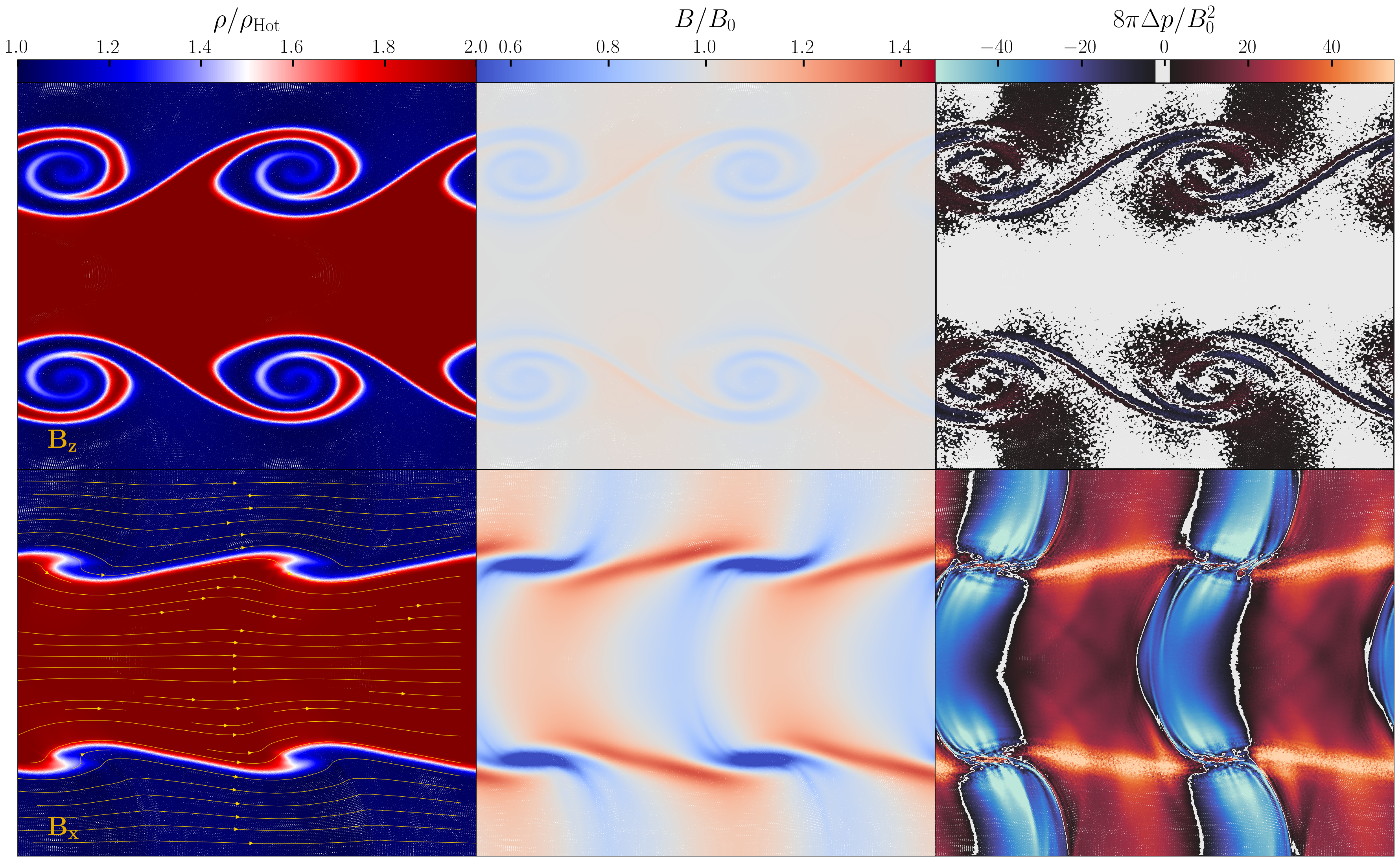}
    \caption{Colormaps in the $xy$-plane of the KHI with anisotropic viscosity without plasma microinstability limiters at $t=1.5\tau_{\rm KH}$. \textit{Top row}: Initial magnetic field in the $\hat{z}$ direction. \textit{Bottom row}: Initial magnetic field in the $\hat{x}$ direction. \textit{Left column}: Density colormap, normalized to the hot gas density; yellow arrows indicate the in-plane magnetic-field direction, i.e. the $(B_x,B_y)$ components, thus they are not visible in the upper panel. \textit{Middle column}: Magnetic field strength colormap, normalized to the initial magnetic field. \textit{Right column}: Pressure anisotropy, normalized to $B_0^2/8\pi$. White regions satisfy the mirror/firehose stability bounds, $-2 < 8\pi\Delta p/B_0^2 < 1$, while blue and orange regions exceed the firehose and mirror thresholds, respectively.}
    \label{fig:Braginskii_colormap}
\end{figure*}

Fig.~\ref{fig:Braginskii_colormap} shows the $xy$-plane of the KHI simulations with Braginskii viscosity without the plasma microinstability limiters after $t = 1.5 \tau_{\rm KH}$ of the simulation with $B_z$ in the top row, and $B_x$ in the bottom row. The left column shows the density colormap normalized to the hot medium density. In the run with initial $B_z$ (upper row), the KHI can grow similarly to the non-viscous case shown in \citet{Marin-Gilabert_2022}, where the characteristic rolls of the KHI are fully developed\footnote{Note that the version of the code used here is different than the one used in \citet{Marin-Gilabert_2022}, therefore the results might be slightly different.}. The magnetic field remains similar to the initial one after $1.5 \tau_{\rm KH}$ (middle panel), while the pressure anisotropy remains close to zero (right panel). The white colors show the regions where the plasma remains within the mirror/firehose stability bounds, which in the case of initial $B_z$ is the majority of the fluid.

In the run with $B_x$ (bottom row of Fig.~\ref{fig:Braginskii_colormap}), although the magnetic field and the velocity gradient are initially perpendicular, the density colormap (left panel) shows a clear suppression of the instability. The reason is the bend of the magnetic field lines due to the growth of the KHI. Due to the fluid motions, the magnetic field lines (indicated by the yellow vector field in the colormap) are bent in the $\hat{y}$ direction (parallel to the velocity gradient), producing a non-zero pressure anisotropy that results in a strong suppression of the growth (note that $\eta \gg \eta_{\rm Crit}$). The plasma motions also lead to an increase in the magnetic field strength in areas where the magnetic field lines are compressed, or decrease where they are decompressed (middle panel). These processes lead to a non-zero pressure anisotropy (right panel), with a positive value in regions where the field strength increases and a negative value where the field strength decreases \citep{Schekochihin_2005, Squire_2023}. The pressure anisotropy colormap shows regions where the firehose instability is exceeded ($8\pi \Delta p / B_0^2 < -2$, blue colors) and regions where the mirror instability is exceeded ($8\pi \Delta p / B_0^2 > 1$, orange colors). Only a small fraction of the gas lies within the limits set by the plasma microinstabilities (white regions). This means that, if we switch on the hard wall limit for plasma microinstabilities, the value of $\Delta p$ of the vast majority of the plasma will be limited by the microinstabilities (see \S \ref{sec:microins}). Due to the weak magnetic field used ($\beta = 10^3$), the magnetic field lines are bent more easily. This has three effects: $i)$ the KHI is able to grow; $ii)$ the bend of the lines in the $\hat{y}$ direction produces a strong effect of anisotropic viscosity, resulting in the partial suppression of the KHI; $iii)$ the plasma microinstabilities are triggered easily.

\begin{figure}
    \centering
	\includegraphics[width=\columnwidth]{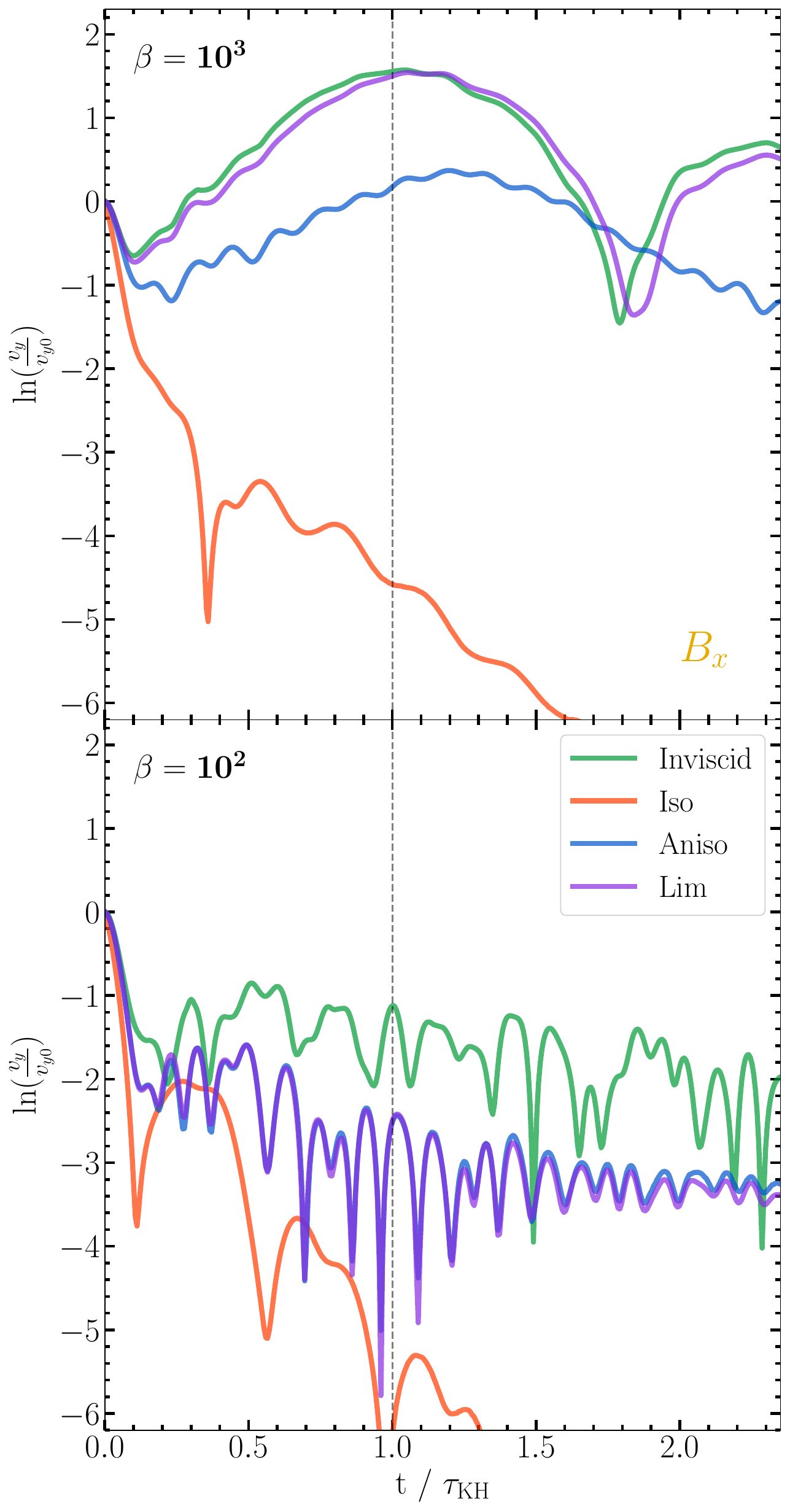}
    \caption{Growth rate of the KHI of a setup with ${\bf B} = B\hat{x}$ for different viscosity treatments: inviscid (green), isotropic (red), anisotropic (blue), and anisotropic with plasma microinstabilities limits (purple). \textit{Top panel}: Initial magnetic field strength of $\beta = 10^3$. \textit{Bottom panel}: Initial magnetic field strength of $\beta = 10^2$.}
    \label{fig:braginskii_McNally_1e3_1e2}
\end{figure}

The growth rate of the instability with $B_x$ can be seen in the upper panel of Fig.~\ref{fig:braginskii_McNally_1e3_1e2}. While the isotropic case (labeled as ``Iso'') strongly suppresses the KHI, the inviscid case (labeled as ``Inviscid'') allows the growth. However, the maximum $y$-velocity reached is lower than the $B_z$ case (see appendix \ref{app:growth_rate_Bz}), since the magnetic field itself slightly suppresses the growth of the instability. In the anisotropic case (labeled as ``Aniso''), there is an initial growth; however, as soon as the magnetic field lines are bent, viscosity suppresses the growth, reaching a much lower amplitude than the non-viscous case. As described above, the pressure anisotropy values exceed the limits set by the plasma microinstabilities. If we switch these limits on (labeled as ``Lim''), the growth of the instability follows a similar growth as the inviscid case, since the $\Delta p$ of the majority of the gas is limited by these microinstabilities.

\begin{figure*}
    \centering
	\includegraphics[width=0.8\textwidth]{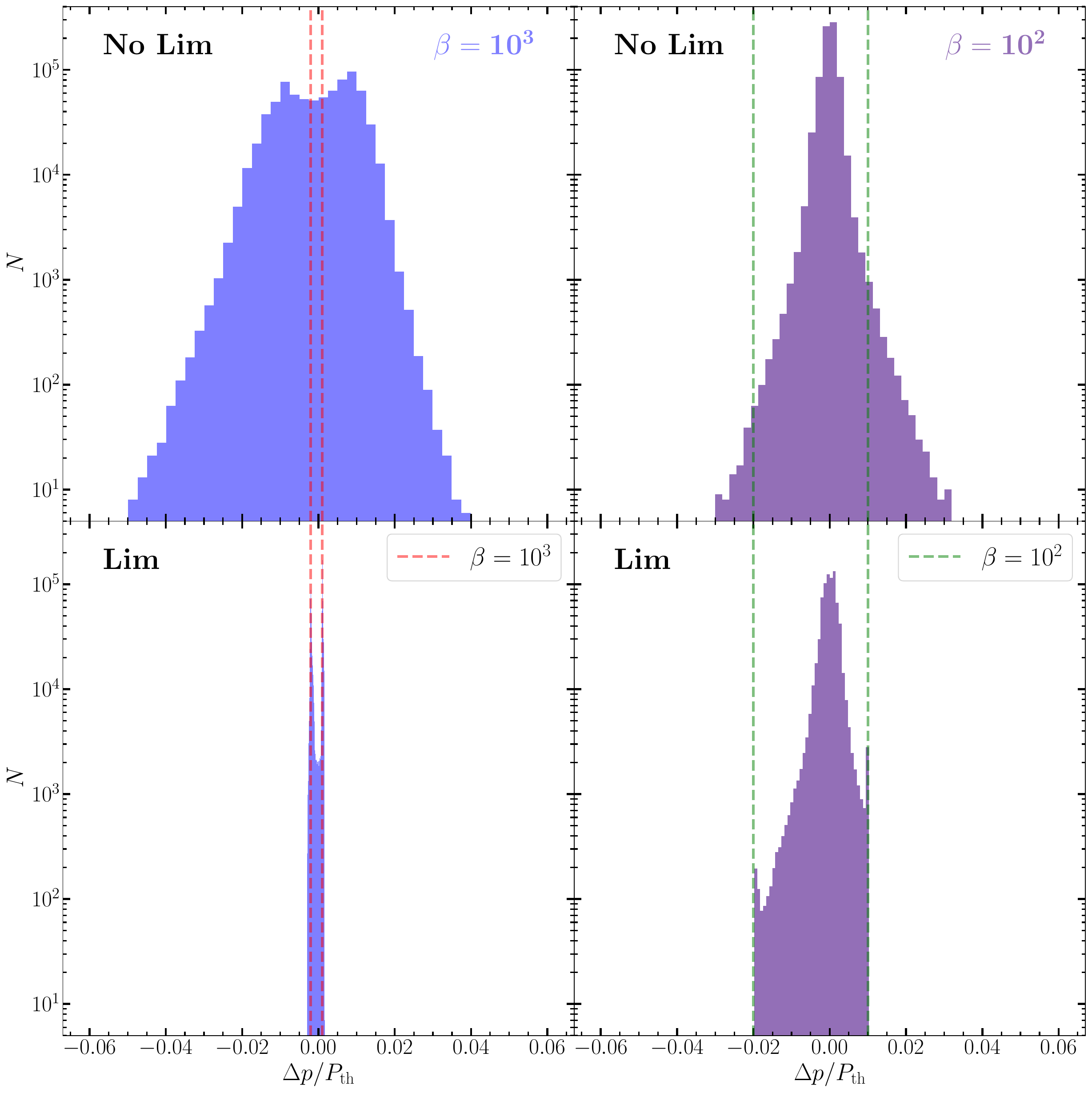}
    \caption{Histogram of the pressure anisotropy of all the particles of the KHI simulation, normalized to the thermal pressure. \textit{Top row}: Runs where $\Delta p$ can evolve without setting the plasma microinstabilities limits. \textit{Bottom row}: Runs where $\Delta p$ is limited to the mirror and firehose instabilities limits, for $\beta = 10^3$ (red-dashed line) and $\beta = 10^2$ (green-dashed line). \textit{Left column}: Results with $\beta = 10^3$. \textit{Right column}: Results with $\beta = 10^2$.}
    \label{fig:braginskii_pres_aniso_hist}
\end{figure*}

To analyze in detail the amount of particles affected if we switch on the plasma microinstability limits, Fig.~\ref{fig:braginskii_pres_aniso_hist} shows a histogram of $\Delta p$ of all the particles in the setup with $\beta = 10^3$ (left column), together with the values of the limits set by the microinstabilities (dashed lines). The top-left panel shows the histogram of the simulation without the limits after $t = 0.5\tau_{\rm KH}$, allowing $\Delta p$ to go beyond the limits. It shows how the vast majority of particles are found outside the limits set for $\beta = 10^3$. If we switch on the limits (bottom-left panel), the range allowed for $\Delta p$ is much smaller; thus, the viscous effect is largely reduced, which explains why the KHI behaves similarly to the inviscid case in Fig.~\ref{fig:braginskii_McNally_1e3_1e2}. 

For comparison, we also ran the KHI test with a stronger magnetic field: $\beta=10^2$. In this case, the magnetic field lines have a larger tension; thus, they are more difficult to bend. This leads to a stronger suppression of the KHI due to magnetic tension, as can be seen in the lower panel of Fig.~\ref{fig:braginskii_McNally_1e3_1e2}. The inviscid case grows (the slope is positive within $1\tau_{\rm KH}$, \citealp{Marin-Gilabert_2022}), but much less than the case with $\beta = 10^3$. Since the field lines are more difficult to bend, this means that the contribution to anisotropic viscosity is also lower than in the case of $\beta = 10^3$, although it still affects the growth of the KHI. The contribution of the magnetic component parallel to $\nabla v$ is smaller compared to the case with $\beta = 10^3$, resulting in a narrower range of values of $\Delta p$ (see top-right panel of Fig.~\ref{fig:braginskii_pres_aniso_hist}), producing that the majority of particles are found within the plasma microinstability limits for $\beta = 10^2$. This translates into a smaller overall effect of anisotropic viscosity. Additionally, the stronger magnetic field sets the plasma microinstability limits to larger values of $\Delta p$. Therefore, if we switch on the limits (bottom-right panel in Fig.~\ref{fig:braginskii_pres_aniso_hist}), the range of $\Delta p$ is mildly affected, and the behavior is similar to the anisotropic case without the limits (see lower panel of Fig.~\ref{fig:braginskii_McNally_1e3_1e2}).

\section{Cosmological simulations including Braginskii viscosity} \label{sec:cosmo_sims}

The main goal of this paper is to show that the Braginskii viscosity implementation can be effectively applied to cosmological simulations of galaxy clusters; therefore, the implementation must be validated in a cosmological context as well. To this end, in this section, we present the first-ever cosmological simulations of galaxy clusters including Braginskii viscosity.

We perform zoom-in simulations of one galaxy cluster of $M_{\rm Vir} = 2\times10^{15}$ M$_{\odot}$, with a particle mass resolution of $m_{\rm gas} = 1.56\times10^8$ M$_{\odot}$ in gas, and $m_{\rm DM} = 8.44\times10^8$ M$_{\odot}$ in dark matter. We picked a low resolution for testing, although in the future we plan to run simulations with the resolution achieved in \citet{Steinwandel_2024}. At such resolutions, the resolved macroscopic scales can approach the local Coulomb mean free path in parts of the ICM. However, below the mean free path, a fluid description is no longer the appropriate limit, and an MHD--kinetic hybrid or fully kinetic treatment would be required. The adopted cosmological parameters are $\Omega_0 = 0.24$, $\Omega_\Lambda = 0.76$, $\Omega_{\rm b} = 0.04$, $h = 0.72$, and $\sigma_8 = 0.8$, starting from an initial redshift of $z_{\rm ini} = 70$. We include magnetic fields based on the implementation of \citet{Bonafede_2011} and \citet{Stasyszyn_2013} with an initial seed of $B_{\rm ini} = 10^{-12}$~G (comoving), $5\%$ of isotropic thermal conduction, and artificial viscosity and conductivity \citep{Balsara_1995, Cullen_2010, Price_2008}. We adopted a Wendland $C^6$ kernel \citep{Wendland_1995, Dehnen_2012} with 295 neighbors. To isolate the effects of Braginskii viscosity, we perform non-radiative simulations, i.e., without including subgrid models like star formation or feedback.

The pressure anisotropy and the viscous stress tensor are calculated in comoving coordinates, using peculiar velocities and comoving spatial gradients \citep{Groth_2023}. The microinstability limiters \eqref{eqn:micro_limiters} in comoving coordinates are evaluated as
\begin{equation}
    -\frac{B_c^2 a}{4\pi} < \Delta p_c < \frac{B_c^2 a}{8\pi} \, ,
\end{equation}
with
\begin{equation}
    \Delta p_c \equiv a^{3\gamma}\Delta p_{\rm phys}
\end{equation}
\begin{equation}
    {\bf B}_c \equiv a^2 {\bf B}_{\rm phys} \, .
\end{equation}
The comoving variables are denoted by the subscript ``$c$'', ``phys'' denotes the physical units, $\gamma = 5/3$ is the adiabatic index, and $a$ is the scale factor. 

In future projects, we will analyze the effect of Braginskii viscosity in detail via key features like turbulence spectrum, velocity structure function, or density fluctuations. However, this paper is focused on qualitatively highlighting the effects of viscosity to show that the implementation is well-behaved, and the best way of doing this is by looking at the magnetic amplification. 

The amplification and evolution of magnetic fields are deeply linked with turbulence resulting from hierarchical cluster formation and mergers of clusters \citep{Schekochihin_2005, Subramanian_2006}. Turbulent motions amplify magnetic fields primarily through a dynamo process, where turbulent flows stretch and fold magnetic field lines, exponentially increasing their strength. This fluctuation dynamo can effectively amplify initially weak fields to observed microgauss strengths within typical cluster evolution timescales ($\sim 5$ Gyr) \citep{Brandenburg_2013}. During the kinematic (early) phase of the dynamo, the magnetic field grows exponentially:
\begin{equation}
    \frac{{\rm d}E_{\rm mag}}{{\rm d} t} = 2 \gamma E_{\rm mag} \, ,
    \label{eqn:dynamo}
\end{equation}
where $\gamma$ is the growth rate of the dynamo, which is correlated with the smallest eddies' turnover time \citep{Subramanian_2006, Steinwandel_2022}. Since viscosity suppresses the turbulence at small scales, it inherently affects the dynamo process, leading to weaker magnetic fields. 

\begin{figure*}
    \centering
	\includegraphics[width=\textwidth]{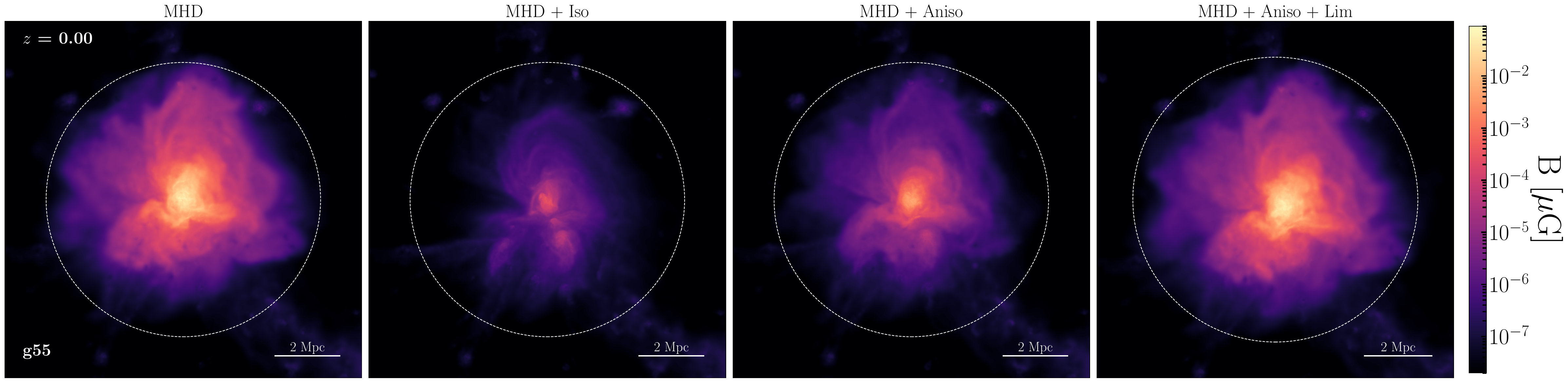}
    \caption{Projected magnetic field strength of the galaxy cluster cosmological simulation at $z=0$. The white circles indicate the $R_{200}$ of the cluster. {\it From left to right}: MHD only; MHD with Spitzer viscosity; MHD with Braginskii viscosity; and MHD with Braginskii viscosity and plasma microinstability limiters.}
    \label{fig:galaxy_cluster}
\end{figure*}

Fig.~\ref{fig:galaxy_cluster} shows the magnetic field strength at $z=0$ for the non-viscous case (first panel), full Spitzer viscosity case (second panel), Braginskii viscosity case without microinstability limiters (third panel), and Braginskii viscosity case with microinstability limiters (fourth panel). The absolute magnetic-field strength in these cosmological runs should not be interpreted quantitatively, because additional numerical dissipation in this setup suppresses the small-scale turbulent cascade and therefore reduces the efficiency of dynamo amplification. Nevertheless, the qualitative differences between the viscosity models can still be seen and still carry valuable information. The same large-scale structures can be seen in all cases; however, there are differences among the models that are worth highlighting:
\begin{enumerate}
    \item We find the biggest amplification of the magnetic field in the non-viscous case, where the dynamo effectively amplifies the magnetic field.
    \item The isotropic viscosity strongly suppresses turbulence at small scales, leading to a much weaker magnetic field at $z=0$ compared to the non-viscous case.
    \item In the anisotropic viscosity run without limiters, the magnetic field is amplified more effectively than in the isotropic case, but the strength reached at $z=0$ is still almost an order of magnitude lower than the non-viscous case.
    \item Finally, in the case with plasma limiters, the magnetic field strength reaches a similar magnitude as the non-viscous case. This is the result of the suppression of $\Delta p$, which translates into a lower viscous effect compared to the run without the limiters. However, it is important to note that the weaker magnetic field found in these simulations triggers the instabilities very easily. We expect that in future simulations with a higher magnetic field amplification, the limits will not be triggered so easily, leading to a larger $\Delta p$ range and a higher effective viscosity. 
\end{enumerate}

\section{Conclusions} \label{sec:conclusions}

In this work, we presented the implementation of Braginskii viscosity in the SPMHD code \textsc{OpenGadget3}. The implementation has been validated following the benchmark tests proposed in \citet{Berlok_2019}, showing great agreement with analytical solutions. We have compared the results with cases without viscosity and isotropic Spitzer viscosity, highlighting the different behavior of plasma in the presence of Braginskii viscosity. Additionally, we have shown cosmological simulations of galaxy clusters, proving the capability of the implementation in properly simulating weakly collisional plasmas in a cosmological context. Our key conclusions are:
\begin{itemize}
    \item In the presence of a constant magnetic field, when a sound wave propagates parallel to the magnetic field, the effect of the anisotropic viscosity is exactly the same as the isotropic viscosity. However, when the propagation is perpendicular to the magnetic field, the effect of the anisotropic viscosity is purely due to compression (or expansion) of the fluid, leading to a weaker effect compared to the isotropic case.
    \item For the second test, we set an $x$-dependent wave propagating in the $\hat{y}$ direction, with a static magnetic field with both parallel and perpendicular components. Under these conditions, the pressure anisotropy introduces a force in the $\hat{x}$ direction, due to the misalignment between the propagating wave and the magnetic field. This leads to significant differences with the isotropic and inviscid cases, where no forces in the $\hat{x}$ direction arise. Our solutions (with the hydro solver off) match exactly the analytical solution and the results shown in \citet{Berlok_2019}. However, when we switch the hydro solver on, the results differ slightly due to the development of acoustic waves. This is in agreement with the results found in \citet{Hopkins_2017}.
    \item We also simulated the propagation of a circularly polarized Alfvén wave. This particular type of MHD wave should not be suppressed in the presence of anisotropic viscosity, since it has a constant density and magnetic field strength, therefore zero pressure anisotropy. Our results show zero suppression in the anisotropic case, even with very high viscosity. In contrast, the run with isotropic viscosity strongly suppresses the amplitude of the wave even when the Spitzer value is $10^3$ lower than the anisotropic case. This contrast shows that the behavior is set by pressure anisotropy, rather than the absolute viscosity value, confirming the robustness of our Braginskii implementation.
    \item In a linearly polarized Alfvén wave, the anisotropic viscosity strongly affects the wave propagation. The wave is interrupted when the firehose instability is triggered, leading to a different evolution compared to the inviscid and isotropic cases. Our results show this interruption, matching the previous results of \citet{Squire_2016, Squire_2017b, Berlok_2019}. 
    \item We also tested our implementation in a fast magnetosonic wave setup, where our results match the expected amplitude damping of the magnetic field and density for the inviscid, isotropic, and anisotropic cases. The velocity, magnetic, and density profiles also fit the analytical solutions, demonstrating the accuracy of our scheme.
    \item In a more complex setup like the KHI, our results show the different behavior depending on the magnetic field direction. The shear motion was in the $\hat{x}$ direction; thus, to avoid magnetic suppression, we ran the simulation with magnetic fields in the $\hat{z}$ and in the $\hat{x}$ directions and an initial $\beta = 10^3$. When ${\bf B}=B\hat{z}$, the Braginskii viscosity has a negligible effect. However, when ${\bf B}=B\hat{x}$, the plasma motions bend the magnetic field in the $\hat{y}$ direction, leading to a non-zero pressure anisotropy and producing viscous effects that damp the growth of the KHI. Due to the weak magnetic field in this setup, the plasma microinstabilities are easily triggered. Therefore, when we switch on the plasma microinstability limits, the KHI is able to grow, reaching a similar amplitude to the inviscid case.
    \item With a stronger magnetic field ($\beta = 10^2$), the KHI growth is reduced due to the magnetic tension, and the magnetic field lines are bent less easily. As a result, the pressure anisotropy grows less than in the case with $\beta = 10^3$. The stronger magnetic tension also leads to larger plasma microinstability limits, producing that the instabilities are triggered less easily, thus the KHI behaves as the case without the limits.
    \item Finally, we tested our implementation in realistic galaxy cluster cosmological simulations. The results show how, in the presence of Spitzer viscosity, the turbulence suppression reduces the dynamo mechanism, leading to a weaker magnetic field compared to the inviscid case. The Braginskii viscosity also produces a weaker magnetic field at $z=0$, although not as weak as the Spitzer case, indicating a lower turbulence suppression than the isotropic case. This shows how the anisotropic nature of the Braginskii model effectively reduces the overall viscosity of the system. With plasma microinstability limiters, the magnetic field strength reached at $z=0$ is similar to the inviscid case, due to the limited pressure anisotropy.
\end{itemize}

In summary, all the tests performed match the expected analytical solutions, probing the robustness of our numerical implementation. The idealized tests highlight the differences between the anisotropic nature of the Braginskii viscosity and the isotropic nature of the Spitzer viscosity. These differences result in a different plasma behavior dependent not only on the viscosity model employed but also on the magnetic field direction and the wave propagation. In highly chaotic and complex scenarios like galaxy clusters, this might lead to very different macroscopic outcomes and observational signatures.

Our cosmological cluster runs demonstrate that the Braginskii module is ready to run realistic cosmological simulations of galaxy clusters. It captures the expected competition between anisotropic viscous transport, turbulence, and small-scale dynamo action. These results indicate that \textsc{OpenGadget3} can robustly evolve weakly collisional plasmas in cosmological settings and resolve the geometry-dependent coupling between viscosity, magnetic fields, and flow.

This work shows the capability of \textsc{OpenGadget3} to run cosmological simulations including Braginskii viscosity. However, future work will perform higher-resolution cosmological zoom-in simulations of massive clusters to better resolve the macroscopic turbulent cascade over the resolved inertial range. These simulations will enable a direct comparison between predicted turbulence levels and the anisotropic behavior of the plasma, and the turbulence velocities and turbulent pressure measured by XRISM, providing observational constraints on the effective viscosity and microinstability limiters in the intracluster medium.

\section*{Acknowledgements}

TM wants to thank Frederick Groth and Ludwig Böss for the intense discussions, and Thomas Berlok and Lorenzo Sironi for the useful comments. The authors also want to thank the referee for their very useful comments. KD and TM acknowledge support by the COMPLEX project from the European Research Council (ERC) under the European Union’s Horizon 2020 research and innovation program grant agreement ERC-2019-AdG 882679. This work has been supported by the Munich Excellence Cluster Origins, funded by the Deutsche Forschungsgemeinschaft (DFG, German Research Foundation) under Germany’s Excellence Strategy EXC-2094 390783311. TM acknowledges the support provided by a Smithsonian Scholar Award.
MV is supported by the Fondazione ICSC (National Recovery and Resilience Plan - PNRR), Project ID CN-00000013 "Italian Research Center on High-Performance Computing, Big Data and Quantum Computing" funded by MUR - Next Generation EU. MV also acknowledges partial financial support from the INFN Indark Grant. Support for JAZ was provided by the {\it Chandra} X-ray
Observatory Center, which is operated by the Smithsonian Astrophysical Observatory for and on behalf of NASA under
contract NAS8-03060.

\section*{Data Availability}
The data underlying this article will be shared on reasonable request to the author.
 



\bibliographystyle{mnras}
\bibliography{example} 

@article{Allen_2011,
   title={Cosmological Parameters from Observations of Galaxy Clusters},
   volume={49},
   ISSN={1545-4282},
   url={http://dx.doi.org/10.1146/annurev-astro-081710-102514},
   DOI={10.1146/annurev-astro-081710-102514},
   number={1},
   journal={Annual Review of Astronomy and Astrophysics},
   publisher={Annual Reviews},
   author={Allen, Steven W. and Evrard, August E. and Mantz, Adam B.},
   year={2011},
   month=sep, pages={409–470} 
}

@article{Arzamasskiy_2023,
  title = {Kinetic Turbulence in Collisionless High-$\ensuremath{\beta}$ Plasmas},
  author = {Arzamasskiy, Lev and Kunz, Matthew W. and Squire, Jonathan and Quataert, Eliot and Schekochihin, Alexander A.},
  journal = {Phys. Rev. X},
  volume = {13},
  issue = {2},
  pages = {021014},
  numpages = {34},
  year = {2023},
  month = {Apr},
  publisher = {American Physical Society},
  doi = {10.1103/PhysRevX.13.021014},
  url = {https://link.aps.org/doi/10.1103/PhysRevX.13.021014}
}

@article{Bale_2009,
  title = {Magnetic Fluctuation Power Near Proton Temperature Anisotropy Instability Thresholds in the Solar Wind},
  author = {Bale, S. D. and Kasper, J. C. and Howes, G. G. and Quataert, E. and Salem, C. and Sundkvist, D.},
  journal = {Phys. Rev. Lett.},
  volume = {103},
  issue = {21},
  pages = {211101},
  numpages = {4},
  year = {2009},
  month = {Nov},
  publisher = {American Physical Society},
  doi = {10.1103/PhysRevLett.103.211101},
  url = {https://link.aps.org/doi/10.1103/PhysRevLett.103.211101}
}

@article{Balsara_1995,
title = {von Neumann stability analysis of smoothed particle hydrodynamics—suggestions for optimal algorithms},
journal = {Journal of Computational Physics},
volume = {121},
number = {2},
pages = {357-372},
year = {1995},
issn = {0021-9991},
doi = {https://doi.org/10.1016/S0021-9991(95)90221-X},
url = {https://www.sciencedirect.com/science/article/pii/S002199919590221X},
author = {Dinshaw S. Balsara}
}

@article{Battaglia_2010,
   title={SIMULATIONS OF THE SUNYAEV-ZEL’DOVICH POWER SPECTRUM WITH ACTIVE GALACTIC NUCLEUS FEEDBACK},
   volume={725},
   ISSN={1538-4357},
   url={http://dx.doi.org/10.1088/0004-637X/725/1/91},
   DOI={10.1088/0004-637x/725/1/91},
   number={1},
   journal={The Astrophysical Journal},
   publisher={American Astronomical Society},
   author={Battaglia, N. and Bond, J. R. and Pfrommer, C. and Sievers, J. L. and Sijacki, D.},
   year={2010},
   month=nov, pages={91–99} 
}

@ARTICLE{Beck_2015,
author = {{Beck}, A.~M. and {Murante}, G. and {Arth}, A. and {Remus}, R. -S. and {Teklu}, A.~F. and {Donnert}, J.~M.~F. and {Planelles}, S. and {Beck}, M.~C. and {F{\"o}rster}, P. and {Imgrund}, M. and {Dolag}, K. and {Borgani}, S.},
title = "{An improved SPH scheme for cosmological simulations}",
journal = {\mnras},
year = 2016,
month = jan,
volume = {455},
number = {2},
pages = {2110-2130},
doi = {10.1093/mnras/stv2443},
archivePrefix = {arXiv},
eprint = {1502.07358},
primaryClass = {astro-ph.CO},
adsurl = {https://ui.adsabs.harvard.edu/abs/2016MNRAS.455.2110B}
}

@article{Berlok_2019,
   title={Braginskii viscosity on an unstructured, moving mesh accelerated with super-time-stepping},
   volume={491},
   ISSN={1365-2966},
   url={http://dx.doi.org/10.1093/mnras/stz3115},
   DOI={10.1093/mnras/stz3115},
   number={2},
   journal={Monthly Notices of the Royal Astronomical Society},
   publisher={Oxford University Press (OUP)},
   author={Berlok, Thomas and Pakmor, Rüdiger and Pfrommer, Christoph},
   year={2019},
   month=nov, pages={2919–2938} 
}

@article{Bonafede_2011,
title={A non-ideal magnetohydrodynamic gadget: simulating massive galaxy clusters: A non-ideal magnetohydrodynamic gadget},
volume={418},
ISSN={0035-8711},
url={http://dx.doi.org/10.1111/j.1365-2966.2011.19523.x},
DOI={10.1111/j.1365-2966.2011.19523.x},
number={4},
journal={Monthly Notices of the Royal Astronomical Society},
publisher={Oxford University Press (OUP)},
author={Bonafede, A. and Dolag, K. and Stasyszyn, F. and Murante, G. and Borgani, S.},
year={2011},
month=nov, pages={2234–2250} }

@article{Bott_2021,
   title={Adaptive Critical Balance and Firehose Instability in an Expanding, Turbulent, Collisionless Plasma},
   volume={922},
   ISSN={2041-8213},
   url={http://dx.doi.org/10.3847/2041-8213/ac37c2},
   DOI={10.3847/2041-8213/ac37c2},
   number={2},
   journal={The Astrophysical Journal Letters},
   publisher={American Astronomical Society},
   author={Bott, A. F. A. and Arzamasskiy, L. and Kunz, M. W. and Quataert, E. and Squire, J.},
   year={2021},
   month=nov, pages={L35} 
}

@ARTICLE{Braginskii_1965,
       author = {{Braginskii}, S.~I.},
        title = "{Transport Processes in a Plasma}",
      journal = {Reviews of Plasma Physics},
         year = 1965,
        month = jan,
       volume = {1},
        pages = {205},
       adsurl = {https://ui.adsabs.harvard.edu/abs/1965RvPP....1..205B}
}

@ARTICLE{Brandenburg_2013,
       author = {{Brandenburg}, A. and {Lazarian}, A.},
        title = "{Astrophysical Hydromagnetic Turbulence}",
      journal = {\ssr},
         year = 2013,
        month = oct,
       volume = {178},
       number = {2-4},
        pages = {163-200},
          doi = {10.1007/s11214-013-0009-3},
archivePrefix = {arXiv},
       eprint = {1307.5496},
 primaryClass = {astro-ph.SR},
       adsurl = {https://ui.adsabs.harvard.edu/abs/2013SSRv..178..163B}
}

@article{Carilli_2002,
   title={Cluster Magnetic Fields},
   volume={40},
   ISSN={1545-4282},
   url={http://dx.doi.org/10.1146/annurev.astro.40.060401.093852},
   DOI={10.1146/annurev.astro.40.060401.093852},
   number={1},
   journal={Annual Review of Astronomy and Astrophysics},
   publisher={Annual Reviews},
   author={Carilli, C. L. and Taylor, G. B.},
   year={2002},
   month=sep, pages={319–348} 
}

@article{Cullen_2010,
title={Inviscid smoothed particle hydrodynamics},
volume={408},
ISSN={0035-8711},
url={http://dx.doi.org/10.1111/j.1365-2966.2010.17158.x},
DOI={10.1111/j.1365-2966.2010.17158.x},
number={2},
journal={Monthly Notices of the Royal Astronomical Society},
publisher={Oxford University Press (OUP)},
author={Cullen, Lee and Dehnen, Walter},
year={2010},
month={Jul},
pages={669–683}
}

@article{Das_2023,
    author = {Das, Hitesh Kishore and Gronke, Max},
    title = "{Magnetic fields in multiphase turbulence: impact on dynamics and structure}",
    journal = {Monthly Notices of the Royal Astronomical Society},
    volume = {527},
    number = {1},
    pages = {991-1013},
    year = {2023},
    month = {10},
    issn = {0035-8711},
    doi = {10.1093/mnras/stad3125},
    url = {https://doi.org/10.1093/mnras/stad3125},
    eprint = {https://academic.oup.com/mnras/article-pdf/527/1/991/52977193/stad3125.pdf},
}

@article{Dehnen_2012,
title={Improving convergence in smoothed particle hydrodynamics simulations without pairing instability},
volume={425},
ISSN={0035-8711},
url={http://dx.doi.org/10.1111/j.1365-2966.2012.21439.x},
DOI={10.1111/j.1365-2966.2012.21439.x},
number={2},
journal={Monthly Notices of the Royal Astronomical Society},
publisher={Oxford University Press (OUP)},
author={Dehnen, Walter and Aly, Hossam},
year={2012},
month={Aug},
pages={1068–1082}
}

@article{Dolag_2005,
title={Turbulent gas motions in galaxy cluster simulations: the role of smoothed particle hydrodynamics viscosity},
volume={364},
ISSN={1365-2966},
url={http://dx.doi.org/10.1111/j.1365-2966.2005.09630.x},
DOI={10.1111/j.1365-2966.2005.09630.x},
number={3},
journal={Monthly Notices of the Royal Astronomical Society},
publisher={Oxford University Press (OUP)},
author={Dolag, K. and Vazza, F. and Brunetti, G. and Tormen, G.},
year={2005},
month=dec, pages={753–772} }

@article{Dolag_2009,
title={An MHD gadget for cosmological simulations},
volume={398},
ISSN={1365-2966},
url={http://dx.doi.org/10.1111/j.1365-2966.2009.15181.x},
DOI={10.1111/j.1365-2966.2009.15181.x},
number={4},
journal={Monthly Notices of the Royal Astronomical Society},
publisher={Oxford University Press (OUP)},
author={Dolag, K. and Stasyszyn, F.},
year={2009},
month=oct, pages={1678–1697} }

@article{Dong_2009,
title={BUOYANT BUBBLES IN INTRACLUSTER GAS: EFFECTS OF MAGNETIC FIELDS AND ANISOTROPIC VISCOSITY},
volume={704},
ISSN={1538-4357},
url={http://dx.doi.org/10.1088/0004-637X/704/2/1309},
DOI={10.1088/0004-637x/704/2/1309},
number={2},
journal={The Astrophysical Journal},
publisher={American Astronomical Society},
author={Dong, Ruobing and Stone, James M.},
year={2009},
month={Oct},
pages={1309–1320}
}

@ARTICLE{Gaspari_2011,
       author = {{Gaspari}, M. and {Melioli}, C. and {Brighenti}, F. and {D'Ercole}, A.},
        title = "{The dance of heating and cooling in galaxy clusters: three-dimensional simulations of self-regulated active galactic nuclei outflows}",
      journal = {\mnras},
         year = 2011,
        month = feb,
       volume = {411},
       number = {1},
        pages = {349-372},
          doi = {10.1111/j.1365-2966.2010.17688.x},
archivePrefix = {arXiv},
       eprint = {1007.0674},
 primaryClass = {astro-ph.CO},
       adsurl = {https://ui.adsabs.harvard.edu/abs/2011MNRAS.411..349G}
}

@article{Groth_2023,
author = {Groth, Frederick and Steinwandel, Ulrich P and Valentini, Milena and Dolag, Klaus},
title = "{The cosmological simulation code OpenGadget3 – implementation of meshless finite mass}",
journal = {Monthly Notices of the Royal Astronomical Society},
volume = {526},
number = {1},
pages = {616-644},
year = {2023},
month = {09},
issn = {0035-8711},
doi = {10.1093/mnras/stad2717},
url = {https://doi.org/10.1093/mnras/stad2717},
eprint = {https://academic.oup.com/mnras/article-pdf/526/1/616/51771150/stad2717.pdf},
}

@article{Groth_2025,
   title={Turbulent pressure support in galaxy clusters: Impact of the hydrodynamical solver},
   volume={693},
   ISSN={1432-0746},
   url={http://dx.doi.org/10.1051/0004-6361/202451803},
   DOI={10.1051/0004-6361/202451803},
   journal={Astronomy \&; Astrophysics},
   publisher={EDP Sciences},
   author={Groth, Frederick and Valentini, Milena and Steinwandel, Ulrich P. and Vallés-Pérez, David and Dolag, Klaus},
   year={2025},
   month=jan, pages={A263} 
}

@article{Hellinger_2000,
    author = {Hellinger, P. and Matsumoto, H.},
    title = {New kinetic instability: Oblique Alfvén fire hose},
    journal = {Journal of Geophysical Research: Space Physics},
    volume = {105},
    number = {A5},
    pages = {10519-10526},
    doi = {https://doi.org/10.1029/1999JA000297},
    url = {https://agupubs.onlinelibrary.wiley.com/doi/abs/10.1029/1999JA000297},
    eprint = {https://agupubs.onlinelibrary.wiley.com/doi/pdf/10.1029/1999JA000297},
    year = {2000}
}

@ARTICLE{Hitomi_2016,
       author = {{Hitomi} and {Aharonian}, Felix and {Akamatsu}, Hiroki and {Akimoto}, Fumie and {Allen}, Steven W. and {Anabuki}, Naohisa and {Angelini}, Lorella and {Arnaud}, Keith and {Audard}, Marc and {Awaki}, Hisamitsu and {Axelsson}, Magnus and {Bamba}, Aya and {Bautz}, Marshall and {Blandford}, Roger and {Brenneman}, Laura and {Brown}, Gregory V. and {Bulbul}, Esra and {Cackett}, Edward and {Chernyakova}, Maria and {Chiao}, Meng and {Coppi}, Paolo and {Costantini}, Elisa and {de Plaa}, Jelle and {den Herder}, Jan-Willem and {Done}, Chris and {Dotani}, Tadayasu and {Ebisawa}, Ken and {Eckart}, Megan and {Enoto}, Teruaki and {Ezoe}, Yuichiro and {Fabian}, Andrew C. and {Ferrigno}, Carlo and {Foster}, Adam and {Fujimoto}, Ryuichi and {Fukazawa}, Yasushi and {Furuzawa}, Akihiro and {Galeazzi}, Massimiliano and {Gallo}, Luigi and {Gandhi}, Poshak and {Giustini}, Margherita and {Goldwurm}, Andrea and {Gu}, Liyi and {Guainazzi}, Matteo and {Haba}, Yoshito and {Hagino}, Kouichi and {Hamaguchi}, Kenji and {Harrus}, Ilana and {Hatsukade}, Isamu and {Hayashi}, Katsuhiro and {Hayashi}, Takayuki and {Hayashida}, Kiyoshi and {Hiraga}, Junko and {Hornschemeier}, Ann and {Hoshino}, Akio and {Hughes}, John and {Iizuka}, Ryo and {Inoue}, Hajime and {Inoue}, Yoshiyuki and {Ishibashi}, Kazunori and {Ishida}, Manabu and {Ishikawa}, Kumi and {Ishisaki}, Yoshitaka and {Itoh}, Masayuki and {Iyomoto}, Naoko and {Kaastra}, Jelle and {Kallman}, Timothy and {Kamae}, Tuneyoshi and {Kara}, Erin and {Kataoka}, Jun and {Katsuda}, Satoru and {Katsuta}, Junichiro and {Kawaharada}, Madoka and {Kawai}, Nobuyuki and {Kelley}, Richard and {Khangulyan}, Dmitry and {Kilbourne}, Caroline and {King}, Ashley and {Kitaguchi}, Takao and {Kitamoto}, Shunji and {Kitayama}, Tetsu and {Kohmura}, Takayoshi and {Kokubun}, Motohide and {Koyama}, Shu and {Koyama}, Katsuji and {Kretschmar}, Peter and {Krimm}, Hans and {Kubota}, Aya and {Kunieda}, Hideyo and {Laurent}, Philippe and {Lebrun}, Fran{\c{c}}ois and {Lee}, Shiu-Hang and {Leutenegger}, Maurice and {Limousin}, Olivier and {Loewenstein}, Michael and {Long}, Knox S. and {Lumb}, David and {Madejski}, Grzegorz and {Maeda}, Yoshitomo and {Maier}, Daniel and {Makishima}, Kazuo and {Markevitch}, Maxim and {Matsumoto}, Hironori and {Matsushita}, Kyoko and {McCammon}, Dan and {McNamara}, Brian and {Mehdipour}, Missagh and {Miller}, Eric and {Miller}, Jon and {Mineshige}, Shin and {Mitsuda}, Kazuhisa and {Mitsuishi}, Ikuyuki and {Miyazawa}, Takuya and {Mizuno}, Tsunefumi and {Mori}, Hideyuki and {Mori}, Koji and {Moseley}, Harvey and {Mukai}, Koji and {Murakami}, Hiroshi and {Murakami}, Toshio and {Mushotzky}, Richard and {Nagino}, Ryo and {Nakagawa}, Takao and {Nakajima}, Hiroshi and {Nakamori}, Takeshi and {Nakano}, Toshio and {Nakashima}, Shinya and {Nakazawa}, Kazuhiro and {Nobukawa}, Masayoshi and {Noda}, Hirofumi and {Nomachi}, Masaharu and {O'Dell}, Steve and {Odaka}, Hirokazu and {Ohashi}, Takaya and {Ohno}, Masanori and {Okajima}, Takashi and {Ota}, Naomi and {Ozaki}, Masanobu and {Paerels}, Frits and {Paltani}, Stephane and {Parmar}, Arvind and {Petre}, Robert and {Pinto}, Ciro and {Pohl}, Martin and {Porter}, F. Scott and {Pottschmidt}, Katja and {Ramsey}, Brian and {Reynolds}, Christopher and {Russell}, Helen and {Safi-Harb}, Samar and {Saito}, Shinya and {Sakai}, Kazuhiro and {Sameshima}, Hiroaki and {Sato}, Goro and {Sato}, Kosuke and {Sato}, Rie and {Sawada}, Makoto and {Schartel}, Norbert and {Serlemitsos}, Peter and {Seta}, Hiromi and {Shidatsu}, Megumi and {Simionescu}, Aurora and {Smith}, Randall and {Soong}, Yang and {Stawarz}, Lukasz and {Sugawara}, Yasuharu and {Sugita}, Satoshi and {Szymkowiak}, Andrew and {Tajima}, Hiroyasu and {Takahashi}, Hiromitsu and {Takahashi}, Tadayuki and {Takeda}, Shin'Ichiro and {Takei}, Yoh and {Tamagawa}, Toru and {Tamura}, Keisuke and {Tamura}, Takayuki and {Tanaka}, Takaaki and {Tanaka}, Yasuo and {Tanaka}, Yasuyuki and {Tashiro}, Makoto and {Tawara}, Yuzuru and {Terada}, Yukikatsu and {Terashima}, Yuichi and {Tombesi}, Francesco and {Tomida}, Hiroshi and {Tsuboi}, Yohko and {Tsujimoto}, Masahiro and {Tsunemi}, Hiroshi and {Tsuru}, Takeshi and {Uchida}, Hiroyuki and {Uchiyama}, Hideki and {Uchiyama}, Yasunobu and {Ueda}, Shutaro and {Ueda}, Yoshihiro and {Ueno}, Shiro and {Uno}, Shin'Ichiro and {Urry}, Meg and {Ursino}, Eugenio and {de Vries}, Cor and {Watanabe}, Shin and {Werner}, Norbert},
        title = "{The quiescent intracluster medium in the core of the Perseus cluster}",
      journal = {\nat},
         year = 2016,
        month = jul,
       volume = {535},
       number = {7610},
        pages = {117-121},
          doi = {10.1038/nature18627},
archivePrefix = {arXiv},
       eprint = {1607.04487},
 primaryClass = {astro-ph.GA},
       adsurl = {https://ui.adsabs.harvard.edu/abs/2016Natur.535..117H}
}

@ARTICLE{Hitomi_2018,
       author = {{Hitomi} and {Aharonian}, Felix and {Akamatsu}, Hiroki and {Akimoto}, Fumie and {Allen}, Steven W. and {Angelini}, Lorella and {Audard}, Marc and {Awaki}, Hisamitsu and {Axelsson}, Magnus and {Bamba}, Aya and {Bautz}, Marshall W. and {Blandford}, Roger and {Brenneman}, Laura W. and {Brown}, Gregory V. and {Bulbul}, Esra and {Cackett}, Edward M. and {Canning}, Rebecca E.~A. and {Chernyakova}, Maria and {Chiao}, Meng P. and {Coppi}, Paolo S. and {Costantini}, Elisa and {de Plaa}, Jelle and {de Vries}, Cor P. and {den Herder}, Jan-Willem and {Done}, Chris and {Dotani}, Tadayasu and {Ebisawa}, Ken and {Eckart}, Megan E. and {Enoto}, Teruaki and {Ezoe}, Yuichiro and {Fabian}, Andrew C. and {Ferrigno}, Carlo and {Foster}, Adam R. and {Fujimoto}, Ryuichi and {Fukazawa}, Yasushi and {Furuzawa}, Akihiro and {Galeazzi}, Massimiliano and {Gallo}, Luigi C. and {Gandhi}, Poshak and {Giustini}, Margherita and {Goldwurm}, Andrea and {Gu}, Liyi and {Guainazzi}, Matteo and {Haba}, Yoshito and {Hagino}, Kouichi and {Hamaguchi}, Kenji and {Harrus}, Ilana M. and {Hatsukade}, Isamu and {Hayashi}, Katsuhiro and {Hayashi}, Takayuki and {Hayashi}, Tasuku and {Hayashida}, Kiyoshi and {Hiraga}, Junko S. and {Hornschemeier}, Ann and {Hoshino}, Akio and {Hughes}, John P. and {Ichinohe}, Yuto and {Iizuka}, Ryo and {Inoue}, Hajime and {Inoue}, Shota and {Inoue}, Yoshiyuki and {Ishida}, Manabu and {Ishikawa}, Kumi and {Ishisaki}, Yoshitaka and {Iwai}, Masachika and {Kaastra}, Jelle and {Kallman}, Tim and {Kamae}, Tsuneyoshi and {Kataoka}, Jun and {Katsuda}, Satoru and {Kawai}, Nobuyuki and {Kelley}, Richard L. and {Kilbourne}, Caroline A. and {Kitaguchi}, Takao and {Kitamoto}, Shunji and {Kitayama}, Tetsu and {Kohmura}, Takayoshi and {Kokubun}, Motohide and {Koyama}, Katsuji and {Koyama}, Shu and {Kretschmar}, Peter and {Krimm}, Hans A. and {Kubota}, Aya and {Kunieda}, Hideyo and {Laurent}, Philippe and {Lee}, Shiu-Hang and {Leutenegger}, Maurice A. and {Limousin}, Olivier and {Loewenstein}, Michael and {Long}, Knox S. and {Lumb}, David and {Madejski}, Greg and {Maeda}, Yoshitomo and {Maier}, Daniel and {Makishima}, Kazuo and {Markevitch}, Maxim and {Matsumoto}, Hironori and {Matsushita}, Kyoko and {McCammon}, Dan and {McNamara}, Brian R. and {Mehdipour}, Missagh and {Miller}, Eric D. and {Miller}, Jon M. and {Mineshige}, Shin and {Mitsuda}, Kazuhisa and {Mitsuishi}, Ikuyuki and {Miyazawa}, Takuya and {Mizuno}, Tsunefumi and {Mori}, Hideyuki and {Mori}, Koji and {Mukai}, Koji and {Murakami}, Hiroshi and {Mushotzky}, Richard F. and {Nakagawa}, Takao and {Nakajima}, Hiroshi and {Nakamori}, Takeshi and {Nakashima}, Shinya and {Nakazawa}, Kazuhiro and {Nobukawa}, Kumiko K. and {Nobukawa}, Masayoshi and {Noda}, Hirofumi and {Odaka}, Hirokazu and {Ohashi}, Takaya and {Ohno}, Masanori and {Okajima}, Takashi and {Ota}, Naomi and {Ozaki}, Masanobu and {Paerels}, Frits and {Paltani}, St{\'e}phane and {Petre}, Robert and {Pinto}, Ciro and {Porter}, Frederick S. and {Pottschmidt}, Katja and {Reynolds}, Christopher S. and {Safi-Harb}, Samar and {Saito}, Shinya and {Sakai}, Kazuhiro and {Sasaki}, Toru and {Sato}, Goro and {Sato}, Kosuke and {Sato}, Rie and {Sawada}, Makoto and {Schartel}, Norbert and {Serlemtsos}, Peter J. and {Seta}, Hiromi and {Shidatsu}, Megumi and {Simionescu}, Aurora and {Smith}, Randall K. and {Soong}, Yang and {Stawarz}, {\L}ukasz and {Sugawara}, Yasuharu and {Sugita}, Satoshi and {Szymkowiak}, Andrew and {Tajima}, Hiroyasu and {Takahashi}, Hiromitsu and {Takahashi}, Tadayuki and {Takeda}, Shin'ichiro and {Takei}, Yoh and {Tamagawa}, Toru and {Tamura}, Takayuki and {Tanaka}, Keigo and {Tanaka}, Takaaki and {Tanaka}, Yasuo and {Tanaka}, Yasuyuki T. and {Tashiro}, Makoto S. and {Tawara}, Yuzuru and {Terada}, Yukikatsu and {Terashima}, Yuichi and {Tombesi}, Francesco and {Tomida}, Hiroshi and {Tsuboi}, Yohko and {Tsujimoto}, Masahiro and {Tsunemi}, Hiroshi and {Tsuru}, Takeshi Go and {Uchida}, Hiroyuki and {Uchiyama}, Hideki and {Uchiyama}, Yasunobu and {Ueda}, Shutaro and {Ueda}, Yoshihiro and {Uno}, Shin'ichiro and {Urry}, C. Megan and {Ursino}, Eugenio and {Wang}, Qian H.~S. and {Watanabe}, Shin and {Werner}, Norbert and {Wilkins}, Dan R. and {Williams}, Brian J. and {Yamada}, Shinya and {Yamaguchi}, Hiroya and {Yamaoka}, Kazutaka and {Yamasaki}, Noriko Y. and {Yamauchi}, Makoto and {Yamauchi}, Shigeo and {Yaqoob}, Tahir and {Yatsu}, Yoichi and {Yonetoku}, Daisuke and {Zhuravleva}, Irina and {Zoghbi}, Abderahmen},
        title = "{Atmospheric gas dynamics in the Perseus cluster observed with Hitomi}",
      journal = {\pasj},
         year = 2018,
        month = mar,
       volume = {70},
       number = {2},
          eid = {9},
        pages = {9},
          doi = {10.1093/pasj/psx138},
archivePrefix = {arXiv},
       eprint = {1711.00240},
 primaryClass = {astro-ph.HE},
       adsurl = {https://ui.adsabs.harvard.edu/abs/2018PASJ...70....9H}
}

@article{Hopkins_2017,
    author = {Hopkins, Philip F.},
    title = {Anisotropic diffusion in mesh-free numerical magnetohydrodynamics},
    journal = {Monthly Notices of the Royal Astronomical Society},
    volume = {466},
    number = {3},
    pages = {3387-3405},
    year = {2016},
    month = {12},
    issn = {0035-8711},
    doi = {10.1093/mnras/stw3306},
    url = {https://doi.org/10.1093/mnras/stw3306},
    eprint = {https://academic.oup.com/mnras/article-pdf/466/3/3387/10904268/stw3306.pdf},
}

@article{Iapichino_2008,
    author = {Iapichino, L. and Adamek, J. and Schmidt, W. and Niemeyer, J. C.},
    title = {Hydrodynamical adaptive mesh refinement simulations of turbulent flows – I. Substructure in a wind},
    journal = {Monthly Notices of the Royal Astronomical Society},
    volume = {388},
    number = {3},
    pages = {1079-1088},
    year = {2008},
    month = {07},
    issn = {0035-8711},
    doi = {10.1111/j.1365-2966.2008.13137.x},
    url = {https://doi.org/10.1111/j.1365-2966.2008.13137.x},
    eprint = {https://academic.oup.com/mnras/article-pdf/388/3/1079/2806039/mnras0388-1079.pdf},
}

@article{Iapichino_2017,
title={Adaptive mesh refinement simulations of a galaxy cluster merger – I. Resolving and modelling the turbulent flow in the cluster outskirts},
volume={469},
ISSN={1365-2966},
url={http://dx.doi.org/10.1093/mnras/stx882},
DOI={10.1093/mnras/stx882},
number={3},
journal={Monthly Notices of the Royal Astronomical Society},
publisher={Oxford University Press (OUP)},
author={Iapichino, L. and Federrath, C. and Klessen, R. S.},
year={2017},
month=apr, pages={3641–3655} }

@ARTICLE{Kingsland_2019,
       author = {{Kingsland}, Matthew and {Yang}, H. -Y. Karen and {Reynolds}, Christopher S. and {Zuhone}, John A.},
        title = "{Effects of Anisotropic Viscosity on the Evolution of Active Galactic Nuclei Bubbles in Galaxy Clusters}",
      journal = {\apjl},
         year = 2019,
        month = sep,
       volume = {883},
       number = {1},
          eid = {L23},
        pages = {L23},
          doi = {10.3847/2041-8213/ab40be},
archivePrefix = {arXiv},
       eprint = {1909.01339},
 primaryClass = {astro-ph.HE},
       adsurl = {https://ui.adsabs.harvard.edu/abs/2019ApJ...883L..23K}
}

@article{Kravtsov_2012,
   author = "Kravtsov, Andrey V. and Borgani, Stefano",
   title = "Formation of Galaxy Clusters", 
   journal= "Annual Review of Astronomy and Astrophysics",
   year = "2012",
   volume = "50",
   number = "Volume 50, 2012",
   pages = "353-409",
   doi = "https://doi.org/10.1146/annurev-astro-081811-125502",
   url = "https://www.annualreviews.org/content/journals/10.1146/annurev-astro-081811-125502",
   publisher = "Annual Reviews",
   issn = "1545-4282",
   type = "Journal Article",
}

@article{Kunz_2011,
    author = {Kunz, Matthew W.},
    title = {Dynamical stability of a thermally stratified intracluster medium with anisotropic momentum and heat transport},
    journal = {Monthly Notices of the Royal Astronomical Society},
    volume = {417},
    number = {1},
    pages = {602-616},
    year = {2011},
    month = {10},
    issn = {0035-8711},
    doi = {10.1111/j.1365-2966.2011.19303.x},
    url = {https://doi.org/10.1111/j.1365-2966.2011.19303.x},
    eprint = {https://academic.oup.com/mnras/article-pdf/417/1/602/3029760/mnras0417-0602.pdf},
}

@article{Kunz_2012,
doi = {10.1088/0004-637X/754/2/122},
url = {https://dx.doi.org/10.1088/0004-637X/754/2/122},
year = {2012},
month = {jul},
publisher = {The American Astronomical Society},
volume = {754},
number = {2},
pages = {122},
author = {Kunz, Matthew W. and Bogdanović, Tamara and Reynolds, Christopher S. and Stone, James M.},
title = {BUOYANCY INSTABILITIES IN A WEAKLY COLLISIONAL INTRACLUSTER MEDIUM},
journal = {The Astrophysical Journal}
}

@article{Kunz_2014,
    author = {Kunz, Matthew and Schekochihin, A. and Stone, James},
    year = {2014},
    month = {01},
    pages = {},
    title = {Firehose and Mirror Instabilities in a Collisionless Shearing Plasma},
    volume = {112},
    journal = {Physical Review Letters},
    doi = {10.1103/PhysRevLett.112.205003}
}

@inbook{Kunz_2022,
    title={Plasma Physics of the Intracluster Medium},
    ISBN={9789811645440},
    url={http://dx.doi.org/10.1007/978-981-16-4544-0_125-1},
    DOI={10.1007/978-981-16-4544-0_125-1},
    booktitle={Handbook of X-ray and Gamma-ray Astrophysics},
    publisher={Springer Nature Singapore},
    author={Kunz, Matthew W. and Jones, Thomas W. and Zhuravleva, Irina},
    year={2022},
    pages={1–42}
}

@ARTICLE{Lau_2009,
       author = {{Lau}, Erwin T. and {Kravtsov}, Andrey V. and {Nagai}, Daisuke},
        title = "{Residual Gas Motions in the Intracluster Medium and Bias in Hydrostatic Measurements of Mass Profiles of Clusters}",
      journal = {\apj},
         year = 2009,
        month = nov,
       volume = {705},
       number = {2},
        pages = {1129-1138},
          doi = {10.1088/0004-637X/705/2/1129},
archivePrefix = {arXiv},
       eprint = {0903.4895},
 primaryClass = {astro-ph.CO},
       adsurl = {https://ui.adsabs.harvard.edu/abs/2009ApJ...705.1129L}
}

@article{Marin-Gilabert_2022,
   title={The role of physical and numerical viscosity in hydrodynamical instabilities},
   volume={517},
   ISSN={1365-2966},
   url={http://dx.doi.org/10.1093/mnras/stac3042},
   DOI={10.1093/mnras/stac3042},
   number={4},
   journal={Monthly Notices of the Royal Astronomical Society},
   publisher={Oxford University Press (OUP)},
   author={Marin-Gilabert, Tirso and Valentini, Milena and Steinwandel, Ulrich P and Dolag, Klaus},
   year={2022},
   month=oct, pages={5971–5991} 
}

@article{Marin-Gilabert_2024,
   title={Density Fluctuations in the Intracluster Medium: An Attempt to Constrain Viscosity with Cosmological Simulations},
   volume={976},
   ISSN={1538-4357},
   url={http://dx.doi.org/10.3847/1538-4357/ad8127},
   DOI={10.3847/1538-4357/ad8127},
   number={1},
   journal={The Astrophysical Journal},
   publisher={American Astronomical Society},
   author={Marin-Gilabert, Tirso and Steinwandel, Ulrich P. and Valentini, Milena and Vallés-Pérez, David and Dolag, Klaus},
   year={2024},
   month=nov, pages={67} 
}

@article{Marin-Gilabert_2025,
   title={The (limited) effect of viscosity in multiphase turbulent mixing},
   volume={710},
   ISSN={1432-0746},
   url={http://dx.doi.org/10.1051/0004-6361/202555233},
   DOI={10.1051/0004-6361/202555233},
   journal={Astronomy &; Astrophysics},
   publisher={EDP Sciences},
   author={Marin-Gilabert, Tirso and Gronke, Max and Peng Oh, S.},
   year={2026},
   month=May, pages={A44} 
}

@ARTICLE{McCourt_2012,
       author = {{McCourt}, Michael and {Sharma}, Prateek and {Quataert}, Eliot and {Parrish}, Ian J.},
        title = "{Thermal instability in gravitationally stratified plasmas: implications for multiphase structure in clusters and galaxy haloes}",
      journal = {\mnras},
         year = 2012,
        month = feb,
       volume = {419},
       number = {4},
        pages = {3319-3337},
          doi = {10.1111/j.1365-2966.2011.19972.x},
archivePrefix = {arXiv},
       eprint = {1105.2563},
 primaryClass = {astro-ph.CO},
       adsurl = {https://ui.adsabs.harvard.edu/abs/2012MNRAS.419.3319M}
}

@ARTICLE{Miniati_2015,
       author = {{Miniati}, Francesco},
        title = "{The Matryoshka Run. II. Time-dependent Turbulence Statistics, Stochastic Particle Acceleration, and Microphysics Impact in a Massive Galaxy Cluster}",
      journal = {\apj},
         year = 2015,
        month = feb,
       volume = {800},
       number = {1},
          eid = {60},
        pages = {60},
          doi = {10.1088/0004-637X/800/1/60},
archivePrefix = {arXiv},
       eprint = {1409.3576},
 primaryClass = {astro-ph.CO},
       adsurl = {https://ui.adsabs.harvard.edu/abs/2015ApJ...800...60M}
}

@article{Nelson_2014,
   title={HYDRODYNAMIC SIMULATION OF NON-THERMAL PRESSURE PROFILES OF GALAXY CLUSTERS},
   volume={792},
   ISSN={1538-4357},
   url={http://dx.doi.org/10.1088/0004-637X/792/1/25},
   DOI={10.1088/0004-637x/792/1/25},
   number={1},
   journal={The Astrophysical Journal},
   publisher={American Astronomical Society},
   author={Nelson, Kaylea and Lau, Erwin T. and Nagai, Daisuke},
   year={2014},
   month=aug, pages={25} 
}

@article{Parrish_2012,
   title={The effects of anisotropic viscosity on turbulence and heat transport in the intracluster medium: Anisotropic viscosity in the ICM},
   volume={422},
   ISSN={0035-8711},
   url={http://dx.doi.org/10.1111/j.1365-2966.2012.20650.x},
   DOI={10.1111/j.1365-2966.2012.20650.x},
   number={1},
   journal={Monthly Notices of the Royal Astronomical Society},
   publisher={Oxford University Press (OUP)},
   author={Parrish, Ian J. and McCourt, Michael and Quataert, Eliot and Sharma, Prateek},
   year={2012},
   month=mar, pages={704–718} 
}

@article{Price_2008,
title = {Modelling discontinuities and Kelvin–Helmholtz instabilities in SPH},
journal = {Journal of Computational Physics},
volume = {227},
number = {24},
pages = {10040-10057},
year = {2008},
issn = {0021-9991},
doi = {https://doi.org/10.1016/j.jcp.2008.08.011},
url = {https://www.sciencedirect.com/science/article/pii/S0021999108004270},
author = {Daniel J. Price}
}

@ARTICLE{Quataert_2008,
       author = {{Quataert}, Eliot},
        title = "{Buoyancy Instabilities in Weakly Magnetized Low-Collisionality Plasmas}",
      journal = {\apj},
         year = 2008,
        month = feb,
       volume = {673},
       number = {2},
        pages = {758-762},
          doi = {10.1086/525248},
archivePrefix = {arXiv},
       eprint = {0710.5521},
 primaryClass = {astro-ph},
       adsurl = {https://ui.adsabs.harvard.edu/abs/2008ApJ...673..758Q}
}

@article{Rappaz_2024,
   title={The effect of pressure-anisotropy-driven kinetic instabilities on magnetic field amplification in galaxy clusters},
   volume={683},
   ISSN={1432-0746},
   url={http://dx.doi.org/10.1051/0004-6361/202347497},
   DOI={10.1051/0004-6361/202347497},
   journal={Astronomy \&; Astrophysics},
   publisher={EDP Sciences},
   author={Rappaz, Y. and Schober, J.},
   year={2024},
   month=mar, pages={A35} }

@article{Rincon_2014,
   title={Non-linear mirror instability},
   volume={447},
   ISSN={1745-3925},
   url={http://dx.doi.org/10.1093/mnrasl/slu179},
   DOI={10.1093/mnrasl/slu179},
   number={1},
   journal={Monthly Notices of the Royal Astronomical Society: Letters},
   publisher={Oxford University Press (OUP)},
   author={Rincon, F. and Schekochihin, A. A. and Cowley, S. C.},
   year={2014},
   month=dec, pages={L45–L49} }

@article{Rosin_2011,
   title={A non-linear theory of the parallel firehose and gyrothermal instabilities in a weakly collisional plasma: Theory of firehose and gyrothermal instabilities},
   volume={413},
   ISSN={0035-8711},
   url={http://dx.doi.org/10.1111/j.1365-2966.2010.17931.x},
   DOI={10.1111/j.1365-2966.2010.17931.x},
   number={1},
   journal={Monthly Notices of the Royal Astronomical Society},
   publisher={Oxford University Press (OUP)},
   author={Rosin, M. S. and Schekochihin, A. A. and Rincon, F. and Cowley, S. C.},
   year={2011},
   month=Mar, pages={7–38} 
}

@article{Sarazin_1986,
    title = {X-ray emission from clusters of galaxies},
    author = {Sarazin, Craig L.},
    journal = {Rev. Mod. Phys.},
    volume = {58},
    issue = {1},
    pages = {1--115},
    numpages = {0},
    year = {1986},
    month = {Jan},
    publisher = {American Physical Society},
    doi = {10.1103/RevModPhys.58.1},
    url = {https://link.aps.org/doi/10.1103/RevModPhys.58.1}
}

@article{Schekochihin_2005,
doi = {10.1086/431202},
url = {https://dx.doi.org/10.1086/431202},
year = {2005},
month = {aug},
publisher = {},
volume = {629},
number = {1},
pages = {139},
author = {Schekochihin, A. A. and Cowley, S. C. and Kulsrud, R. M. and Hammett, G. W. and Sharma, P.},
title = {Plasma Instabilities and Magnetic Field Growth in Clusters of Galaxies},
journal = {The Astrophysical Journal}
}

@article{Schekochihin_2006,
    author = {Schekochihin, A. A. and Cowley, S. C.},
    title = {Turbulence, magnetic fields, and plasma physics in clusters of galaxies},
    journal = {Physics of Plasmas},
    volume = {13},
    number = {5},
    pages = {056501},
    year = {2006},
    month = {05},
    issn = {1070-664X},
    doi = {10.1063/1.2179053},
    url = {https://doi.org/10.1063/1.2179053},
    eprint = {https://pubs.aip.org/aip/pop/article-pdf/doi/10.1063/1.2179053/15807160/056501_1_online.pdf},
}

@ARTICLE{Schekochihin_2008,
       author = {{Schekochihin}, A.~A. and {Cowley}, S.~C. and {Kulsrud}, R.~M. and {Rosin}, M.~S. and {Heinemann}, T.},
        title = "{Nonlinear Growth of Firehose and Mirror Fluctuations in Astrophysical Plasmas}",
      journal = {\prl},
         year = 2008,
        month = feb,
       volume = {100},
       number = {8},
          eid = {081301},
        pages = {081301},
          doi = {10.1103/PhysRevLett.100.081301},
archivePrefix = {arXiv},
       eprint = {0709.3828},
 primaryClass = {astro-ph},
       adsurl = {https://ui.adsabs.harvard.edu/abs/2008PhRvL.100h1301S}
}

@article{Schekochihin_2010,
    author = {Schekochihin, A. A. and Cowley, S. C. and Rincon, F. and Rosin, M. S.},
    title = {Magnetofluid dynamics of magnetized cosmic plasma: firehose and gyrothermal instabilities},
    journal = {Monthly Notices of the Royal Astronomical Society},
    volume = {405},
    number = {1},
    pages = {291-300},
    year = {2010},
    month = {06},
    issn = {0035-8711},
    doi = {10.1111/j.1365-2966.2010.16493.x},
    url = {https://doi.org/10.1111/j.1365-2966.2010.16493.x},
    eprint = {https://academic.oup.com/mnras/article-pdf/405/1/291/3364446/mnras0405-0291.pdf},
}

@ARTICLE{Schmidt_2014,
       author = {{Schmidt}, W. and {Almgren}, A.~S. and {Braun}, H. and {Engels}, J.~F. and {Niemeyer}, J.~C. and {Schulz}, J. and {Mekuria}, R.~R. and {Aspden}, A.~J. and {Bell}, J.~B.},
        title = "{Cosmological fluid mechanics with adaptively refined large eddy simulations}",
      journal = {\mnras},
         year = 2014,
        month = jun,
       volume = {440},
       number = {4},
        pages = {3051-3077},
          doi = {10.1093/mnras/stu501},
archivePrefix = {arXiv},
       eprint = {1309.3996},
 primaryClass = {astro-ph.CO},
       adsurl = {https://ui.adsabs.harvard.edu/abs/2014MNRAS.440.3051S}
}

@article{Schmidt_2016,
   title={Hot and turbulent gas in clusters},
   volume={459},
   ISSN={1365-2966},
   url={http://dx.doi.org/10.1093/mnras/stw632},
   DOI={10.1093/mnras/stw632},
   number={1},
   journal={Monthly Notices of the Royal Astronomical Society},
   publisher={Oxford University Press (OUP)},
   author={Schmidt, W. and Engels, J. F. and Niemeyer, J. C. and Almgren, A. S.},
   year={2016},
   month=mar, pages={701–719} 
}

@article{Sijacki_2007,
   title={A unified model for AGN feedback in cosmological simulations of structure formation: AGN feedback in cosmological simulations},
   volume={380},
   ISSN={0035-8711},
   url={http://dx.doi.org/10.1111/j.1365-2966.2007.12153.x},
   DOI={10.1111/j.1365-2966.2007.12153.x},
   number={3},
   journal={Monthly Notices of the Royal Astronomical Society},
   publisher={Oxford University Press (OUP)},
   author={Sijacki, Debora and Springel, Volker and Di Matteo, Tiziana and Hernquist, Lars},
   year={2007},
   month=aug, pages={877–900} 
}

@ARTICLE{Southwood_1993,
       author = {{Southwood}, David J. and {Kivelson}, Margaret G.},
        title = "{Mirror instability. I - Physical mechanism of linear instability}",
      journal = {\jgr},
         year = 1993,
        month = jun,
       volume = {98},
       number = {A6},
        pages = {9181-9187},
          doi = {10.1029/92JA02837},
       adsurl = {https://ui.adsabs.harvard.edu/abs/1993JGR....98.9181S}
}

@BOOK{Spitzer_1962,
       author = {{Spitzer}, L.},
        title = "{Physics of Fully Ionized Gases}",
         year = 1962,
       adsurl = {https://ui.adsabs.harvard.edu/abs/1962pfig.book.....S}
}

@article{Springel_2005,
author = {Springel, Volker},
title = "{The cosmological simulation code gadget-2}",
journal = {Monthly Notices of the Royal Astronomical Society},
volume = {364},
number = {4},
pages = {1105-1134},
year = {2005},
month = {12},
issn = {0035-8711},
doi = {10.1111/j.1365-2966.2005.09655.x},
url = {https://doi.org/10.1111/j.1365-2966.2005.09655.x},
eprint = {https://academic.oup.com/mnras/article-pdf/364/4/1105/18657201/364-4-1105.pdf},
}

@article{Squire_2016,
doi = {10.3847/2041-8205/830/2/L25},
url = {https://doi.org/10.3847/2041-8205/830/2/L25},
year = {2016},
month = {oct},
publisher = {The American Astronomical Society},
volume = {830},
number = {2},
pages = {L25},
author = {Squire, J. and Quataert, E. and Schekochihin, A. A.},
title = {A STRINGENT LIMIT ON THE AMPLITUDE OF ALFVÉNIC PERTURBATIONS IN HIGH-BETA LOW-COLLISIONALITY PLASMAS},
journal = {The Astrophysical Journal Letters}
}

@article{Squire_2017,
doi = {10.1088/1367-2630/aa6bb1},
url = {https://dx.doi.org/10.1088/1367-2630/aa6bb1},
year = {2017},
month = {may},
publisher = {IOP Publishing},
volume = {19},
number = {5},
pages = {055005},
author = {Squire, J and Schekochihin, A A and Quataert, E},
title = {Amplitude limits and nonlinear damping of shear-Alfvén waves in high-beta low-collisionality plasmas},
journal = {New Journal of Physics}
}

@article{Squire_2017b,
   title={Kinetic Simulations of the Interruption of Large-Amplitude Shear-Alfvén Waves in a High-beta Plasma},
   volume={119},
   ISSN={1079-7114},
   url={http://dx.doi.org/10.1103/PhysRevLett.119.155101},
   DOI={10.1103/physrevlett.119.155101},
   number={15},
   journal={Physical Review Letters},
   publisher={American Physical Society (APS)},
   author={Squire, J. and Kunz, M. W. and Quataert, E. and Schekochihin, A. A.},
   year={2017},
   month=oct 
}

@article{Squire_2023, 
    title={Pressure anisotropy and viscous heating in weakly collisional plasma turbulence}, 
    volume={89}, 
    DOI={10.1017/S0022377823000727}, 
    number={4}, 
    journal={Journal of Plasma Physics}, 
    author={Squire, J. and Kunz, M.W. and Arzamasskiy, L. and Johnston, Z. and Quataert, E. and Schekochihin, A.A.}, 
    year={2023}, 
    pages={905890417}
}

@ARTICLE{Stasyszyn_2013,
author = {{Stasyszyn}, F.~A. and {Dolag}, K. and {Beck}, A.~M.},
title = "{A divergence-cleaning scheme for cosmological SPMHD simulations}",
journal = {\mnras},
year = 2013,
month = jan,
volume = {428},
number = {1},
pages = {13-27},
doi = {10.1093/mnras/sts018},
archivePrefix = {arXiv},
eprint = {1205.4169},
primaryClass = {astro-ph.IM},
adsurl = {https://ui.adsabs.harvard.edu/abs/2013MNRAS.428...13S}
}

@article{Steinwandel_2022,
title={On the Small-scale Turbulent Dynamo in the Intracluster Medium: A Comparison to Dynamo Theory*},
volume={933},
ISSN={1538-4357},
url={http://dx.doi.org/10.3847/1538-4357/ac715c},
DOI={10.3847/1538-4357/ac715c},
number={2},
journal={The Astrophysical Journal},
publisher={American Astronomical Society},
author={Steinwandel, Ulrich P. and Böss, Ludwig M. and Dolag, Klaus and Lesch, Harald},
year={2022},
month=jul, pages={131} }

@article{Steinwandel_2024,
doi = {10.3847/1538-4357/ad39ee},
url = {https://dx.doi.org/10.3847/1538-4357/ad39ee},
year = {2024},
month = {may},
publisher = {The American Astronomical Society},
volume = {967},
number = {2},
pages = {125},
author = {Ulrich P. Steinwandel and Klaus Dolag and Ludwig M. Böss and Tirso Marin-Gilabert},
title = {Toward Cosmological Simulations of the Magnetized Intracluster Medium with Resolved Coulomb Collision Scale},
journal = {The Astrophysical Journal}
}

@article{Subramanian_2006,
author = {Subramanian, Kandaswamy and Shukurov, Anvar and Haugen, Nils Erland L.},
title = "{Evolving turbulence and magnetic fields in galaxy clusters}",
journal = {Monthly Notices of the Royal Astronomical Society},
volume = {366},
number = {4},
pages = {1437-1454},
year = {2006},
month = {03},
issn = {0035-8711},
doi = {10.1111/j.1365-2966.2006.09918.x},
url = {https://doi.org/10.1111/j.1365-2966.2006.09918.x},
eprint = {https://academic.oup.com/mnras/article-pdf/366/4/1437/3915441/366-4-1437.pdf},
}

@article{Suzuki_2013,
doi = {10.1088/0004-637X/768/2/175},
url = {https://dx.doi.org/10.1088/0004-637X/768/2/175},
year = {2013},
month = {apr},
publisher = {The American Astronomical Society},
volume = {768},
number = {2},
pages = {175},
author = {Kentaro Suzuki and Takayuki Ogawa and Yosuke Matsumoto and Ryoji Matsumoto},
title = {MAGNETOHYDRODYNAMIC SIMULATIONS OF THE FORMATION OF COLD FRONTS IN CLUSTERS OF GALAXIES: EFFECTS OF ANISOTROPIC VISCOSITY},
journal = {The Astrophysical Journal}
}

@article{Tevlin_2025,
   title={Magnetic dynamos in galaxy clusters: The crucial role of galaxy formation physics at high redshifts},
   volume={701},
   ISSN={1432-0746},
   url={http://dx.doi.org/10.1051/0004-6361/202452823},
   DOI={10.1051/0004-6361/202452823},
   journal={Astronomy &amp; Astrophysics},
   publisher={EDP Sciences},
   author={Tevlin, L. and Berlok, T. and Pfrommer, C. and Talbot, R. Y. and Whittingham, J. and Puchwein, E. and Pakmor, R. and Weinberger, R. and Springel, V.},
   year={2025},
   month=sep, pages={A114} 
}

@article{Vazza_2012,
   title={Turbulence in the ICM from mergers, cool-core sloshing, and jets: results from a new multi-scale filtering approach},
   volume={544},
   ISSN={1432-0746},
   url={http://dx.doi.org/10.1051/0004-6361/201118688},
   DOI={10.1051/0004-6361/201118688},
   journal={Astronomy \&; Astrophysics},
   publisher={EDP Sciences},
   author={Vazza, F. and Roediger, E. and Brüggen, M.},
   year={2012},
   month=aug, pages={A103} }

@ARTICLE{Vazza_2018,
       author = {{Vazza}, F. and {Brunetti}, G. and {Br{\"u}ggen}, M. and {Bonafede}, A.},
        title = "{Resolved magnetic dynamo action in the simulated intracluster medium}",
      journal = {\mnras},
         year = 2018,
        month = feb,
       volume = {474},
       number = {2},
        pages = {1672-1687},
          doi = {10.1093/mnras/stx2830},
archivePrefix = {arXiv},
       eprint = {1711.02673},
 primaryClass = {astro-ph.CO},
       adsurl = {https://ui.adsabs.harvard.edu/abs/2018MNRAS.474.1672V}
}

@article{Vazza_2018b,
   title={The turbulent pressure support in galaxy clusters revisited},
   volume={481},
   ISSN={1745-3933},
   url={http://dx.doi.org/10.1093/mnrasl/sly172},
   DOI={10.1093/mnrasl/sly172},
   number={1},
   journal={Monthly Notices of the Royal Astronomical Society: Letters},
   publisher={Oxford University Press (OUP)},
   author={Vazza, F and Angelinelli, M and Jones, T W and Eckert, D and Brüggen, M and Brunetti, G and Gheller, C},
   year={2018},
   month=sep, pages={L120–L124} 
}

@ARTICLE{Vazza_2025,
       author = {{Vazza}, F. and {Brunetti}, G.},
        title = "{On the interpretation of XRISM X-ray measurements of turbulence in the intracluster medium: a comparison with cosmological simulations}",
      journal = {arXiv e-prints},
         year = 2025,
        month = jul,
          eid = {arXiv:2507.04727},
        pages = {arXiv:2507.04727},
          doi = {10.48550/arXiv.2507.04727},
archivePrefix = {arXiv},
       eprint = {2507.04727},
 primaryClass = {astro-ph.CO},
       adsurl = {https://ui.adsabs.harvard.edu/abs/2025arXiv250704727V}
}

@article{Vikhlinin_2001,
title={[ITAL]Chandra[/ITAL] Estimate of the Magnetic Field Strength near the Cold Front in A3667},
volume={549},
ISSN={0004-637X},
url={http://dx.doi.org/10.1086/319126},
DOI={10.1086/319126},
number={1},
journal={The Astrophysical Journal},
publisher={American Astronomical Society},
author={Vikhlinin, A. and Markevitch, M. and Murray, S. S.},
year={2001},
month={Mar},
pages={L47–L50}
}

@article{Wendland_1995,
title={Piecewise polynomial, positive definite and compactly supported radial functions of minimal degree},
author={H. Wendland},
journal={Advances in Computational Mathematics},
year={1995},
volume={4},
pages={389-396}
}

@article{XRISM_2025_centaurus,
   title={The bulk motion of gas in the core of the Centaurus galaxy cluster},
   volume={638},
   ISSN={1476-4687},
   url={http://dx.doi.org/10.1038/s41586-024-08561-z},
   DOI={10.1038/s41586-024-08561-z},
   number={8050},
   journal={Nature},
   publisher={Springer Science and Business Media LLC},
   author={{XRISM Collaboration} and Audard, Marc and Awaki, Hisamitsu and Ballhausen, Ralf and Bamba, Aya and Behar, Ehud and Boissay-Malaquin, Rozenn and Brenneman, Laura and Brown, Gregory V. and Corrales, Lia and Costantini, Elisa and Cumbee, Renata and Done, Chris and Dotani, Tadayasu and Ebisawa, Ken and Eckart, Megan E. and Eckert, Dominique and Enoto, Teruaki and Eguchi, Satoshi and Ezoe, Yuichiro and Foster, Adam and Fujimoto, Ryuichi and Fujita, Yutaka and Fukazawa, Yasushi and Fukushima, Kotaro and Furuzawa, Akihiro and Gallo, Luigi and García, Javier A. and Gu, Liyi and Guainazzi, Matteo and Hagino, Kouichi and Hamaguchi, Kenji and Hatsukade, Isamu and Hayashi, Katsuhiro and Hayashi, Takayuki and Hell, Natalie and Hodges-Kluck, Edmund and Hornschemeier, Ann and Ichinohe, Yuto and Ishida, Manabu and Ishikawa, Kumi and Ishisaki, Yoshitaka and Kaastra, Jelle and Kallman, Timothy and Kara, Erin and Katsuda, Satoru and Kanemaru, Yoshiaki and Kelley, Richard and Kilbourne, Caroline and Kitamoto, Shunji and Kobayashi, Shogo and Kohmura, Takayoshi and Kubota, Aya and Leutenegger, Maurice and Loewenstein, Michael and Maeda, Yoshitomo and Markevitch, Maxim and Matsumoto, Hironori and Matsushita, Kyoko and McCammon, Dan and McNamara, Brian and Mernier, François and Miller, Eric D. and Miller, Jon M. and Mitsuishi, Ikuyuki and Mizumoto, Misaki and Mizuno, Tsunefumi and Mori, Koji and Mukai, Koji and Murakami, Hiroshi and Mushotzky, Richard and Nakajima, Hiroshi and Nakazawa, Kazuhiro and Ness, Jan-Uwe and Nobukawa, Kumiko and Nobukawa, Masayoshi and Noda, Hirofumi and Odaka, Hirokazu and Ogawa, Shoji and Ogorzalek, Anna and Okajima, Takashi and Ota, Naomi and Paltani, Stephane and Petre, Robert and Plucinsky, Paul and Porter, Frederick S. and Pottschmidt, Katja and Sato, Kosuke and Sato, Toshiki and Sawada, Makoto and Seta, Hiromi and Shidatsu, Megumi and Simionescu, Aurora and Smith, Randall and Suzuki, Hiromasa and Szymkowiak, Andrew and Takahashi, Hiromitsu and Takeo, Mai and Tamagawa, Toru and Tamura, Keisuke and Tanaka, Takaaki and Tanimoto, Atsushi and Tashiro, Makoto and Terada, Yukikatsu and Terashima, Yuichi and Trigo, María Díaz and Tsuboi, Yohko and Tsujimoto, Masahiro and Tsunemi, Hiroshi and Tsuru, Takeshi G. and Uchida, Hiroyuki and Uchida, Nagomi and Uchida, Yuusuke and Uchiyama, Hideki and Ueda, Yoshihiro and Uno, Shinichiro and Vink, Jacco and Watanabe, Shin and Williams, Brian J. and Yamada, Satoshi and Yamada, Shinya and Yamaguchi, Hiroya and Yamaoka, Kazutaka and Yamasaki, Noriko Y. and Yamauchi, Makoto and Yamauchi, Shigeo and Yaqoob, Tahir and Yoneyama, Tomokage and Yoshida, Tessei and Yukita, Mihoko and Zhuravleva, Irina and Kondo, Marie and Werner, Norbert and Plšek, Tomáš and Sun, Ming and Hosogi, Kokoro and Majumder, Anwesh},
   year={2025},
   month={feb}, 
   pages={365–369} 
}

@article{XRISM_2025_abell,
    doi = {10.3847/2041-8213/ada7cd},
    url = {https://doi.org/10.3847/2041-8213/ada7cd},
    year = {2025},
    month = {mar},
    publisher = {The American Astronomical Society},
    volume = {982},
    number = {1},
    pages = {L5},
    author = {{XRISM Collaboration} and Audard, Marc and Awaki, Hisamitsu and Ballhausen, Ralf and Bamba, Aya and Behar, Ehud and Boissay-Malaquin, Rozenn and Brenneman, Laura and Brown, Gregory V. and Corrales, Lia and Costantini, Elisa and Cumbee, Renata and Diaz Trigo, Maria and Done, Chris and Dotani, Tadayasu and Ebisawa, Ken and Eckart, Megan E. and Eckert, Dominique and Eguchi, Satoshi and Enoto, Teruaki and Ezoe, Yuichiro and Foster, Adam and Fujimoto, Ryuichi and Fujita, Yutaka and Fukazawa, Yasushi and Fukushima, Kotaro and Furuzawa, Akihiro and Gallo, Luigi and García, Javier A. and Gu, Liyi and Guainazzi, Matteo and Hagino, Kouichi and Hamaguchi, Kenji and Hatsukade, Isamu and Hayashi, Katsuhiro and Hayashi, Takayuki and Hell, Natalie and Hodges-Kluck, Edmund and Hornschemeier, Ann and Ichinohe, Yuto and Ishida, Manabu and Ishikawa, Kumi and Ishisaki, Yoshitaka and Kaastra, Jelle and Kallman, Timothy and Kara, Erin and Katsuda, Satoru and Kanemaru, Yoshiaki and Kelley, Richard and Kilbourne, Caroline and Kitamoto, Shunji and Kobayashi, Shogo and Kohmura, Takayoshi and Kubota, Aya and Leutenegger, Maurice and Loewenstein, Michael and Maeda, Yoshitomo and Markevitch, Maxim and Matsumoto, Hironori and Matsushita, Kyoko and McCammon, Dan and McNamara, Brian and Mernier, François and Miller, Eric D. and Miller, Jon M. and Mitsuishi, Ikuyuki and Mizumoto, Misaki and Mizuno, Tsunefumi and Mori, Koji and Mukai, Koji and Murakami, Hiroshi and Mushotzky, Richard and Nakajima, Hiroshi and Nakazawa, Kazuhiro and Ness, Jan-Uwe and Nobukawa, Kumiko and Nobukawa, Masayoshi and Noda, Hirofumi and Odaka, Hirokazu and Ogawa, Shoji and Ogorzalek, Anna and Okajima, Takashi and Ota, Naomi and Paltani, Stephane and Petre, Robert and Plucinsky, Paul and Porter, Frederick S. and Pottschmidt, Katja and Sato, Kosuke and Sato, Toshiki and Sawada, Makoto and Seta, Hiromi and Shidatsu, Megumi and Simionescu, Aurora and Smith, Randall and Suzuki, Hiromasa and Szymkowiak, Andrew and Takahashi, Hiromitsu and Takeo, Mai and Tamagawa, Toru and Tamura, Keisuke and Tanaka, Takaaki and Tanimoto, Atsushi and Tashiro, Makoto and Terada, Yukikatsu and Terashima, Yuichi and Tsuboi, Yohko and Tsujimoto, Masahiro and Tsunemi, Hiroshi and Tsuru, Takeshi and Uchida, Hiroyuki and Uchida, Nagomi and Uchida, Yuusuke and Uchiyama, Hideki and Ueda, Yoshihiro and Uno, Shinichiro and Vink, Jacco and Watanabe, Shin and Williams, Brian J. and Yamada, Satoshi and Yamada, Shinya and Yamaguchi, Hiroya and Yamaoka, Kazutaka and Yamasaki, Noriko and Yamauchi, Makoto and Yamauchi, Shigeo and Yaqoob, Tahir and Yoneyama, Tomokage and Yoshida, Tessei and Yukita, Mihoko and Zhuravleva, Irina and Bartalesi, Tommaso and Ettori, Stefano and Kosarzycki, Roman and Lovisari, Lorenzo and Rose, Tom and Sarkar, Arnab and Sun, Ming and Tamhane, Prathamesh},
    title = {XRISM Reveals Low Nonthermal Pressure in the Core of the Hot, Relaxed Galaxy Cluster A2029},
    journal = {The Astrophysical Journal Letters}
}

@misc{XRISM_2025_Coma,
      title={XRISM forecast for the Coma cluster: stormy, with a steep power spectrum}, 
      author={{XRISM Collaboration} and Marc Audard and Hisamitsu Awaki and Ralf Ballhausen and Aya Bamba and Ehud Behar and Rozenn Boissay-Malaquin and Laura Brenneman and Gregory V. Brown and Lia Corrales and Elisa Costantini and Renata Cumbee and Maria Diaz Trigo and Chris Done and Tadayasu Dotani and Ken Ebisawa and Megan E. Eckart and Dominique Eckert and Satoshi Eguchi and Teruaki Enoto and Yuichiro Ezoe and Adam Foster and Ryuichi Fujimoto and Yutaka Fujita and Yasushi Fukazawa and Kotaro Fukushima and Akihiro Furuzawa and Luigi Gallo and Javier A. Garcia and Liyi Gu and Matteo Guainazzi and Kouichi Hagino and Kenji Hamaguchi and Isamu Hatsukade and Katsuhiro Hayashi and Takayuki Hayashi and Natalie Hell and Edmund Hodges-Kluck and Ann Hornschemeier and Yuto Ichinohe and Daiki Ishi and Manabu Ishida and Kumi Ishikawa and Yoshitaka Ishisaki and Jelle Kaastra and Timothy Kallman and Erin Kara and Satoru Katsuda and Yoshiaki Kanemaru and Richard Kelley and Caroline Kilbourne and Shunji Kitamoto and Shogo Kobayashi and Takayoshi Kohmura and Aya Kubota and Maurice Leutenegger and Michael Loewenstein and Yoshitomo Maeda and Maxim Markevitch and Hironori Matsumoto and Kyoko Matsushita and Dan McCammon and Brian McNamara and Francois Mernier and Eric D. Miller and Jon M. Miller and Ikuyuki Mitsuishi and Misaki Mizumoto and Tsunefumi Mizuno and Koji Mori and Koji Mukai and Hiroshi Murakami and Richard Mushotzky and Hiroshi Nakajima and Kazuhiro Nakazawa and Jan-Uwe Ness and Kumiko Nobukawa and Masayoshi Nobukawa and Hirofumi Noda and Hirokazu Odaka and Shoji Ogawa and Anna Ogorzalek and Takashi Okajima and Naomi Ota and Stephane Paltani and Robert Petre and Paul Plucinsky and Frederick S. Porter and Katja Pottschmidt and Kosuke Sato and Toshiki Sato and Makoto Sawada and Hiromi Seta and Megumi Shidatsu and Aurora Simionescu and Randall Smith and Hiromasa Suzuki and Andrew Szymkowiak and Hiromitsu Takahashi and Mai Takeo and Toru Tamagawa and Keisuke Tamura and Takaaki Tanaka and Atsushi Tanimoto and Makoto Tashiro and Yukikatsu Terada and Yuichi Terashima and Yohko Tsuboi and Masahiro Tsujimoto and Hiroshi Tsunemi and Takeshi Tsuru and Aysegul Tumer and Hiroyuki Uchida and Nagomi Uchida and Yuusuke Uchida and Hideki Uchiyama and Shutaro Ueda and Yoshihiro Ueda and Shinichiro Uno and Jacco Vink and Shin Watanabe and Brian J. Williams and Satoshi Yamada and Shinya Yamada and Hiroya Yamaguchi and Kazutaka Yamaoka and Noriko Yamasaki and Makoto Yamauchi and Shigeo Yamauchi and Tahir Yaqoob and Tomokage Yoneyama and Tessei Yoshida and Mihoko Yukita and Irina Zhuravleva and Andrew Fabian and Dylan Nelson and Nobuhiro Okabe and Annalisa Pillepich and Cicely Potter and Manon Regamey and Kosei Sakai and Mona Shishido and Nhut Truong and Daniel R. Wik and John ZuHone},
      year={2025},
      month = {apr},
      eprint={2504.20928},
      archivePrefix={arXiv},
      primaryClass={astro-ph.HE},
      url={https://arxiv.org/abs/2504.20928}, 
}

@article{XRISM_2025_Abell2029_profile,
    author = {{XRISM Collaboration} and Audard, Marc and Awaki, Hisamitsu and Ballhausen, Ralf and Bamba, Aya and Behar, Ehud and Boissay-Malaquin, Rozenn and Brenneman, Laura and Brown, Gregory V and Corrales, Lia and Costantini, Elisa and Cumbee, Renata and Díaz Trigo, Maria and Done, Chris and Dotani, Tadayasu and Ebisawa, Ken and Eckart, Megan E and Eckert, Dominique and Eguchi, Satoshi and Enoto, Teruaki and Ezoe, Yuichiro and Foster, Adam and Fujimoto, Ryuichi and Fujita, Yutaka and Fukazawa, Yasushi and Fukushima, Kotaro and Furuzawa, Akihiro and Gallo, Luigi C and García, Javier A and Gu, Liyi and Guainazzi, Matteo and Hagino, Kouichi and Hamaguchi, Kenji and Hatsukade, Isamu and Hayashi, Katsuhiro and Hayashi, Takayuki and Hell, Natalie and Hodges-Kluck, Edmund and Hornschemeier, Ann and Ichinohe, Yuto and Ishi, Daiki and Ishida, Manabu and Ishikawa, Kumi and Ishisaki, Yoshitaka and Kaastra, Jelle and Kallman, Timothy and Kara, Erin and Katsuda, Satoru and Kanemaru, Yoshiaki and Kelley, Richard and Kilbourne, Caroline and Kitamoto, Shunji and Kobayashi, Shogo and Kohmura, Takayoshi and Kubota, Aya and Leutenegger, Maurice A and Loewenstein, Michael and Maeda, Yoshitomo and Markevitch, Maxim and Matsumoto, Hironori and Matsushita, Kyoko and McCammon, Dan and McNamara, Brian and Mernier, François and Miller, Eric and Miller, Jon M and Mitsuishi, Ikuyuki and Mizumoto, Misaki and Mizuno, Tsunefumi and Mori, Koji and Mukai, Koji and Murakami, Hiroshi and Mushotzky, Richard and Nakajima, Hiroshi and Nakazawa, Kazuhiro and Ness, Jan-Uwe and Nobukawa, Kumiko and Nobukawa, Masayoshi and Noda, Hirofumi and Odaka, Hirokazu and Ogawa, Shoji and Ogorzalek, Anna and Okajima, Takashi and Ota, Naomi and Paltani, Stephane and Petre, Robert and Plucinsky, Paul and Porter, Frederick S and Pottschmidt, Katja and Sato, Kosuke and Sato, Toshiki and Sawada, Makoto and Seta, Hiromi and Shidatsu, Megumi and Simionescu, Aurora and Smith, Randall and Suzuki, Hiromasa and Szymkowiak, Andrew and Takahashi, Hiromitsu and Takeo, Mai and Tamagawa, Toru and Tamura, Keisuke and Tanaka, Takaaki and Tanimoto, Atsushi and Tashiro, Makoto and Terada, Yukikatsu and Terashima, Yuichi and Tsuboi, Yohko and Tsujimoto, Masahiro and Tsunemi, Hiroshi and Tsuru, Takeshi Go and Uchida, Hiroyuki and Uchida, Nagomi and Uchida, Yuusuke and Uchiyama, Hideki and Ueda, Yoshihiro and Uno, Shinichiro and Vink, Jacco and Watanabe, Shin and Williams, Brian J and Yamada, Satoshi and Yamada, Shinya and Yamaguchi, Hiroya and Yamaoka, Kazutaka and Yamasaki, Noriko and Yamauchi, Makoto and Yamauchi, Shigeo and Yaqoob, Tahir and Yoneyama, Tomokage and Yoshida, Tessei and Yukita, Mihoko and Zhuravleva, Irina and Bartalesi, Tommaso and Ettori, Stefano and Kosarzycki, Roman and Lovisari, Lorenzo and Rose, Tom and Sarkar, Arnab and Sun, Ming and Tamhane, Prathamesh},
    title = {Constraining gas motion and non-thermal pressure beyond the core of the Abell 2029 galaxy cluster with XRISM},
    journal = {Publications of the Astronomical Society of Japan},
    volume = {77},
    number = {Supplement_1},
    pages = {S242-S253},
    year = {2025},
    month = {08},
    issn = {2053-051X},
    doi = {10.1093/pasj/psaf055},
    url = {https://doi.org/10.1093/pasj/psaf055},
    eprint = {https://academic.oup.com/pasj/article-pdf/77/Supplement_1/S242/64074754/psaf055.pdf},
}

@misc{XRISM_2025_LoS,
      title={Disentangling Multiple Gas Kinematic Drivers in the Perseus Galaxy Cluster}, 
      author={{XRISM Collaboration} and Marc Audard and Hisamitsu Awaki and Ralf Ballhausen and Aya Bamba and Ehud Behar and Rozenn Boissay-Malaquin and Laura Brenneman and Gregory V. Brown and Lia Corrales and Elisa Costantini and Renata Cumbee and Maria Diaz Trigo and Chris Done and Tadayasu Dotani and Ken Ebisawa and Megan E. Eckart and Dominique Eckert and Satoshi Eguchi and Teruaki Enoto and Yuichiro Ezoe and Adam Foster and Ryuichi Fujimoto and Yutaka Fujita and Yasushi Fukazawa and Kotaro Fukushima and Akihiro Furuzawa and Luigi Gallo and Javier A. Garcia and Liyi Gu and Matteo Guainazzi and Kouichi Hagino and Kenji Hamaguchi and Isamu Hatsukade and Katsuhiro Hayashi and Takayuki Hayashi and Natalie Hell and Edmund Hodges-Kluck and Ann Hornschemeier and Yuto Ichinohe and Daiki Ishi and Manabu Ishida and Kumi Ishikawa and Yoshitaka Ishisaki and Jelle Kaastra and Timothy Kallman and Erin Kara and Satoru Katsuda and Yoshiaki Kanemaru and Richard Kelley and Caroline Kilbourne and Shunji Kitamoto and Shogo Kobayashi and Takayoshi Kohmura and Aya Kubota and Maurice Leutenegger and Michael Loewenstein and Yoshitomo Maeda and Maxim Markevitch and Hironori Matsumoto and Kyoko Matsushita and Dan McCammon and Brian McNamara and Francois Mernier and Eric D. Miller and Jon M. Miller and Ikuyuki Mitsuishi and Misaki Mizumoto and Tsunefumi Mizuno and Koji Mori and Koji Mukai and Hiroshi Murakami and Richard Mushotzky and Hiroshi Nakajima and Kazuhiro Nakazawa and Jan-Uwe Ness and Kumiko Nobukawa and Masayoshi Nobukawa and Hirofumi Noda and Hirokazu Odaka and Shoji Ogawa and Anna Ogorzalek and Takashi Okajima and Naomi Ota and Stephane Paltani and Robert Petre and Paul Plucinsky and Frederick S. Porter and Katja Pottschmidt and Kosuke Sato and Toshiki Sato and Makoto Sawada and Hiromi Seta and Megumi Shidatsu and Aurora Simionescu and Randall Smith and Hiromasa Suzuki and Andrew Szymkowiak and Hiromitsu Takahashi and Mai Takeo and Toru Tamagawa and Keisuke Tamura and Takaaki Tanaka and Atsushi Tanimoto and Makoto Tashiro and Yukikatsu Terada and Yuichi Terashima and Yohko Tsuboi and Masahiro Tsujimoto and Hiroshi Tsunemi and Takeshi G. Tsuru and Aysegul Tumer and Hiroyuki Uchida and Nagomi Uchida and Yuusuke Uchida and Hideki Uchiyama and Yoshihiro Ueda and Shinichiro Uno and Jacco Vink and Shin Watanabe and Brian J. Williams and Satoshi Yamada and Shinya Yamada and Hiroya Yamaguchi and Kazutaka Yamaoka and Noriko Yamasaki and Makoto Yamauchi and Shigeo Yamauchi and Tahir Yaqoob and Tomokage Yoneyama and Tessei Yoshida and Mihoko Yukita and Irina Zhuravleva and Elena Bellomi and Ian Drury and Annie Heinrich and Julie Hlavacek-Larrondo and Julian Meunier and Kostas Migkas and Lior Shefler and Phillip C. Stancil and Nhut Truong and Shutaro Ueda and Benjamin Vigneron and Congyao Zhang and John ZuHone},
      year={2025},
      eprint={2509.04421},
      archivePrefix={arXiv},
      primaryClass={astro-ph.HE},
      url={https://arxiv.org/abs/2509.04421}, 
}

@article{XRISM_2025_sims,
    doi = {10.3847/2041-8213/ae100c},
    url = {https://doi.org/10.3847/2041-8213/ae100c},
    year = {2025},
    month = {oct},
    publisher = {The American Astronomical Society},
    volume = {993},
    number = {1},
    pages = {L11},
    author = {{XRISM Collaboration} and Audard, Marc and Awaki, Hisamitsu and Ballhausen, Ralf and Bamba, Aya and Behar, Ehud and Boissay-Malaquin, Rozenn and Brenneman, Laura and Brown, Gregory V. and Corrales, Lia and Costantini, Elisa and Cumbee, Renata and Diaz Trigo, Maria and Done, Chris and Dotani, Tadayasu and Ebisawa, Ken and Eckart, Megan E. and Eckert, Dominique and Eguchi, Satoshi and Enoto, Teruaki and Ezoe, Yuichiro and Foster, Adam and Fujimoto, Ryuichi and Fujita, Yutaka and Fukazawa, Yasushi and Fukushima, Kotaro and Furuzawa, Akihiro and Gallo, Luigi and García, Javier A. and Gu, Liyi and Guainazzi, Matteo and Hagino, Kouichi and Hamaguchi, Kenji and Hatsukade, Isamu and Hayashi, Katsuhiro and Hayashi, Takayuki and Hell, Natalie and Hodges-Kluck, Edmund and Hornschemeier, Ann and Ichinohe, Yuto and Ishi, Daiki and Ishida, Manabu and Ishikawa, Kumi and Ishisaki, Yoshitaka and Kaastra, Jelle and Kallman, Timothy and Kara, Erin and Katsuda, Satoru and Kanemaru, Yoshiaki and Kelley, Richard and Kilbourne, Caroline and Kitamoto, Shunji and Kobayashi, Shogo and Kohmura, Takayoshi and Kubota, Aya and Leutenegger, Maurice and Loewenstein, Michael and Maeda, Yoshitomo and Markevitch, Maxim and Matsumoto, Hironori and Matsushita, Kyoko and McCammon, Dan and McNamara, Brian and Mernier, François and Miller, Eric D. and Miller, Jon M. and Mitsuishi, Ikuyuki and Mizumoto, Misaki and Mizuno, Tsunefumi and Mori, Koji and Mukai, Koji and Murakami, Hiroshi and Mushotzky, Richard and Nakajima, Hiroshi and Nakazawa, Kazuhiro and Ness, Jan-Uwe and Nobukawa, Kumiko and Nobukawa, Masayoshi and Noda, Hirofumi and Odaka, Hirokazu and Ogawa, Shoji and Ogorzałek, Anna and Okajima, Takashi and Ota, Naomi and Paltani, Stephane and Petre, Robert and Plucinsky, Paul and Porter, Frederick S. and Pottschmidt, Katja and Sato, Kosuke and Sato, Toshiki and Sawada, Makoto and Seta, Hiromi and Shidatsu, Megumi and Simionescu, Aurora and Smith, Randall and Suzuki, Hiromasa and Szymkowiak, Andrew and Takahashi, Hiromitsu and Takeo, Mai and Tamagawa, Toru and Tamura, Keisuke and Tanaka, Takaaki and Tanimoto, Atsushi and Tashiro, Makoto and Terada, Yukikatsu and Terashima, Yuichi and Tsuboi, Yohko and Tsujimoto, Masahiro and Tsunemi, Hiroshi and Tsuru, Takeshi and Tümer, Ayşegül and Uchida, Hiroyuki and Uchida, Nagomi and Uchida, Yuusuke and Uchiyama, Hideki and Ueda, Shutaro and Ueda, Yoshihiro and Uno, Shinichiro and Vink, Jacco and Watanabe, Shin and Williams, Brian J. and Yamada, Satoshi and Yamada, Shinya and Yamaguchi, Hiroya and Yamaoka, Kazutaka and Yamasaki, Noriko and Yamauchi, Makoto and Yamauchi, Shigeo and Yaqoob, Tahir and Yoneyama, Tomokage and Yoshida, Tessei and Yukita, Mihoko and Zhuravleva, Irina and Cui, Weiguang and Ettori, Stefano and Grayson, Skylar and Heinrich, Annie and McCall, Hannah and Nelson, Dylan and Okabe, Nobuhiro and Omiya, Yuki and Sarkar, Arnab and Scannapieco, Evan and Sun, Ming and Tanaka, Keita and Truong, Nhut and Wik, Daniel R. and Zhang, Congyao and ZuHone, John},
    title = {Comparing XRISM Cluster Velocity Dispersions with Predictions from Cosmological Simulations: Are Feedback Models Too Ejective?},
    journal = {The Astrophysical Journal Letters}
}

@article{XRISM_2025_Ophiuchus,
   title={XRISM observation of the Ophiuchus galaxy cluster: Quiescent velocity structure in the dynamically disturbed core},
   volume={77},
   ISSN={2053-051X},
   url={http://dx.doi.org/10.1093/pasj/psaf089},
   DOI={10.1093/pasj/psaf089},
   number={Supplement_1},
   journal={Publications of the Astronomical Society of Japan},
   publisher={Oxford University Press (OUP)},
   author={Fujita, Yutaka and Fukushima, Kotaro and Sato, Kosuke and Fukazawa, Yasushi and Kondo, Marie},
   year={2025},
   month=aug, pages={S270–S275} 
}

@ARTICLE{Zuhone_2015,
       author = {{ZuHone}, J.~A. and {Kunz}, M.~W. and {Markevitch}, M. and {Stone}, J.~M. and {Biffi}, V.},
        title = "{The Effect of Anisotropic Viscosity on Cold Fronts in Galaxy Clusters}",
      journal = {Astrophysical Journal},
         year = 2015,
        month = jan,
       volume = {798},
       number = {2},
          eid = {90},
        pages = {90},
          doi = {10.1088/0004-637X/798/2/90},
archivePrefix = {arXiv},
       eprint = {1406.4031},
 primaryClass = {astro-ph.CO},
       adsurl = {https://ui.adsabs.harvard.edu/abs/2015ApJ...798...90Z}
}

@article{ZuHone_2016,
   title={Cold fronts: probes of plasma astrophysics in galaxy clusters},
   volume={82},
   ISSN={1469-7807},
   url={http://dx.doi.org/10.1017/S0022377816000544},
   DOI={10.1017/s0022377816000544},
   number={3},
   journal={Journal of Plasma Physics},
   publisher={Cambridge University Press (CUP)},
   author={ZuHone, John A. and Roediger, E.},
   year={2016},
   month=jun 
}



\appendix

\section{Damping Rates Sound Wave I} \label{app:gamma_soundwaveI}

Considering small-amplitude plane waves, we can use linear perturbation theory to derive the damping rate of a wave due to viscosity. We can express the waves as
\begin{equation}
    v_x = \delta v \mathrm{e}^{i(kx - \omega t)}
\end{equation}
\begin{equation}
    \rho = \rho_0 + \delta \rho \mathrm{e}^{i(kx - \omega t)}
\end{equation}
\begin{equation}
    p = p_0 + c_s^2\delta \rho \mathrm{e}^{i(kx - \omega t)} \, ,
\end{equation}
with $c_s^2 = (\partial p / \partial \rho)_s$. Replacing in the continuity and momentum (eq. \eqref{eqn:mass_eq} and eq. \eqref{eqn:momentum_eq} respectively), we get
\begin{equation}
    -i \omega \delta \rho + i k \rho_0 \delta v = 0 \, ,
\end{equation}
\begin{equation}
    -i \omega \rho_0 \delta v = -i k c_s^2 \delta \rho + (\nabla \cdot {\bf \Pi})_x \, .
\end{equation}

To get the damping rate in each case, we compute the force due to viscosity in the isotropic, anisotropic parallel, and anisotropic perpendicular cases.

In the isotropic case, the viscous stress tensor is defined as
\begin{equation}
    \boldsymbol{\Pi}_{\rm iso} = \eta \, \left( \nabla {\bf v} + \nabla {\bf v}^{\rm T} - \frac{2}{3} \nabla \cdot {\bf v} \right) \, .
\end{equation}

For a wave propagating in the $\hat{x}$ direction:
\begin{equation}
    \partial_x v_x = ik\delta v, \hspace{2cm} \nabla \cdot v = ik\delta v \, .
    \label{eqn:solution_grad_div}
\end{equation}
Thus, the viscous stress tensor is
\begin{equation}
    \mathbf{\Pi}_{xx} = \eta (2ik\delta v - \frac{2}{3}ik\delta v) = \frac{4}{3}\eta \, i k \delta v \, .
\end{equation}
The viscous force is given by 
\begin{equation}
    (\nabla \cdot \mathbf{\Pi})_x = \partial_x \mathbf{\Pi}_{xx} = \frac{4}{3}\eta \, ik \, \partial_x \delta v = \frac{4}{3} \eta \, i k (i k \delta v) = - \frac{4}{3} \eta k^2 \delta v \, .
    \label{eqn:iso_visc_force}
\end{equation}

Plugging the viscous force in and inserting $\delta \rho = (\rho_0 k / \omega) \delta v$ from the continuity eq. \eqref{eqn:momentum_eq}, we get the dispersion relation
\begin{equation}
    - i \omega \rho_0 \delta v = - i k c_s^2 \left( \frac{\rho_0 k}{\omega} \delta v \right) - \frac{4}{3} \eta k^2 \delta v
\end{equation}
\begin{equation}
    \omega^2 + i \frac{4}{3} \nu k^2 \omega - c_s^2 k^2 = 0 \, ,
\end{equation}
with $\nu = \eta / \rho_0$. Solving for $\omega$, we get
\begin{equation}
    \omega = \pm \sqrt{c_s^2k^2 - \frac{4}{9}\nu^2k^4} - i \frac{2}{3} \nu k^2 \, .
\end{equation}
The damping rate is given by the imaginary solution; therefore
\begin{equation}
    \gamma_{\rm Iso} = \frac{2}{3} \nu k^2 \, .
\end{equation}

For the anisotropic case where the magnetic field is parallel to the velocity gradient, the magnetic field has an $x$-component. $\hat{b}_x = 1$, therefore $\hat{b} \hat{b} : \nabla {\bf v} = \partial_x v_x = i k \delta v$, and the velocity divergence is given by eq. \eqref{eqn:solution_grad_div}. The pressure anisotropy is given by
\begin{equation}
    \Delta p = \eta (3 ik \delta v - ik\delta v) = 2 i \eta k \delta v \, .
\end{equation}
Only the $x$-derivative is non-zero, therefore
\begin{equation}
    \partial_x \Delta p = \partial_x (2 i \eta k \delta v) = 2i\eta k \partial_x v_x = 2 i \eta k (i k \delta v) = -2 \eta k^2 \delta v \, .
\end{equation}
The projection term in the viscous stress tensor needs to have only $x$-component, $\hat{b}_x \hat{b}_x - 1/3 \,\delta_{xx} = 1-1/3 = 2/3$. Thus, the viscous force is given by 
\begin{equation}
    (\nabla \cdot \mathbf{\Pi})_x = \partial_x \Delta p \left( \hat{b}_x \hat{b}_x - \frac{1}{3} \,\delta_{xx} \right) = -\frac{4}{3} \eta k^2 \delta v \, .
\end{equation}
We get the same result as the isotropic case (eq. \eqref{eqn:iso_visc_force}), leading to the same damping rate
\begin{equation}
    \gamma_{\parallel} = \frac{2}{3} \nu k^2 \, ,
\end{equation}
showing that in the direction of the magnetic field, the Braginskii viscosity behaves as the isotropic case.

In the anisotropic case with a perpendicular magnetic field component, the magnetic field has only a $y$-component (or $z$-component). The only non-zero gradient is $\partial_x v_x$. However, the $x$-component of the magnetic field is zero, therefore $\hat{b} \hat{b} : \nabla {\bf v} = 0$, and the pressure anisotropy is driven by the velocity divergence \eqref{eqn:solution_grad_div}, leading to 
\begin{equation}
    \Delta p = -i \eta k \delta v \, .
\end{equation}
Same as in the parallel case, only the $x$-derivative is non-zero
\begin{equation}
    \partial_x (\Delta p) = \partial_x (-i \eta k \delta v) = -i \eta k \partial_x v_x = -i \eta k (i k \delta v) = \eta k^2 \delta v \, . 
\end{equation}
The projection term is $\hat{b}_x \hat{b}_x - 1/3 \,\delta_{xx} = -1/3$. The viscous force is given by 
\begin{equation}
    (\nabla \cdot \mathbf{\Pi})_x = \partial_x \Delta p \left( \hat{b}_x \hat{b}_x - \frac{1}{3} \,\delta_{xx} \right) = -\frac{1}{3} \eta k^2 \delta v \, .
\end{equation}
Thus, the dispersion relation reads
\begin{equation}
    - i \omega \rho_0 \delta v = - i k c_s^2 \left( \frac{\rho_0 k}{\omega} \delta v \right) - \frac{1}{3} \eta k^2 \delta v
\end{equation}
\begin{equation}
    \omega^2 + i \frac{1}{3} \nu k^2 \omega - c_s^2 k^2 = 0 \, ,
\end{equation}
with solution
\begin{equation}
    \omega = \pm \sqrt{c_s^2k^2 - \frac{1}{36}\nu^2k^4} - i \frac{1}{6} \nu k^2 \, .
\end{equation}
The damping rate for the perpendicular case is given by
\begin{equation}
    \gamma_{\perp} = \frac{1}{6} \nu k^2 \, .
\end{equation}

\section{Damping Rates Sound Wave II} \label{app:damping_ratesII}

In this case, we have a wave initialized along the $\hat{y}$ direction, propagating in the $\hat{x}$ direction. Thus, since $\delta v_x = 0$ and $\delta v_y \neq 0$, the spacial derivatives are given by
\begin{equation}
    \partial_xv_x = ik\delta v_x = 0 \, , \hspace{0.6cm} \partial_x v_y = i k \delta v_y \, , \hspace{0.6cm} \nabla \cdot v = i k \delta v_x = 0 \, .
\end{equation}

The isotropic viscous stress tensor is
\begin{equation}
    \mathbf{\Pi}_{xy} = \mathbf{\Pi}_{yx} = \eta \,ik\delta v_y \, ,
\end{equation}
and the viscous force
\begin{equation}
    (\nabla \cdot \mathbf{\Pi})_x = \partial_x \mathbf{\Pi}_{xy} = i \eta k \, \partial_x \delta v_y = -\eta k^2 \delta v_y \, .
\end{equation}
Since there is no pressure gradient in the $\hat{y}$ direction, the momentum eq. \eqref{eqn:momentum_eq} becomes
\begin{equation}
    -i \omega \rho_0 \delta v_y = (\nabla \cdot \mathbf{\Pi})_x = -\eta k^2 \delta v_y \, .
\end{equation}
Solving for $\omega$,
\begin{equation}
    \omega = i \nu k^2 \, ,
\end{equation}
leads to a viscous damping rate
\begin{equation}
    \gamma_{\rm Iso} = \nu k^2 \, .
\end{equation}

The derivation for the anisotropic case can be found in \citet{Berlok_2019}, which leads to a damping rate of
\begin{equation}
    \gamma_{\rm Aniso} = \frac{5}{6} \nu k^2 \, .
\end{equation}

\section{Sound Wave II result with the hydro solver on} \label{app:soundwaveII}

To isolate the effects of anisotropic viscosity, the simulations presented in \S \ref{sec:soundwave_II} were performed with the hydro solver off. If we keep the hydro solver on, the results differ slightly from the ones presented in \S \ref{sec:soundwave_II}.

\begin{figure}
    \centering
	\includegraphics[width=\columnwidth]{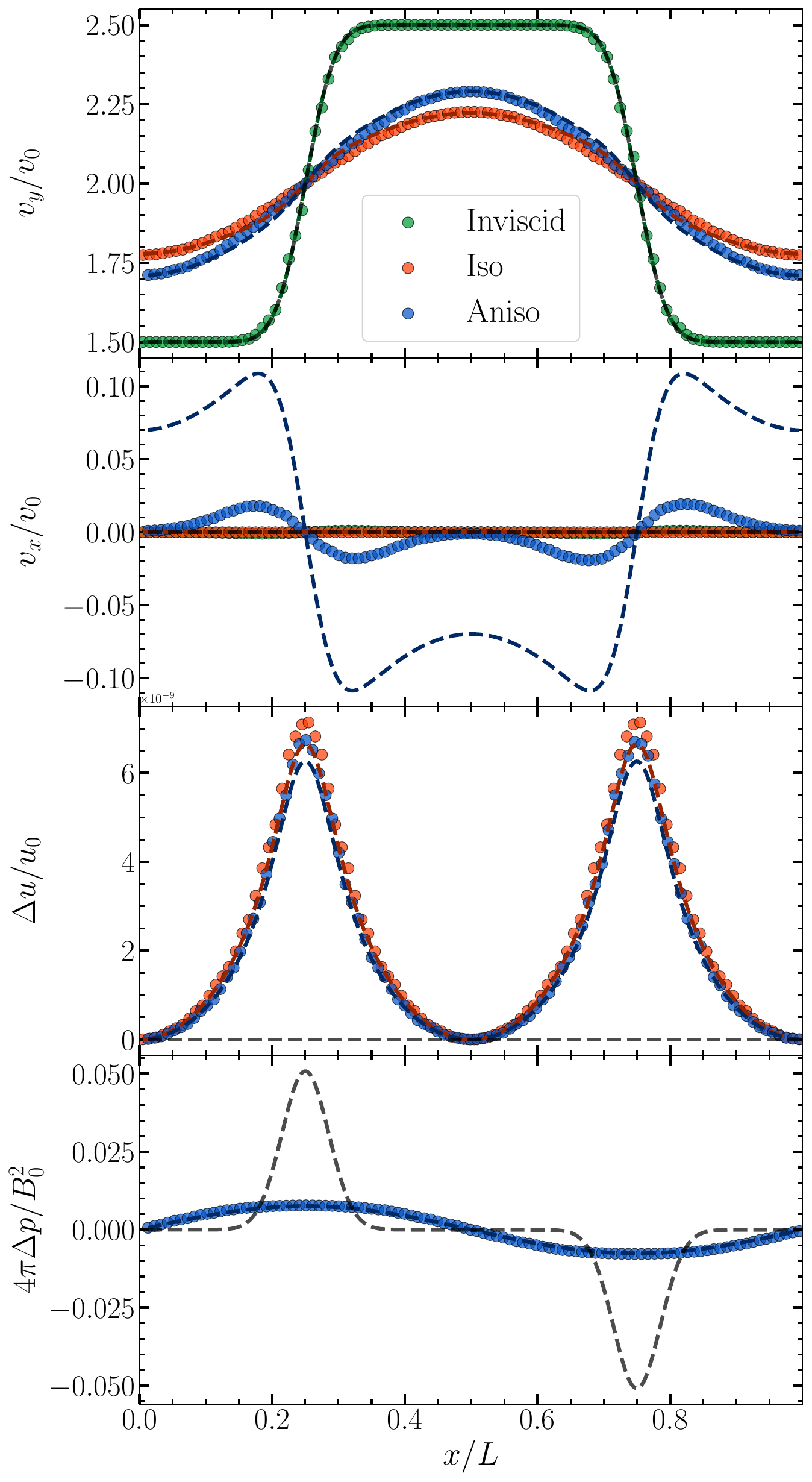}
    \caption{Results of the sound wave of test \ref{sec:soundwave_II} with the hydro solver on after $ct/L=1$. {\it From top to bottom}: $v_y$ profile; $v_x$ profile;  cumulative viscous heating profile; pressure anisotropy profile.}
    \label{fig:soundwave_II_hydro}
\end{figure}

Fig.~\ref{fig:soundwave_II_hydro} shows the results with the hydro solver on. Although the profile of $v_y$ matches the results with the hydro solver off (first panel), the development of acoustic waves injects a sine-phase component into $v_x$. This sine component has nodes at $x/L = 0, 0.5, 1$, a similar result to the one found in \citet{Hopkins_2017} (second panel), deviating from the analytical solution. The non-viscous wave is not damped at all, keeping the initial profile, while the isotropic case is damped, matching the damping rate predicted in appendix~\ref{app:damping_ratesII}. In the isotropic case, there is no kinetic energy conversion from $v_y$ to $v_x$, since there is no pressure anisotropy.

The viscous heating is shown in the third panel, where the dissipation due to isotropic viscosity is larger than the one due to anisotropic viscosity (see appendix \ref{app:damping_ratesII} for the derivation of the different dissipation rates). The numerical data lies $\sim 5\%$ above the theoretical expected result, which is the result of having the hydro solver on. Finally, the pressure anisotropy looks the same as the case with the hydro solver off, matching the analytical solution.

\section{Alfvén Wave with Isotropic Viscosity}\label{app:alfven_iso_visc}

We consider small-amplitude, linearly polarized shear–Alfvén perturbations in a uniform medium with background magnetic field $\mathbf{B} = B_0\hat{\mathbf{x}}$. As above, perturbations depend only on $x$ and are transverse, so $\mathbf{v}_{\perp} \perp \hat{x}$ and $\mathbf{B}_{\perp} \perp \hat{x}$. Because the mode is incompressible, $\nabla \cdot \mathbf{v}=0$. Adopting the same plane–wave form as in \S\ref{app:gamma_soundwaveI}, we write
\begin{equation}
    v_\perp = \delta v_\perp \mathrm{e}^{i(kx-\omega t)}, \hspace{2cm}
    B_\perp = \delta B_\perp \, \mathrm{e}^{i(kx-\omega t)},
\end{equation}
with $\delta v_x=\delta B_x=0$. Using the momentum eq. \eqref{eqn:momentum_eq} (with the isotropic viscous stress already defined in \S\ref{app:gamma_soundwaveI}) together with the induction equation, we obtain
\begin{equation}
    (-i\omega + \nu k^2 ) \delta v_\perp = i k\frac{v_A^2}{B_0}\delta B_\perp \,,
    \qquad
    -i\omega \delta{B}_\perp = i k B_0 \delta v_\perp\, ,
\end{equation}
where $v_A \equiv B_0/\sqrt{4\pi\rho_0}$. This gives us the viscous shear–Alfvén dispersion relation
\begin{equation}
    \omega^2 + i\nu k^2\omega - k^2 v_A^2 = 0\,.
\end{equation}
Solving for $\omega$,
\begin{equation}
    \omega_\pm = \pm\Omega - i\gamma_{\rm Iso}\,
    \qquad
    \gamma_{\rm Iso} = \frac{\nu k^2}{2}\,,
    \qquad
    \Omega = \sqrt{k^2 v_A^2 - \gamma_{\rm Iso}^2}\,.
\end{equation}
Thus, isotropic viscosity introduces an exponential amplitude decay at rate $\gamma=\nu k^2/2$. In the inviscid case where $\gamma_{\rm Iso} = 0$, $\Omega = k v_A = \omega_0$.

The inviscid evolution in \S \ref{sec:lin_alfven} is given in eq. \eqref{eqn:evo_lin_alfven_B} and \eqref{eqn:evo_lin_alfven_v}, 
\begin{equation}
    \delta B_\perp(x,t) = \mathcal{B}(t)\cos(kx)\,,
    \label{eqn:B_evo_app}
\end{equation}
\begin{equation}
    \delta v_\perp(x,t) = \mathcal{V}(t)\sin(kx)\, ,
    \label{eqn:v_evo_app}
\end{equation}
where $\mathcal{B}(t)$ and $\mathcal{V}(t)$ are the time-dependent amplitudes.

The linearized momentum and induction equations with isotropic viscosity (and incompressible geometry) are
\begin{equation}
\rho_0\partial_t \delta v_\perp = \frac{v_A^2}{B_0} \partial_x \delta B_\perp + \rho_0 \nu \partial_x^2 \delta v_\perp \, ,
\end{equation}
\begin{equation}
    \partial_t \delta B_\perp = B_0\partial_x \delta v_\perp \,.
\end{equation}
The derivatives of eq. \eqref{eqn:B_evo_app} and \eqref{eqn:v_evo_app} are given by
\begin{equation}
    \partial_x \delta B_\perp = -k\mathcal{B}(t) \sin(kx)\,,
\end{equation}
\begin{equation}
    \partial_t \delta v_\perp = \mathcal{V}’(t) \sin(kx)\,,
\end{equation}
\begin{equation}
    \partial_x \delta v_\perp = k \,\mathcal{V}(t) \cos(kx)\,,
\end{equation}
\begin{equation}
    \partial_x^2 \delta v_\perp = -k^2 \,\mathcal{V}(t)\sin(kx)\,,
\end{equation}
\begin{equation}
    \partial_t \delta B_\perp = \mathcal{B}’(t) \cos(kx)\,.
\end{equation}

Differentiating the induction equation and substituting the momentum equation gives us the equations of a damped oscillator:
\begin{equation}
    \mathcal{B}'' + \nu k^2\mathcal{B}' + \omega_0^2\mathcal{B} = 0\,,
    \hspace{2cm}
    \mathcal{V} = \frac{\mathcal{B}'}{B_0 k}\,,
\end{equation}
with $\omega_0 \equiv k v_A$.

Considering the initial conditions in \S \ref{sec:lin_alfven}, $\mathcal{B}(0)=\delta B_0$ and $\mathcal{V}(0)=0$ (so $\mathcal{B}'(0)=0$), the viscous generalization is
\begin{equation}
    \mathcal{B}(t) = -B_0 A \mathrm{e}^{-\gamma t} \left[\cos(\Omega t)+\frac{\gamma}{\Omega}\sin(\Omega t)\right]\,,
\end{equation}
\begin{equation}
    \mathcal{V}(t) = \frac{\omega_0}{\Omega} v_A A \,\mathrm{e}^{-\gamma t}\sin(\Omega t)\,.
\end{equation}
In the limit $\nu\to 0$, these reduce exactly to eqs. \eqref{eqn:evo_lin_alfven_B} and \eqref{eqn:evo_lin_alfven_v}.

\section{Scaling estimate for viscosity in the ICM} \label{app:ratio_visc_icm}

To test the Braginskii viscosity implementation under realistic conditions in \S \ref{sec:circ_alfven}, we estimate the $\nu / cL$ expected in the ICM. To do so, we first express $\nu$ as a function of the temperature and density using that $\nu = \eta / \rho$ and the definition of the Spitzer viscosity coefficient, eq. \eqref{eqn:viscosity}. The soundspeed can also be expressed in terms of the temperature
\begin{equation}
    c = \sqrt{\frac{\gamma k_{\rm B} T}{\mu m_{\rm p}}} \, .
\end{equation}
Thus, the ratio $\nu / cL$ can be expressed as a function of the density, temperature, and scale as:
\begin{equation}
    \frac{\nu}{c L} = 0.406 \, \frac{k_B^2} {\mu_e e^4 \ln \Lambda} \sqrt{\frac{\mu}{\gamma}} \, \frac{T^2}{n_e L} \approx 1.953 \times10^3 \, \frac{T^2}{n_e L} \, .
    \label{eqn:nu_c_L}
\end{equation}

\begin{figure}
    \centering
	\includegraphics[width=\columnwidth]{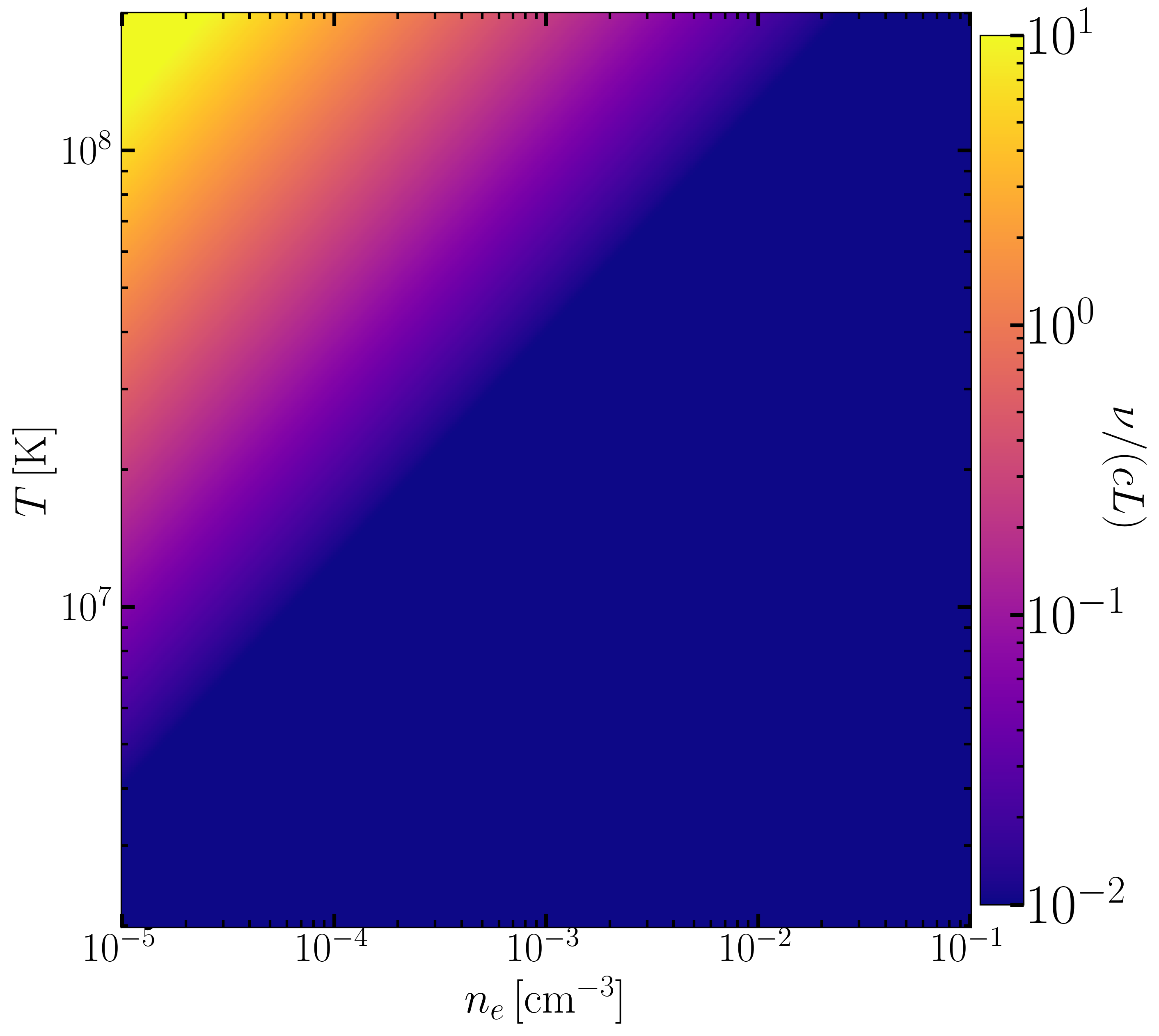}
    \caption{Colormap showing the solution of eq. \eqref{eqn:nu_c_L}, where we have estimated the expected $\nu /cL$ ratio in the ICM.}
    \label{fig:ratio_nu_cs_L}
\end{figure}

We assume $L = 100$~kpc, and we take a range of temperatures and densities typical in the ICM. Fig.~\ref{fig:ratio_nu_cs_L} shows the solution of eq. \eqref{eqn:nu_c_L}. In very hot and very diffuse plasmas, $\nu / cL \approx 10$. Therefore, we consider this extreme case in our setup to study stability of the Braginskii implementation in \S \ref{sec:circ_alfven}.

\section{KHI Growth when $\mathbf{B} = B \hat{z}$} \label{app:growth_rate_Bz}

\begin{figure}
    \centering
	\includegraphics[width=\columnwidth]{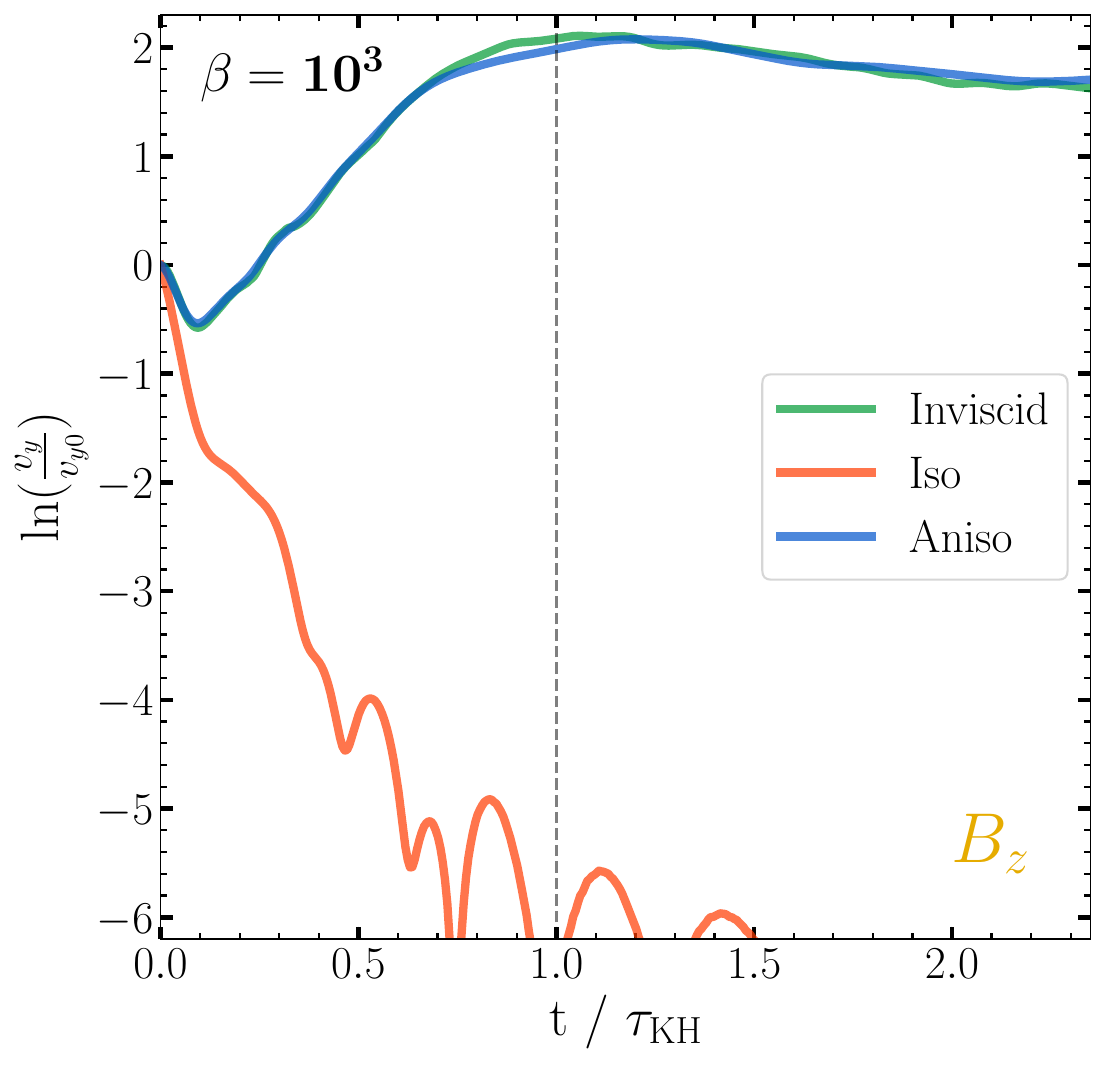}
    \caption{Growth rate of the KHI of a setup with $\beta = 10^3$ and initial magnetic field ${\bf B} = B\hat{z}$ for different viscosity treatments: inviscid (green), isotropic (red), and anisotropic (blue).}
    \label{fig:braginskii_McNally_1e3_Bz}
\end{figure}

Fig.~\ref{fig:braginskii_McNally_1e3_Bz} shows the growth rate of the instability for the different cases when ${\bf B} = B\hat{z}$. The isotropic case strongly suppresses the growth of the KHI, while the inviscid and anisotropic cases behave very similarly, allowing the growth of the instability. This kind of behavior is expected, since $B_z \perp \nabla v$, therefore, the effect of the anisotropic viscosity should be almost zero. In contrast to the case with ${\bf B} = B\hat{x}$, in this case the growth of the KHI does not bend the magnetic field lines, thus there is never a $y$-component of the magnetic field, and the anisotropic viscosity is always close to zero. The case including plasma microinstability limiters is not included, since no effect is expected.


\bsp	
\label{lastpage}
\end{document}